%% file: main.tex
\documentclass{article}
\usepackage{amsmath, amsthm, amssymb}
\usepackage{graphicx}
\usepackage{verbatim}
\usepackage{natbib}
\usepackage{caption}
\usepackage{subcaption}
\usepackage{fancyvrb}
\usepackage{enumerate}
\usepackage{relsize}
\usepackage{multirow}
\usepackage[section]{placeins}

\usepackage{hyperref}
\usepackage[margin=1in]{geometry}
\hypersetup{colorlinks,citecolor=blue,urlcolor=blue,linkcolor=blue}

\usepackage{stefan_tex}
\graphicspath{{figs/}}

\theoremstyle{plain}
\newtheorem{prop}{Proposition}

\newtheorem{lemm}[prop]{Lemma}
\newtheorem{theo}[prop]{Theorem}

\theoremstyle{definition}
\newtheorem{exam}{Example}
\newtheorem{defi}[exam]{Definition}
\newtheorem{assumption}{Assumption}

\theoremstyle{remark}

\newtheorem{rema}[prop]{Remark}

\author{Lihua Lei \footnote{Authors listed alphabetically.}\\
  \texttt{lihualei@stanford.edu}
  \and 
  Roshni Sahoo\\
  \texttt{rsahoo@stanford.edu}
  \and 
  Stefan Wager\\
  \texttt{swager@stanford.edu}
}

\title{Policy Learning under Biased Sample Selection}

\date{Stanford University}

\begin{document}

\maketitle

\begin{abstract}
Practitioners often use data from a randomized controlled trial to learn a treatment assignment policy that can be deployed on a target population. A recurring concern in doing so is that, even if the randomized trial was well-executed (i.e., internal validity holds), the study participants may not represent a random sample of the target population (i.e., external validity fails) – and this may lead to policies that perform suboptimally on the target population. We consider a model where observable attributes can impact sample selection probabilities arbitrarily but the effect of unobservable attributes is bounded by a constant, and we aim to learn policies with the best possible performance guarantees that hold under any sampling bias of this type. In particular, we derive the partial identification result for the worst-case welfare in the presence of sampling bias and show that the optimal max-min, max-min gain, and minimax regret policies depend on both the conditional average treatment effect (CATE) and the conditional value-at-risk (CVaR) of potential outcomes given covariates. To avoid finite-sample inefficiencies of plug-in estimates, we further provide an end-to-end procedure for learning the optimal max-min and max-min gain policies that does not require the separate estimation of nuisance parameters.
\end{abstract}



\begin{section}{Introduction}
\label{sec:intro}

\input{01a-intro}
\begin{subsection}{Related Work}
\input{01b-related-work}
\end{subsection}
\end{section}

\begin{section}{Identification}
\label{sec:identification}
\input{03a-identification}

\end{section}

\begin{section}{Equivalence}
\label{sec:equivalence}
\input{03b-equivalence}
\end{section}

\begin{section}{End-to-End Policy Learning via RU Regression}
\label{sec:learning}
\input{03c-learning}

\end{section}

\begin{section}{Experiments}
\label{sec:experiments}
\input{04-experiments}

\end{section}

\bibliographystyle{plainnat} 
\bibliography{ref.bib}

\appendix

\begin{section}{Additional Experimental Results}
\label{sec:add_exp}
\input{04b-additional-results}

\end{section}

\begin{section}{Experimental Details}
\label{sec:exp_details}
\input{05-experiment-details}
\end{section}

\begin{section}{Standard Results}
\label{sec:standard_results}
\input{06-standard-results}
\end{section}

\begin{section}{Proofs of Main Results}
\label{sec:proof}
\input{07-main-proofs}

\end{section}

\begin{section}{Proofs of Technical Results}
\label{sec:technical_proof}
\input{08-technical-proofs}
\end{section}

\end{document}

%% file: 01a-intro.tex
Practitioners often use data from a randomized controlled trial (RCT) to a learn treatment assignment policy that can be deployed on a target population. Formally, in the study population, each unit $i$ has covariates $X_{i} \in \mathcal{X}$ and potential outcomes $Y_{i}(0), Y_{i}(1) \in \mathcal{Y}$ under control and treatment, respectively, that are distributed according to a study potential outcome distribution $P$. After administering the trial, the practitioner has access to samples $(X_{i}, Y_{i}, W_{i}) \sim P_{\text{obs}}$, where $P_{\text{obs}}$ is the observed data distribution that is consistent with $P$ under an internally-valid RCT, meaning that the treatments $W_{i} \in \{0, 1\}$ are randomly assigned and the observed outcome $Y_{i}$ equals $Y_{i}(W_{i}),$ the potential outcome under the prescribed treatment.

A common criticism of RCTs is that they lack external validity. External validity captures the extent to which conclusions drawn from the RCT study population can generalize to a broader population or other target populations. An RCT may lack external validity for a variety of reasons. For example, \cite{bell2016estimates} demonstrates that non-random site selection in the evaluation of the Reading First educational program would have led to lower impact estimates than representative site selection. In addition, \citet{wang2018efficacy} discuss that randomized trials for measuring the effects of anti-depressants rely on volunteers to opt in to participate, so the findings of these studies may not apply to non-volunteers. In addition, the time lag between data collection and deployment may cause the target population to differ from the study population due to temporal shift.

Standard policy learning approaches ignore the concern that the randomized control trial may lack external validity. These methods use data from $P_{\text{obs}}$ to learn the policy that maximizes the mean outcome over the study population \citep{athey2021policy,bhattacharya2012inferring, kitagawa2018should, manski2004statistical}:
\begin{equation} \label{eq:standard} \pi^{*} = \argmax_{\pi \in \Pi} \EE[P]{Y(\pi(X))},\end{equation}
where $\pi: \mathcal{X}\mapsto \{0, 1\}$ denotes a treatment assignment policy and $\Pi$ denote a policy class. The implicit assumption of these methods is that the target population is the same as the study population. The learned policy in \eqref{eq:standard} does not have performance guarantees for target potential outcome distributions $Q$ that differ from $P$.

In this work, we aim to identify and learn policies that are robust to certain failures of external validity, namely sampling bias in the selection of study participants. As before, we assume that the practitioner has access to data from $P_{\text{obs}}$, but we instead aim to learn a policy $\pi$ that has performance guarantees on an unknown target distribution $Q$, while the study potential outcome distribution $P$ may be biased relative to $Q$ due to sampling bias.
Following \citet{manski2003partial}, we model sample selection using a binary selection indicator.
We quantify the strength of bias in sample selection via the $\Gamma$-biased sampling model
\citep{aronow2013interval, nie2021covariate, sahoo2022learning}.
This model allows the probability of sample selection for each unit to depend arbitrarily on covariates but only a bounded amount on unobservables. The parameter $\Gamma \geq 1$ captures the strength of the sampling bias due to unobservables, where larger values of $\Gamma$ permit larger amounts of bias.
Note that, when $\Gamma=1$, this model corresponds to ``unconfounded sample selection,'' which is studied in the literature on generalizability \citep[e.g.,][]{stuart2011use, tipton2013improving, tipton2014generalizable}.

\begin{defi}
\label{def:sampling_bias}
Let $\Gamma \geq 1$. For any pair of distributions $P$ and $Q$ over $(X, Y(0), Y(1))$, we say that
$Q$ can generate $P$ under $\Gamma$-biased sampling if there exists a distribution $\tilde{Q}$
over $(X, Y(0), Y(1), S)$, where $S \in \{0, 1\}$ is a ``selection indicator'' that satisfies the
following properties: The $(X,Y(0), Y(1))$-marginal of $\tilde{Q}$ is equal to $Q$,
the $(X,Y(0), Y(1))$-marginal of $\tilde{Q}$ conditionally on $S = 1$ is equal to $P$, and
\begin{equation}
\label{eq:sampling_bias_intro}
\frac{\PP[\tilde{Q}]{S=1 \mid X=x, Y(0)=y_{0}, Y(1)=y_{1}}}{\PP[\tilde{Q}]{S=1 \mid X=x}} \in [\Gamma^{-1}, \Gamma] \quad \forall x \in \mathcal{X}, y_{0}, y_{1} \in \mathcal{Y}.
\end{equation} 
\end{defi}

\begin{figure}
\centering
\includegraphics[width=0.75\textwidth]{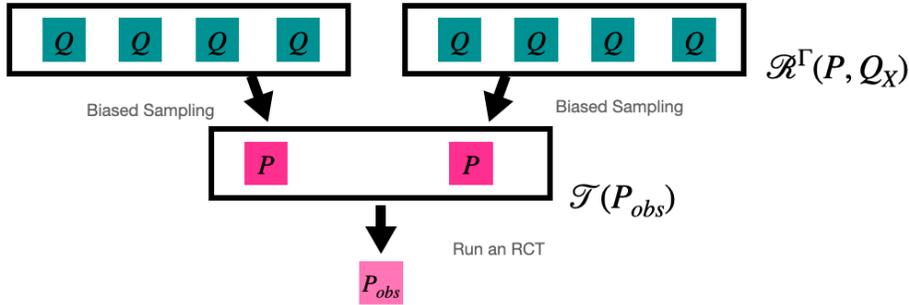}
\caption{We visualize the set of plausible target distributions $\mathcal{S}^{\Gamma}(P_{\text{obs}}, Q_{X})$. The set $\mathcal{T}(P_{\text{obs}})$ is the set of study potential outcome distributions $P$ that are consistent with $P_{\text{obs}}$ under an internally-valid RCT. The set $\mathcal{R}^{\Gamma}(P, Q_{X})$ is the set of target potential outcome distributions that can generate $P$ under $\Gamma$-biased sampling and have covariate distribution $Q_{X}$.}
\label{fig:challenges}
\end{figure}

There are two key challenges when learning policies that are robust to biased sample selection. The first challenge is the ``missing data problem'' that arises in the standard policy learning setting. In a randomized controlled trial, units are assigned to either control or treatment, so $Y_{i}(0), Y_{i}(1)$ are not simultaneously observed for any unit. Although we can identify the marginal distributions of the study potential outcome distribution $P_{X, Y(0)}, P_{X, Y(1)}$ from $P_{\text{obs}}$, the joint distribution $P$ cannot be identified from $P_{\text{obs}}$. So, the true study potential outcome distribution is unknown, and there are many potential outcome distributions $P$ that yield $P_{\text{obs}}$ under an internally-valid RCT. Let $\mathcal{T}(P_{\text{obs}})$ be the set of study potential outcome distributions that yield $P_{\text{obs}}$ under an internally-valid RCT, i.e.,
\[\mathcal{T}(P_{\text{obs}}) = \{P: P_{X, Y(0)} = P_{\text{obs}, X, Y \mid W=0}, P_{X, Y(1)} = P_{\text{obs}, X, Y \mid W=1} \}.\]
The second challenge arises from biased sampling. We note that the true target potential outcome distribution is also unknown, and given a particular study potential outcome distribution $P$, there are many possible target distributions $Q$ that can generate $P$ under $\Gamma$-biased sampling for $\Gamma > 1$. Let $\mathcal{R}^{\Gamma}(P, Q_{X})$ be the set of potential outcome distributions $Q$ that have covariate distribution $Q_{X}$ and can generate $P$ via $\Gamma$-biased sampling. Thus, we can define $\mathcal{S}^{\Gamma}(P_{\text{obs}}, Q_{X})$ to be the robustness set, the set of all plausible target potential outcome distributions consistent with $P_{\text{obs}}$ under $\Gamma$-biased sampling and an internally-valid RCT:
\[ \mathcal{S}^{\Gamma}(P_{\text{obs}}, Q_{X}) = \{ Q \mid Q \in \mathcal{R}^{\Gamma}(P, Q_{X}) \text{ where } P \in \mathcal{T}(P_{\text{obs}}) \}.\]
This set is visualized in Figure \ref{fig:challenges}.

To obtain performance guarantees on the unknown target population, we apply the distributionally robust optimization (DRO) framework \citep{ben2013robust} to learn a policy that is robust to all plausible target distributions in the robustness set. As in \citet{manski2011choosing}, we consider a number of different ways to quantify good performance of a policy that account for ambiguity under partial identification.
The max-min objective considers worst-case welfare over the partially identified set, and the max-min policy is
\begin{equation} 
\label{eq:intro_maxmin}
\pi_{\Gamma, \text{maxmin}}^{*}  \in \argmax_{\pi  \in \Pi} \inf_{Q \in \mathcal{S}^{\Gamma}(P_{\text{obs}}, Q_{X})} \EE[Q]{Y(\pi(X))}.
\end{equation}
Optimizing max-min guarantees is conceptually simple; however, in many cases, the max-min objective is viewed as too pessimistic \citep{manski2011choosing,savage1951theory}.
To avoid this pessimism we also consider two other objectives. When a natural baseline
policy $\pi_{0}$ is available (e.g., there is a clear status quo), we can optimize max-min improvements
over the baseline; the max-min gain policy is
\begin{equation}
\label{eq:intro_gain}
\pi_{\Gamma, \text{gain}}^{*}  \in \argmax_{\pi  \in \Pi} \inf_{Q \in \mathcal{S}^{\Gamma}(P_{\text{obs}}, Q_{X})} \EE[Q]{Y(\pi(X))} - \EE[Q]{Y(\pi_{0}(X))}.
\end{equation}
As discussed in \citet{kallus2021minimax}, when the status quo policy is already considered to be a
reasonably good decision rule, then the max-min gain approach may be considered desirable in that it
provides ``safe'' improvements over the status quo, i.e., we will remain with the status quo unless
the data lets us unequivocally prefer another action.
Finally, we also consider a minimax regret criterion; the minimax regret policy is
\begin{equation}
\label{eq:intro_regret}
\pi_{\Gamma, \text{regret}}^{*} \in \argmin_{\pi  \in \Pi} \sup_{Q \in \mathcal{S}^{\Gamma}(P_{\text{obs}}, Q_{X})} R_{Q}(\pi(X)),
\end{equation}
where regret under distribution $Q$ is the gap between the mean outcome of a policy and that of the best-performing policy under distribution $Q$:
\begin{equation}
R_{Q}(\pi(X)) = \sup_{\pi' \in \Pi} \EE[Q]{Y(\pi'(X))} -  \EE[Q]{Y(\pi(X))}.
\end{equation}
Several authors have argued in favor of minimax regret is a relevant criterion for guiding
action under ambiguity \citep{manski2011choosing,savage1951theory}; this criterion does not
suffer from pessimistic failure modes of the max-min criterion, and also can be used in the
absence of a strong baseline policy.

Our work has two main contributions. In our first contribution, we demonstrate that the optimal policies defined in \eqref{eq:intro_maxmin}, \eqref{eq:intro_gain}, and \eqref{eq:intro_regret} are identifiable under $P_{\text{obs}}$ and give closed-form expressions for these policies when the policy class $\Pi$ consists of deterministic, unconstrained, binary-valued functions. In our second contribution, we give end-to-end procedures for learning the optimal max-min and max-min gain policies using data from $P_{\text{obs}}$ that do not require separate estimation nuisance parameters.

In Section \ref{sec:identification}, we demonstrate that when $\Pi$ consists of deterministic, unconstrained, binary-valued functions, the optimal policy under the max-min, max-min gain, and minimax regret objectives are identifiable using only data from $P_{\text{obs}}$, and we give closed-form expressions for these policies. In Section \ref{sec:equivalence}, we give an equivalence result that is essential for demonstrating that our learning procedures yield the optimal policies. The result shows that robust optimization of a potential outcome value function $v^{*}(\pi(X); X, Y(0), Y(1))$ over $\mathcal{S}^{\Gamma}(P_{\text{obs}}, Q_{X})$ is equivalent to robust optimization of a observed data value function $v(\pi(X); X, Y, W)$ over a robustness set of observed data distributions when the mean of the observed data value function under $P_{\text{obs}}$ is equal to the mean of the potential outcome value function under any potential outcome distribution $P \in \mathcal{T}(P_{\text{obs}})$. In Section \ref{sec:learning}, we demonstrate that the robust optimization of an observed data value function over a robustness set of observed data distributions can be solved with an extension of Rockafellar-Uryasev (RU) Regression \citep{sahoo2022learning}. We combine this result with the equivalence result from the previous section to give a procedure for learning the optimal max-min and max-min gain policies. In Section \ref{sec:experiments}, we evaluate our learning procedure empirically in simulations and a semi-synthetic experiment with the voting dataset of \citet{gerber2008social}.






%% file: 01b-related-work.tex
The study of optimal treatment allocation has received attention in economics \citep{athey2021policy, bhattacharya2012inferring, kitagawa2018should, manski2004statistical}, statistics \citep{qian2011performance, zhao2012estimating}, and computer science \citep{swaminathan2015batch}. The standard setting in this literature does not address the concern that the study population may differ from the target population of interest. The distinctive feature of our setting is that we permit the study potential outcome distribution $P$ to differ from the target potential outcome distribution $Q$ in the sense of $\Gamma$-biased sampling (Definition \ref{def:sampling_bias}). \citet{sahoo2022learning} introduces the $\Gamma$-biased sampling model for the supervised learning setting and focuses on statistical and algorithmic considerations for learning robust regression models. In this work, we consider optimal treatment choice using the potential outcomes framework and identify optimal policies under various robust objectives, including minimax regret.

Our contribution of policy learning under biased sample selection is part of the literature on policy learning under partial identification \citep{adjaho2022externally,christensen2022optimal,ben2021safe, hansen2001robust,hatt2022generalizing,higbee2022policy, kallus2022doubly, kallus2021minimax, mufactored, manski2000identification, manski2007minimax, si2020distributional, watson2016approximate}. The main challenge in this area is that the optimal policy is not identifiable because it depends quantities that are unknown, such as missing outcome data \citep{ben2021safe,higbee2022policy,manski2007minimax}, unknown propensity weights \citep{kallus2021minimax}, or an unknown target distribution \citep{adjaho2022externally, hatt2022generalizing, si2020distributional,mufactored}. These works take a robust optimization approach to solve this problem; they define a plausible set of values for the unknown quantities and find a policy that performs well when the unknown quantities take on their worst-case values. Our work is in line with these previous works because we define a set of plausible values for target potential outcome distributions that is consistent with our observed data and the assumed model of sampling bias and identify policies that are robust to the worst-case target distribution. 

Of this literature, our work is most related to \citet{adjaho2022externally} and \citet{si2020distributional}. Similar to our work, they use a robust optimization approach to develop policies that have welfare guarantees for target populations that may differ from the study population. Nonetheless, the robustness set we consider in this paper has a different form than the other works (e.g., Wasserstein balls \citep{adjaho2022externally} and KL-divergence balls \citep{si2020distributional} about the study distribution). Unlike \citet{adjaho2022externally} and \citet{si2020distributional} which only discuss the max-min policy, we are able to derive the closed-form expressions for the max-min gain policy and the minimax regret policy that are arguably more relevant in practice \citep{manski2000identification}. 
 
Another difference between our work and that of \citet{adjaho2022externally} and \citet{si2020distributional} is that they both define their robustness set by placing constraints on the joint distribution over $X, Y(0), Y(1)$, while the robustness sets that we consider place constraints on the conditional distribution $Y(0), Y(1) \mid X$. The approach of constraining the joint distribution over outcomes and covariates is overly conservative for a number of reasons. First, we typically observe the shift in covariate distribution at test-time ($Q_{X}$ is observed at test-time), so it is often not necessary to protect against arbitary covariate shifts. Second, when the policy class $\Pi$ consists of unconstrained binary-valued functions, the optimal policy solves the optimal treatment assignment problem for every $x \in \mathcal{X}$, so the covariate shift does not impact the optimal policy. Third, empirical evaluations have revealed that constraining the joint distribution over outcomes and covariates can yield conservative results compared to handling covariate shift and conditional shift separately \citep{mufactored}.

%% file: 03a-identification.tex
First, we define the notation and setup of our policy learning problem and define the set of target distributions that we aim to be robust to. Second, we define various robustness criteria that we will consider in this work. Finally, we identify the optimal policies under each robustness criterion and discuss connections between them.

\begin{subsection}{Robustness Set}
  We consider the following policy learning problem. In our study population, each unit $i$ is associated with covariates $X_{i} \in \mathcal{X}$ and potential outcomes $Y_{i}(0), Y_{i}(1) \in \mathcal{Y}$ which are drawn independently from an (unknown) study distribution $P$. Throughout the paper we assume that
    \[P_{(Y(0), Y(1))\mid X = x}\text{ is absolutely continuous with respect to the Lebesgue measure for every }x\in \mathcal{X}.\]
  We will not mention this assumption again for simplicity. We have access to data $(X_{i}, Y_{i}, W_{i}) \sim P_{\text{obs}}$ from a randomized controlled trial, where $X_{i} \in \mathcal{X}$ are covariates, $W_{i} \in \{0, 1\}$ is the treatment assignment, and $Y_{i} \in \mathcal{Y}$ is the observed outcome. We assume that $P_{\text{obs}}$ is consistent with $P$ under an randomized control trial with treatment probability $e$. This means that treatments are randomly assigned according to $\text{Bernoulli}(e)$ and the observed outcomes satisfy SUTVA (stable unit treatment value assignment), meaning that $Y_{i} = Y_{i}(W_{i}).$ In other words, if $P_{\text{obs}}$ is generated via an internally-valid RCT, then
\begin{align} P_{\text{obs}, X, Y \mid W=w} &= P_{X, Y(w)} \quad \forall w \in \{0, 1\}, \label{eq:rct_marginals} \\
P_{\text{obs}, W \mid X=x} &= \text{Bernoulli}(e) \quad \forall x \in \mathcal{X}.
\end{align}

Let $Q$ be the target potential outcome distribution. We seek to learn a policy $\pi: \mathcal{X} \rightarrow \{0, 1\}$ such that the mean outcome $\EE[Q]{Y(\pi(X))}$ is large when $(X, Y(0), Y(1))$ are sampled from $Q$. In this work, we assume that $\pi \in \Pi$ where
\[\Pi = \{\pi: \mathcal{X}\mapsto \{0, 1\}: \pi \text{ is measurable and deterministic}\}.\]

The two key challenges of this setting include that the true study potential outcome distribution $P$ cannot be identified from the observed data, and furthermore, the target potential outcome distribution $Q$ is not the same as $P$. In particular, we assume $Q$ is unknown and generates $P$ under $\Gamma$-biased sampling, in the sense of Definition \ref{def:sampling_bias}. 

For any marginal covariate distribution $Q_{X},$ we define $\mathcal{S}^{\Gamma}(P_{\text{obs}}, Q_{X})$ to be the set of all target potential outcome distributions that can generate $P_{\text{obs}}$ under $\Gamma$-biased sampling and an internally-valid RCT with treatment probability $e$, have covariate distribution $Q_{X}$, and are absolutely continuous with respect to Lebesgue measure. Let $\mathcal{T}(P_{\text{obs}})$ be the set of potential outcome distributions that can generate $P_{\text{obs}}$ under an internally-valid RCT with treatment probability $e$ and are absolutely continuous with respect to Lebesgue measure. Essentially, this is the set of couplings of $P_{X, Y(0)}$ and $P_{X, Y(1)}$ that satisfy \eqref{eq:rct_marginals}. In addition, we can define $\mathcal{R}^{\Gamma}(P, Q_{X})$ as the set of potential outcome distributions that can generate $P$ under $\Gamma$-biased sampling and have covariate distribution $Q_{X}$. While it is possible to place additional restrictions on the possible values of $P$ and $Q$, we focus on the most general setting where 
\begin{equation} \label{eq:robustness_set} \mathcal{S}^{\Gamma}(P_{\text{obs}}, Q_{X}) = \{ Q \mid Q \in \mathcal{R}^{\Gamma}(P, Q_{X}) \text{ where } P \in \mathcal{T}(P_{\text{obs}})\}.\end{equation}

We aim to learn policies that are robust to all target potential outcome distributions in $\mathcal{S}^{\Gamma}(P_{\text{obs}}, Q_{X}).$ In the next subsection, we define objectives that yield different performance guarantees when optimized over $\mathcal{S}^{\Gamma}(P_{\text{obs}}, Q_{X}).$

 \end{subsection}

 \begin{subsection}{Objectives}
We examine the robust policies given by three different objectives: max-min, max-min gain, and minimax regret. We refer to \citet{manski2011choosing} for a detailed discussion of the max-min and minimax regret objectives.

When learning robust policies, a natural first step is to consider the max-min policy:
\begin{equation}
\label{eq:maxmin}
\pi_{\Gamma, \text{maxmin}}^{*} \in \argmax_{\pi \in \Pi} \inf_{Q \in \mathcal{S}^{\Gamma}(P_{\text{obs}}, Q_{X})} \EE[Q]{Y(\pi(X))}. 
\end{equation}
The max-min policy yields the greatest lower bound on the mean outcome across all states of nature (under all distributions $Q \in \mathcal{S}^{\Gamma}(P_{\text{obs}}, Q_{X})$). The max-min objective is the objective that is studied most often in the literature on policy learning under partial identification \citep{adjaho2022externally, mufactored, si2020distributional}. However, it is known to often be ultra-pessimistic \citep{savage1951theory} because there may be distributions of $Q$ for which all policies perform poorly--and the max-min objective will focus on these instances. In some cases, a decision maker may want to protect against these worst-case scenarios and for that reason, the max-min objective may be reasonable choice \citep{manski2011choosing}.

As an alternative to the max-min objective, we also consider the policy that maximizes the worst-case gain over a baseline $\pi_{0}$, which is given by
\begin{equation}
\label{eq:gain}
\pi_{\Gamma, \text{gain}}^{*} \in \argmax_{\pi \in \Pi} \inf_{Q \in \mathcal{S}^{\Gamma}(P_{\text{obs}}, Q_{X})} \EE[Q]{Y(\pi(X))} - \EE[Q]{Y(\pi_{0}(X))}. 
\end{equation}
The max-min gain policy is a natural choice if we aim to improve robustness relative to a status quo policy. This type of objective has been recently considered by \citet{ben2021safe,kallus2021minimax}.

The minimax regret objective, introduced by \citet{savage1951theory}, is closely related to the max-min gain objective. In minimax regret, we essentially aim to maximize the worst-case gain relative to the best possible baseline policy, a baseline policy that obtains the maximum mean outcome for every choice of $Q$. 

The minimax regret policy is defined to be
\begin{equation}
\label{eq:minmax_regret}
\pi^{*}_{\Gamma, \text{regret}} \in \argmin_{\pi \in \Pi} \sup_{Q \in \mathcal{S}^{\Gamma}(P_{\text{obs}}, Q_{X})} R_{Q}(\pi(X)),
\end{equation}
where the regret of a policy $\pi(X)$ under a distribution $Q$ is given by
\[ R_{Q}(\pi(X)) = \sup_{\pi' \in \Pi} \EE[Q]{Y(\pi'(X))} - \EE[Q]{Y(\pi(X))}.\]
When considering policy classes that include non-deterministic policies, we often find that the minimax regret policy is often non-deterministic \citep{manski2011choosing}. We emphasize that in this work, we define the minimax regret objective with respect to $\Pi$, the class of deterministic, unconstrained, binary-valued functions, so we aim to find the minimax regret policy among the class of deterministic functions.

The minimax regret rule yields the least upper bound on the loss in mean outcome that results from not knowing $Q$ and is viewed as less pessimistic than the max-min rule \citep{manski2011choosing, savage1951theory}. Nevertheless, due to the complex structure of the minimax regret policy, it can often be difficult to identify and learn the minimax regret policy. A key contribution of our work is that we identify a closed-form expression for the minimax regret policy under the class of deterministic, unconstrained, binary-valued policies.

\end{subsection}

\begin{subsection}{Results}
Our identification results rely on the conditional average treatment effect (CATE) function and the conditional value-at-risk \citep{rockafellar2000optimization} of the potential outcomes. We define the CATE below
\begin{equation}
\label{eq:cate}
\tau(x) = \EE[P]{Y(1)- Y(0) \mid X=x}.
\end{equation}
We consider the conditional value-at-risk at a particular quantile \begin{equation} \label{eq:eta} \zeta(\Gamma) = \frac{1}{\Gamma + 1}.\end{equation} Recall that for a continuous random variable $Z$ with quantile function (inverse cdf) $q_{Z}$ and $\zeta \in (0, 1)$
\[ \text{CVaR}_{\zeta}(Z) = \EE[]{Z \mid Z \geq q_{\zeta}(Z)}.\]

In addition, we recall that in the absence of sampling bias, the optimal policy to deploy is
\begin{equation} \label{eq:nonrobust} \pi_{\text{non-robust}}(x) = \mathbb{I}(\tau(x) \geq 0),\end{equation} which treats units that have nonnegative conditional average treatment effect under $P$ \citep{kitagawa2018should}. Note that this policy is identified under $P_{\text{obs}}$ because the CATE is identified under $P_{\text{obs}}.$ 

\begin{theo}
\label{theo:opt_maxmin}
Define
\begin{equation} \label{eq:h} H_{\Gamma}(x) = (1 - \Gamma^{-1}) \cdot \left(\mathrm{CVaR}_{\zeta(\Gamma)}(Y(1) \mid x) - \mathrm{CVaR}_{\zeta(\Gamma)}(Y(0) \mid x)\right). \end{equation}
The policy
\begin{equation} \label{eq:opt_maxmin} \pi^{*}_{\Gamma, \mathrm{maxmin}}(x)=\mathbb{I}(\tau(x) \geq H_{\Gamma}(x)) \end{equation}
solves \eqref{eq:maxmin} for any $Q_{X}$ such that $Q_{X} \ll P_{\mathrm{obs}, X}$ and $\sup_{x \in \mathcal{X}} \frac{dP_{\mathrm{obs}, X}(x)}{dQ_{X}(x)} < \infty.$
\hyperref[subsec:opt_maxmin]{Proof in Appendix \ref{subsec:opt_maxmin}.}
\end{theo}
 We recall that $\Gamma=1$ corresponds to the case where there is no variation in the probability of sample selection due to unobservables. We find that \eqref{eq:opt_maxmin} reduces to $\pi_{\text{non-robust}}$ when $\Gamma=1.$ Interestingly, we note that when $\Gamma > 1$, $H_{\Gamma}(x)$ is not necessarily positive, meaning that \eqref{eq:opt_maxmin} may not necessarily yield a higher threshold for treatment than $\pi_{\text{non-robust}}.$ So, depending on the tail behavior of $Y(1), Y(0)$, the max-min rule may treat more units than $\pi_{\text{non-robust}}$ which assumes no sampling bias due to unobservables.

We note that another interpretation of the max-min policy is that it compares the lower bounds on $\EE[Q]{Y(1) \mid X=x}$ and $\EE[Q]{Y(0) \mid X=x}$ over the robustness set and selects that treatment that yields the higher lower bound.

 \begin{lemm}
\label{lemm:maxmin_lower_bounds}
The max-min policy $\pi_{\Gamma, \mathrm{maxmin}}^{*}$ can be written as
\begin{equation}
\label{eq:lower_bounds}
\mathbb{I}\left(\inf_{Q \in \mathcal{S}^{\Gamma}(P_{\mathrm{obs}}, Q_{X})} \EE[Q]{Y(1) \mid X=x} \geq \inf_{Q \in \mathcal{S}^{\Gamma}(P_{\mathrm{obs}}, Q_{X})} \EE[Q]{Y(0) \mid X=x})\right).
\end{equation}
\hyperref[subsec:maxmin_lower_bounds]{Proof in Appendix \ref{subsec:maxmin_lower_bounds}.}
 \end{lemm}

We compare and contrast the max-min policy to the policy that maximizes the worst-case gain over a baseline policy and the minimax regret policy.

\begin{theo}
\label{theo:opt_gain}
Define
\begin{align}
H_{\Gamma}^{+}(x) &= (1 - \Gamma^{-1}) \cdot \left(\mathrm{CVaR}_{\zeta(\Gamma)}(Y(1) \mid x) + \mathrm{CVaR}_{\zeta(\Gamma)}(-Y(0)\mid x)\right) \label{eq:h_plus}, \\
H_{\Gamma}^{-}(x) &= (1 - \Gamma^{-1}) \cdot \left(-\mathrm{CVaR}_{\zeta(\Gamma)}(-Y(1) \mid x) - \mathrm{CVaR}_{\zeta(\Gamma)}(Y(0) \mid x)\right). \label{eq:h_minus}
\end{align}
The policy
\begin{equation} 
\label{eq:opt_gain}
\pi^{*}_{\Gamma, \mathrm{gain}}(x) = \mathbb{I}(\pi_{0}(x) = 0) \cdot \mathbb{I}(\tau(x) \geq H^{+}_{\Gamma}(x)) + \mathbb{I}(\pi_{0}(x) = 1) \cdot \mathbb{I}(\tau(x) \geq H^{-}_{\Gamma}(x))\end{equation}
solves \eqref{eq:gain} for any $Q_{X}$ such that $Q_{X} \ll P_{\mathrm{obs}, X}$ and $\sup_{x \in \mathcal{X}} \frac{dP_{\mathrm{obs}, X}(x)}{dQ_{X}(x)} < \infty.$
\hyperref[subsec:opt_gain]{Proof in Appendix \ref{subsec:opt_gain}.}
\end{theo}

Note that when $\Gamma=1$, the optimal max-min gain policy defaults to $\pi_{\text{non-robust}}.$ When $\Gamma > 1$, the optimal max-min gain policy in \eqref{eq:opt_gain} has different thresholds for treatment depending on whether the baseline policy recommends treatment or control. Intuitively, we would expect \eqref{eq:opt_gain} to have a higher threshold for treatment when the baseline policy recommends control $(\pi_{0}(x) = 0)$ than when the the baseline policy recommends treatment $(\pi_{0}(x) = 1)$. The following lemma confirms this intuition by demonstrating that $H_{\Gamma}^{+}(x) \geq H_{\Gamma}^{-}(x).$ We also note that threshold for treatment of the max-min policy \eqref{eq:opt_maxmin} falls between $H^{-}_{\Gamma}(x)$ and $H^{+}_{\Gamma}(x).$

\begin{lemm}
\label{lemm:threshold_comparison}
For $x \in \mathrm{supp}(P_{\mathrm{obs}, X}),$
\begin{equation} \label{eq:threshold_comparison} H_{\Gamma}^{-}(x) \leq H_{\Gamma}(x) \leq H_{\Gamma}^{+}(x).\end{equation}
\hyperref[subsec:threshold_comparison]{Proof in Appendix \ref{subsec:threshold_comparison}.}
\end{lemm}

An additional consequence of the above lemma is that if the baseline policy is given by  $\pi_{0}(x) = \mathbb{I}(\tau(x) \geq b(x))$ where $H^{-}_{\Gamma}(x) \leq b(x) \leq H^{+}_{\Gamma}(x)$ for all $x \in \mathcal{X}$, then the baseline policy $\pi_{0}$ maximizes the max-min gain objective, i.e.,
\[ \pi_{\Gamma, \text{gain}}^{*} =\pi_{0}.\]

Lastly, we consider the minimax regret rule. 

\begin{theo}
\label{theo:opt_regret}
Define $H_{\Gamma}^{+}(\cdot), H_{\Gamma}^{-}(\cdot)$ as in \eqref{eq:h_plus}, \eqref{eq:h_minus}. The policy 
\begin{equation} \label{eq:opt_regret} \pi^{*}_{\Gamma, \mathrm{regret}}(x) = \mathbb{I}\Big(\tau(x) \geq \frac{H^{+}_{\Gamma}(x) + H^{-}_{\Gamma}(x)}{2}\Big) \end{equation}
solves \eqref{eq:minmax_regret} for any $Q_{X} \ll P_{\mathrm{obs}, X}$ and $\sup_{x \in \mathcal{X}} \frac{dP_{\mathrm{obs}, X}(x)}{dQ_{X}(x)} < \infty.$
\hyperref[subsec:opt_regret]{Proof in Appendix \ref{subsec:opt_regret}.}
\end{theo}

We give another characterization of the minimax regret policy that is related to the standard policy learning problem. In the absence of sampling bias, solving the following optimization problem 
\begin{equation}
\sup_{\pi \in \Pi} \EE[P]{(2\pi(X) - 1) \cdot (Y(1) - Y(0))}
\end{equation}
yields $\pi_{\text{non-robust}}$, which is the optimal policy when there is no sampling bias \citep{zhao2012estimating}. In the following theorem, we see that the maximizer of this objective over the robustness set yields the minimax regret policy. We find that robust optimization of the objective $\EE[Q]{v^{*}(\pi(X); X, Y(0), Y(1))} = \EE[Q]{(2\pi(X) - 1) \cdot (Y(1) - Y(0))}$ over  the robustness set $\mathcal{S}^{\Gamma}(P_{\text{obs}}, Q_{X})$ yields the minimax regret policy. 

\begin{theo}
\label{theo:po_regret}
Let \[\tilde{v}_{\mathrm{regret}}(z; x, y_{0}, y_{1}) = (2z - 1) \cdot (y_{1}- y_{0}).\] The policy $\pi$ that solves 
\begin{equation}
\label{eq:wc_value_regret}
\sup_{\pi \in \Pi} \inf_{Q \in \mathcal{S}^{\Gamma}(P_{\mathrm{obs}}, Q_{X})} \EE[Q]{\tilde{v}_{\mathrm{regret}}(\pi(X); X, Y(0), Y(1))}
\end{equation}
is equal to the optimal policy under the minimax regret objective \eqref{eq:opt_regret}.
\hyperref[subsec:po_regret]{Proof in Appendix \ref{subsec:po_regret}.}
\end{theo}

Note that the policies in \eqref{eq:opt_maxmin}, \eqref{eq:opt_gain}, \eqref{eq:opt_regret} can be estimated by first estimating the nuisance parameters $\tau(\cdot), H_{\Gamma}(\cdot), H_{\Gamma}^{+}(\cdot), H_{\Gamma}^{-}(\cdot)$ and then forming the defined policies. While this is one potential approach for learning policies, implementing this method may be onerous and give poor empirical performance. In the following sections, we consider an alternative approach for learning max-min and max-min gain policies. 
\end{subsection}

%% file: 03b-equivalence.tex
In this section, we prove an equivalence result that will link robust optimization over potential outcome distributions with robust optimization over observed data distribution. Such an equivalence is useful when we aim to learn robust policies using only data from $P_{\text{obs}}$ without knowledge of the true study population $P$.

Robust optimization over potential outcome distributions can generally be formulated as the following problem 
\begin{equation}
\label{eq:potential_outcomes_problem}
\sup_{\pi \in \Pi} \inf_{Q \in \mathcal{S}^{\Gamma}(P_{\text{obs}}, Q_{X})} \EE[Q]{v^{*}(\pi(X); X, Y(0), Y(1))}.\end{equation}
Note that $v^{*}$ is a potential outcome value function, which is defined in terms of $X, Y(0), Y(1).$
Now, we consider robust optimization over observed data distributions. Let $\mathcal{S}_{\text{obs}}^{\Gamma}(P_{\text{obs}}, Q_{X})$ be a robustness set that consists of observed data distributions. A distribution $Q_{\text{obs}} \in \mathcal{S}_{\text{obs}}^{\Gamma}(P_{\text{obs}}, Q_{X})$ if $Q_{\text{obs}, X} = Q_{X}$ and 
\begin{equation} \label{eq:likelihood_ratio_observed_data} \Gamma^{-1} \leq \frac{dQ_{\text{obs}, Y \mid W=w, X=x}(y)}{dP_{\text{obs}, Y \mid W=w, X=x}(y)} \leq \Gamma \quad \forall w \in \{0, 1\}, x \in \mathcal{X}, y \in \mathcal{Y}.
\end{equation}
We consider \begin{equation} \label{eq:observed_data_problem} \sup_{\pi \in \Pi} \inf_{Q_{\text{obs}} \in \mathcal{S}^{\Gamma}_{\text{obs}}(P_{\text{obs}}, Q_{X})} \EE[Q_{\text{obs}}]{v(\pi(X); X, Y, W)}.\end{equation}
Note that $v$ is an observed data value function, which is defined in terms of $X, Y, W.$

We require the following assumption to establish the link between robust optimization over potential outcome distributions and robust optimization over observed data distributions.
\begin{assumption}
\label{assumption:equal_value}
We assume that $v$ is defined so that 
\[\EE[P_{\text{obs}}]{v(\pi(X); X, Y, W)} = \EE[P]{v^{*}(\pi(X); X, Y(0), Y(1))}, \]
for any $P \in \mathcal{T}(P_{\text{obs}})$ and arbitrary $\pi$.
\end{assumption}

\begin{theo}
\label{theo:equivalence}
Suppose $v$ is a value function that satisfies Assumption \ref{assumption:equal_value} for $v^{*}$. Then a policy $\pi$ that solves \eqref{eq:observed_data_problem} also solves \eqref{eq:potential_outcomes_problem} for any $Q_{X} \ll P_{\text{obs}, X}$ and $\sup_{x \in \mathcal{X}} \frac{dP_{\mathrm{obs}, X}(x)}{dQ_{X}(x)}< \infty.$
\hyperref[subsec:equivalence]{Proof in Appendix \ref{subsec:equivalence}.}
\end{theo}

Theorem \ref{theo:equivalence} is a nontrivial and somewhat surprising result. While $Q\in \mathcal{S}^{\Gamma}(P_{\text{obs}}, Q_{X})$ implies $Q_{\text{obs}}\in \mathcal{S}_{\text{obs}}^{\Gamma}(P_{\text{obs}}, Q_{X})$, the converse is not true generally because $Q_{\text{obs}}$ does not carry any information on the copula between potential outcomes (conditional on $X$). Nonetheless, we show that the worse-case copula for this optimization problem yields a joint distribution in $\mathcal{S}^{\Gamma}(P_{\text{obs}}, Q_{X})$ using an argument that is analogous to Neyman-Pearson Lemma.

We note that Theorem \ref{theo:equivalence} also holds when we optimize over continuous-valued functions $h$. Let $\mathcal{H} = L^{2}(P_{\text{obs}, X}, \mathcal{X}).$ We can extend Assumption \ref{assumption:equal_value}: we assume that $v, v^{*}$ are defined so that
\[ \EE[P_{\text{obs}}]{v(h(X); X, Y, W)} = \EE[P]{v^{*}(h(X); X, Y(0), Y(1))}, \]
for $P \in \mathcal{T}(P_{\text{obs}})$ and $h$ arbitrary. Under this assumption, a function $h$ that solves
\begin{equation} \label{eq:continuous_observed_data} \sup_{h \in \mathcal{H}} \inf_{Q_{\text{obs}} \in \mathcal{S}_{\text{obs}}^{\Gamma}(P_{\text{obs}}, Q_{X})} \EE[Q_{\text{obs}}]{v(h(X); X, Y, W)}.
\end{equation}
also solves 
\begin{equation}
\label{eq:continuous_potential_outcomes}
\sup_{h \in \mathcal{H}} \inf_{Q \in \mathcal{S}^{\Gamma}(P_{\text{obs}}, Q_{X})} \EE[Q]{v^{*}(h(X); X, Y(0), Y(1))}.
\end{equation}

%% file: 03c-learning.tex
In this section, we first demonstrate that \eqref{eq:continuous_observed_data} can be solved via RU Regression \citep{sahoo2022learning}. Next, we define smooth observed data value functions can be used in combination with the RU Regression procedure to learn the optimal max-min and max-min gain policies. We also define an observed data value function to learn the optimal minimax regret policy, however it is nonconcave, which results in a nonconvex RU Regression problem.


First, we demonstrate that RU Regression can be used to solve \eqref{eq:continuous_observed_data}.


\begin{theo}
\label{theo:ru_reg}
Let $v(z;x, y, w)$ be a value function. For any $\Gamma > 1$, we define the following augmented loss function 
\begin{equation}
\label{eq:ru_loss}
L_{\mathrm{RU}}^{\Gamma}(z, a; x, y, w) = -\Gamma^{-1} v(z; x, y, w) + (1 - \Gamma^{-1}) \cdot a + (\Gamma - \Gamma^{-1})(-v(z;x, y, w) - a)_{+}.
\end{equation}
Then, any solution
\begin{equation}
\label{eq:ru_regression}
\{ h_{\Gamma}^{*}, \alpha_{\Gamma}^{*} \} \in \argmin_{h, \alpha} \EE[P_{\mathrm{obs}}]{L_{\text{RU}}^{\Gamma}(h(X), \alpha(X, W); X, Y, W)}
\end{equation}
is also a solution to \eqref{eq:continuous_observed_data} for any $Q_{\mathrm{obs}} \ll P_{\mathrm{obs}}$.
\hyperref[subsec:ru_reg]{Proof in Appendix \ref{subsec:ru_reg}.}
\end{theo}

\begin{rema}
If $v(z; x, y, w)$ is concave in $z$ for any $(x, y, w) \in \mathcal{X} \times \mathcal{Y} \times \{0, 1\}$, then the augmented loss $L_{\mathrm{RU}}^{\Gamma}(z, a; x, y, w)$ is jointly convex in $(z, a)$ for any $(x, y, w) \in \mathcal{X} \times \mathcal{Y} \times \{0, 1\}.$ 
\end{rema}

Now, it remains to define the appropriate observed data value functions $v$ that when plugged into the RU Regression procedure yield the optimal max-min and optimal max-min gain policies.

\begin{subsection}{Observed Data Value Functions for Policy Learning}
First, we define potential outcome value functions that when maximized yield the max-min policy, the max-min gain policy, and the minimax regret policy.
\begin{theo}
\label{theo:po_maxmin}
Define $\mathcal{H} = \{h \in L^{2}(P_{\mathrm{obs}, X}, \mathcal{X}) \mid 0 \leq h(x) \leq 1 \}.$ Let \[v^{*}_{\text{maxmin}}(z; x, y_{0}, y_{1}) =  \log(1 + \exp(2z -1)) \cdot y_{1} + \log (1 + \exp(-2z + 1)) \cdot y_{0}.\]
Then the policy $\pi(x) = \mathbb{I}\left(h^{*}(x) \geq \frac{1}{2}\right)$, where
\begin{equation} \label{eq:wc_value_maxmin} h^{*} \in \argsup_{h \in \mathcal{H}} \inf_{Q \in \mathcal{S}_{\Gamma}(P_{\mathrm{obs}}, Q_{X})} \EE[Q]{v^{*}(h(X); X, Y(0), Y(1))}, \end{equation}
is the optimal policy under the max-min objective \eqref{eq:opt_maxmin}.
\hyperref[subsec:po_maxmin]{Proof in Appendix \ref{subsec:po_maxmin}.}
\end{theo}

\begin{theo}
\label{theo:po_gain}
Define $\mathcal{H} = \{h \in L^{2}(P_{\mathrm{obs}, X}, \mathcal{X}) \mid 0 \leq h(x) \leq 1 \}.$ Let \[v^{*}_{\text{gain}}(z; x, y_{0}, y_{1}) = (1 - \pi_{0}(x)) \log(1 + \exp(2z-1)) \cdot (y_{1} - y_{0}) + \pi_{0}(x) \log (1 + \exp(-2z + 1)) \cdot (y_{0} - y_{1}).\] Then $\pi(x) = \mathbb{I}\left(h^{*}(x) \geq \frac{1}{2}\right)$, where
\begin{equation}
\label{eq:wc_value_gain} h^{*} \in \argsup_{h \in \mathcal{H}} \inf_{Q \in \mathcal{S}_{\Gamma}(P_{\mathrm{obs}}, Q_{X})} \EE[Q]{v^{*}_{\text{gain}}(\pi(X); X, Y(0), Y(1))} 
\end{equation}
is the optimal policy under the max-min gain objective \eqref{eq:opt_gain}.
\hyperref[subsec:po_gain]{Proof in Appendix \ref{subsec:po_gain}.}
\end{theo}

\begin{theo}
\label{theo:po_regret_loss}
Define $\mathcal{H} = \{h \in L^{2}(P_{\mathrm{obs}, X}, \mathcal{X}) \mid 0 \leq h(x) \leq 1 \}.$ Let \[v^{*}_{\text{regret}}(z; x, y_{0}, y_{1}) = (2z-1)(y_{1} - y_{0}) + \log\left(\left|2z- 1\right|\right).\] Then $\pi(x) = \mathbb{I}(h^{*}(x) \geq \frac{1}{2})$, where
\begin{equation}
\label{eq:wc_value_regret_loss} h^{*} \in \argsup_{h \in \mathcal{H}} \inf_{Q \in \mathcal{S}_{\Gamma}(P_{\mathrm{obs}}, Q_{X})} \EE[Q]{v^{*}_{\text{regret}}(\pi(X); X, Y(0), Y(1))} 
\end{equation}
is the optimal policy under the minimax regret objective \eqref{eq:opt_regret}.
\hyperref[subsec:po_regret_loss]{Proof in Appendix \ref{subsec:po_regret_loss}.}
\end{theo}

In the following lemma, we define observed data value functions that satisfy Assumption \ref{assumption:equal_value} for $v^{*}_{\text{maxmin}}, v^{*}_{\text{gain}}, v^{*}_{\text{regret}}.$

\begin{lemm}
\label{lemm:equal_value}
Define 
\begin{equation}
v_{\text{maxmin}}(z; x, y, w) = \log(1 + \exp((2z - 1) \cdot (2w-1))) \cdot \Big( \frac{y}{w e + (1-w)(1-e)} \Big). \label{eq:v_maxmin}
\end{equation}
\begin{equation}
\begin{aligned}
v_{\text{gain}}(z; x, y, w) &= (1 - \pi_{0}(x)) \log(1 + \exp(2z-1)) \cdot \Big(\frac{y \cdot w}{e} - \frac{y \cdot (1 - w)}{1 -e}  \Big) \\
&\indent+ \pi_{0}(x) \log(1 + \exp(-2z+1)) \cdot \Big(\frac{y \cdot (1-w)}{1 -e} - \frac{y \cdot w}{e}\Big). \label{eq:v_gain}
\end{aligned}
\end{equation}
\begin{equation}
v_{\text{regret}}(z; x, y, w) = (2z-1) \cdot \Big(\frac{y \cdot w}{e} - \frac{y \cdot (1 - w)}{1 -e}\Big) + \log\left(\left|2z - 1\right|\right) . \label{eq:v_regret}
\end{equation}
The value functions $v_{\text{maxmin}}, v_{\text{gain}}, v_{\text{regret}}$ satisfy Assumption \ref{assumption:equal_value} with $v^{*}_{\text{maxmin}}, v^{*}_{\text{gain}}, v^{*}_{\text{regret}}$, respectively.
\hyperref[subsec:equal_value]{Proof in Appendix \ref{subsec:equal_value}.}
\end{lemm}

Since the the observed data value functions $v_{\text{maxmin}}, v_{\text{gain}}$ are smooth and concave in $z$, we can combine the results from Section \ref{sec:equivalence} and Theorem \ref{theo:ru_reg}, to show that finding the optimal max-min policy and optimal max-min gain policy amounts to solving convex RU Regression problems. This allows us to learn these policies directly without needing to separately estimating multiple nuisance parameters. We propose to parametrize the functions $h, \alpha$ using neural networks and train them with the RU loss \eqref{eq:ru_loss} jointly.

In contrast, we realize that $v_{\text{regret}}$ is nonconcave in $z$ and has a singularity at $z=\frac{1}{2}$. So, RU Regression with $v_{\text{regret}}$ as the observed data value function is a nonconvex problem and may be difficult to solve via continuous optimization methods. We leave the development of an algorithm to solve the minimax regret problem for future work.

\end{subsection}

%% file: 04-experiments.tex
In this section, we evaluate the learning procedures proposed in Section \ref{sec:learning}. First, we describe how to implement RU Regression for policy learning. Second, we visualize the max-min policy and the max-min gain policy learned via RU Regression on a one-dimensional synthetic dataset and compare these policies to the true policies given by \eqref{eq:opt_maxmin} and \eqref{eq:opt_gain}. We also examine the true minimax regret policy given by \eqref{eq:opt_regret}. Third, in a semi-synthetic experiment, we compare the behavior of the max-min and max-min gain policies to the non-robust policy \eqref{eq:nonrobust}.

\begin{subsection}{Implementation of RU Regression for Policy Learning}
\label{subsec:ru_reg_implementation}

\begin{figure}
\centering
\includegraphics[width=0.5\textwidth]{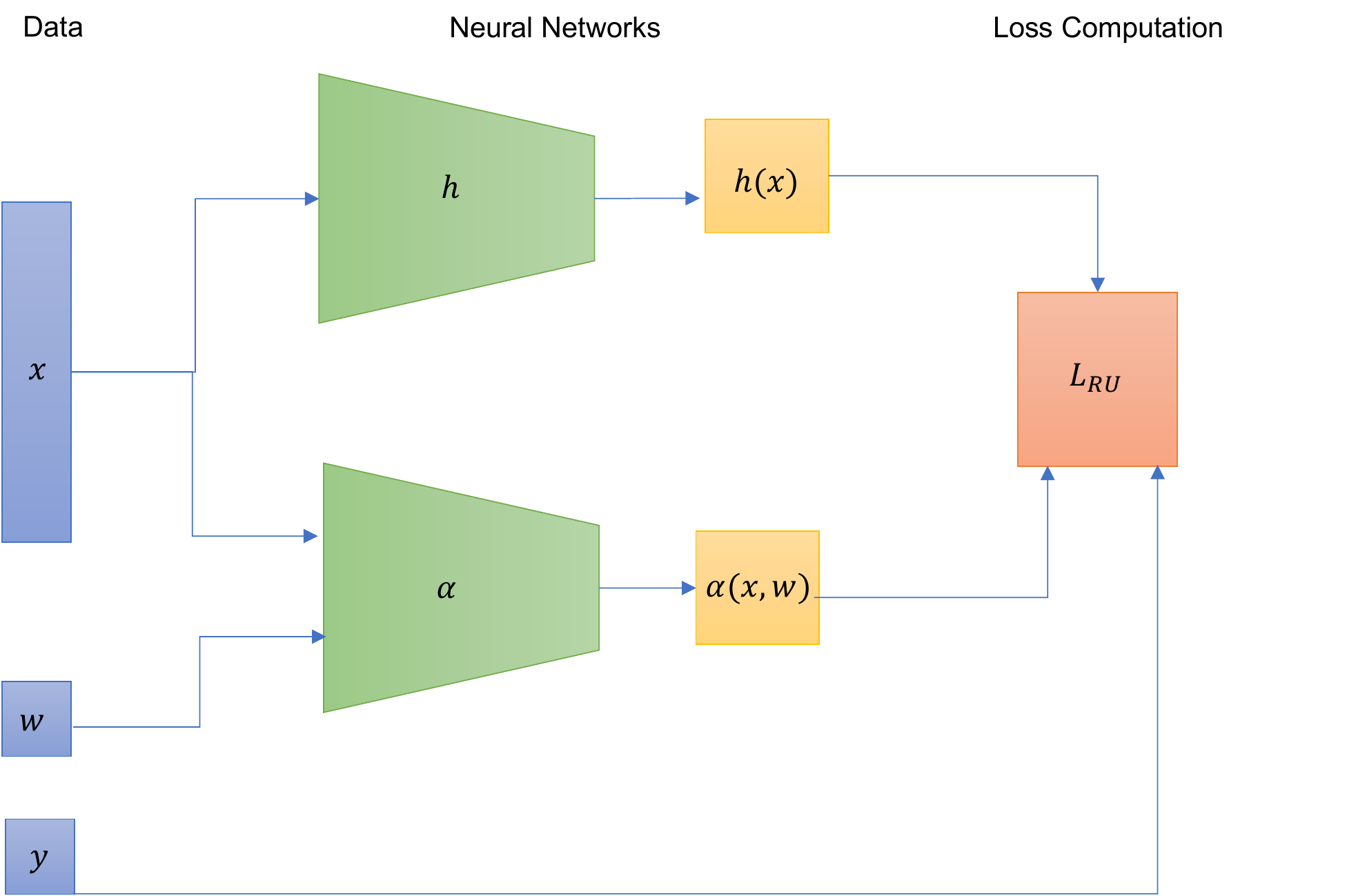}
\caption{RU Regression architecture for policy learning. We note that the function $h$ takes $X$ as input and the auxiliary function $\alpha$ takes $X, W$ as input.}
\label{fig:architecture}
\end{figure}
There are a few subtle differences between the RU Regression implementation in this work and that of \cite{sahoo2022learning}. From Theorem \ref{theo:ru_reg}, the auxiliary function $\alpha$ in the policy learning setting depends on covariates $X$ and the treatment assignment $W$, while in the regression setting of \cite{sahoo2022learning}, $\alpha$ only depends on $X$. This is reflected in our architecture for learning $h, \alpha$ (Figure \ref{fig:architecture}).

We note that the value functions $v_{\text{maxmin}}$ in \eqref{eq:v_maxmin} and $v_{\text{gain}}$ in \eqref{eq:v_gain} are unbounded above, so they do not have maximizers over unbounded function classes. To ensure that the RU loss \eqref{eq:ru_loss} is not unbounded below when $v_{\text{maxmin}}$ or $v_{\text{gain}}$ are plugged into \eqref{eq:ru_loss}, we perform the optimization of $h$ over a class of bounded functions. In particular, we aim to learn a regression function $h_{\Gamma}(x): \mathcal{X} \rightarrow [0, 1].$ Then we determine treatment assignment by computing
\[ \pi_{\Gamma}(x) = \mathbb{I}\Big(h_{\Gamma}(x) \geq \frac{1}{2}\Big).\] Since the policy only depends on whether $h$ is greater or less than $\frac{1}{2}$, the restriction of $h$ to the class of bounded functions does not impact the optimal policy value. To enforce that $h$ must have outputs in $[0, 1]$, we add a sigmoid activation as the final activation function of the neural network that represents $h$, which ensures that the network outputs values in $[0, 1].$
\end{subsection}

\begin{subsection}{Toy Experiment}
First, we consider a simple one-dimensional experiment, so that we can visualize the data distributions and compare the learned policies to the optimal policies computed under the data model.

\begin{subsubsection}{Data}

\begin{figure}
\includegraphics[width=\textwidth]{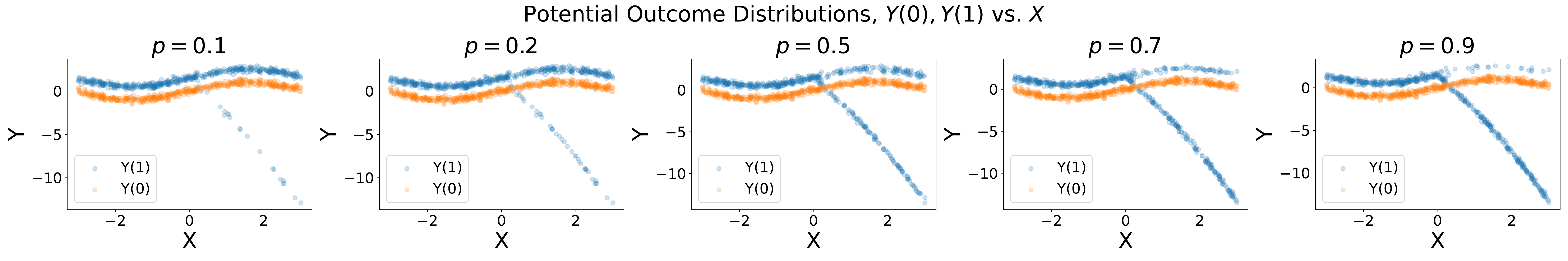}
\caption{We visualize the potential outcome distribution as $p$ varies. When $p$ is low, our potential outcome distribution mostly consists of units for which $Y_{i}(1) > Y_{i}(0)$. As $p$ increases, a larger fraction of units have $Y_{i}(0) > Y_{i}(1).$ }
\label{fig:sim_data}
\end{figure}

We consider a toy example where the study and target potential outcome distribution are generated as follows. Let $U_{i}$ be an unobserved variable that impacts $Y_{i}(1).$
\begin{equation}
\label{eq:dgp_1d}
\begin{split}
X_{i} &\sim \text{Uniform}[-3, 3] \\
U_{i} &\sim \text{Bernoulli}(p) \\
Y_{i}(0) \mid X_{i} &\sim N(\sin(X_{i}) \,, \sigma^{2}) \\ 
Y_{i}(1) \mid X_{i} &\sim N(\sin(X_{i}) + 1.5 - U_{i} \cdot (5X_{i})_{+} \,, \sigma^{2}),
\end{split}
\end{equation}
where $\sigma=0.2.$ We visualize the potential outcome distributions in Figure \ref{fig:sim_data}. We generate study and target potential outcome distributions by varying the Bernoulli parameter $p$ of the distribution of $U_{i}$. Let the study potential outcome distribution $P$ correspond to the potential outcome distribution where $p_{\text{study}}=0.2$. Given $P$,  we generate the observed data distribution $P_{\text{obs}}$ over $(X_{i}, Y_{i}, W_{i})$ as follows
\[ W_{i} \sim \text{Bernoulli}(0.5), \quad Y_{i} = Y_{i}(W_{i}).\]
The target potential outcome distributions $Q$ are generated by varying $p_{\text{target}} \in [0, 1].$

\end{subsubsection}

\begin{subsubsection}{Policies}
We use RU Regression, with the implementation described in Section \ref{subsec:ru_reg_implementation}, to learn the optimal max-min and max-min gain policies with access only to data from the observed data distribution. Since this is a one-dimensional synthetic example, we also use the data model \eqref{eq:dgp_1d} and the closed-form formula \eqref{eq:opt_regret} to compute the minimax regret policy.

We evaluate the following 4 following policies.
\begin{enumerate}
	\item Max-Min Policy (learned via RU Regression).
	\item Max-Min Gain Policy with Always Control Baseline (learned via RU Regression).
	\item Max-Min Gain Policy with Always Treat Baseline (learned via RU Regression).
	\item Minimax Regret (computed using \eqref{eq:dgp_1d} and \eqref{eq:opt_regret}).
\end{enumerate}

\end{subsubsection}

\begin{subsubsection}{Results}

\begin{figure}
\includegraphics[width=\textwidth]{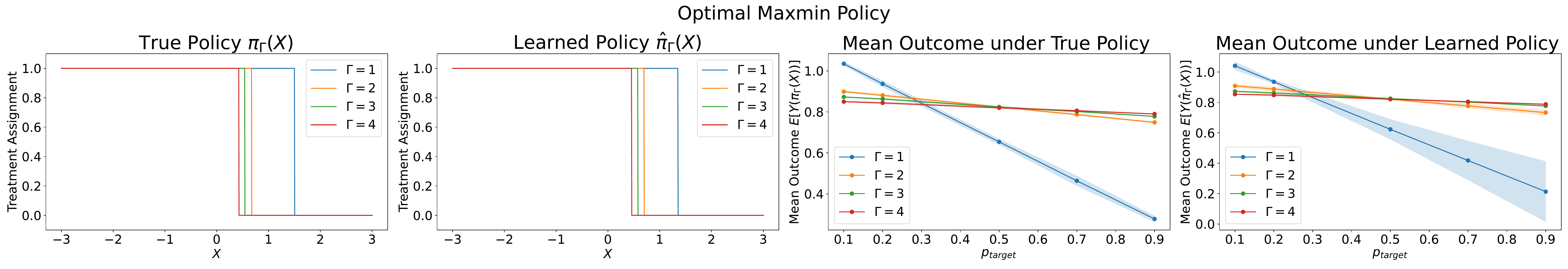}
\caption{Max-min policy results for one-dimensional toy example. \textbf{Left}: We visualize the true max-min policies, computed using the data model \eqref{eq:dgp_1d}. \textbf{Middle Left}: For one random seed, we visualize the learned maxmin policy. \textbf{Middle Right}: We visualize the mean outcome of the true policy for target distributions with different $p_{\text{target}}$, aggregated over 6 random trials, where the randomness is over dataset generation. \textbf{Right}: We visualize the mean outcome of the learned policy for target distributions with different $p_{\text{target}}$, aggregated over 6 random trials, where the randomness is over dataset generation.}
\label{fig:maxmin_1d}
\end{figure}

\begin{figure}
\includegraphics[width=\textwidth]{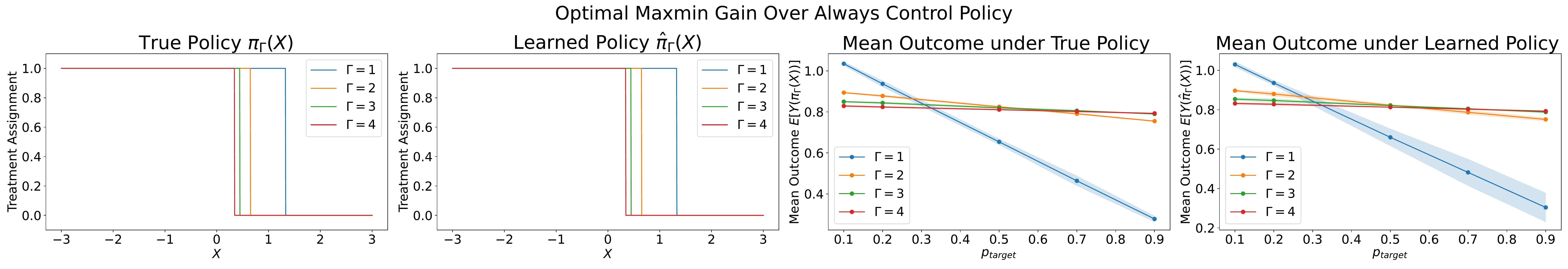}
\caption{Max-min gain (over always control) policy results for one-dimensional toy example. \textbf{Left}: We visualize the true max-min gain policies where the baseline is $\pi_{0}(X)= 0$ (always control), computed using the data model \eqref{eq:dgp_1d}. \textbf{Middle Left}: For one random seed, we visualize the learned maxmin gain over always control policy. \textbf{Middle Right}: We visualize the mean outcome of the true policy for target distributions with different $p_{\text{target}}$, aggregated over 6 random trials, where the randomness is over dataset generation. \textbf{Right}: We visualize the mean outcome of the learned policy for target distributions with different $p_{\text{target}}$, aggregated over 6 random trials, where the randomness is over dataset generation.}
\label{fig:gain_control_1d}
\end{figure}

\begin{figure}
\includegraphics[width=\textwidth]{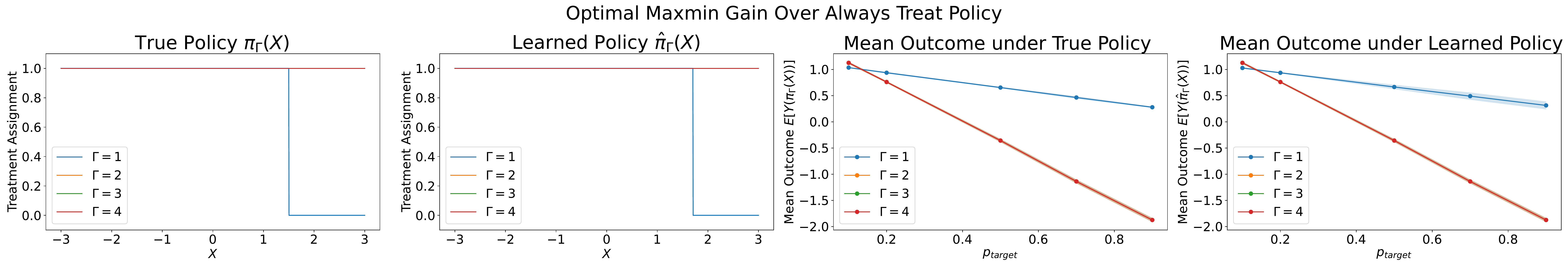}
\caption{Max-min gain (over always treat) policy results for one-dimensional toy example. \textbf{Left}: We visualize the true max-min gain policies where the baseline is $\pi_{0}(X)= 1$ (always treat), computed using the data model \eqref{eq:dgp_1d}. \textbf{Middle Left}: For one random seed, we visualize the learned maxmin gain over always control policy. \textbf{Middle Right}: We visualize the mean outcome of the true policy for target distributions with different $p_{\text{target}}$, aggregated over 6 random trials, where the randomness is over dataset generation. \textbf{Right}: We visualize the mean outcome of the learned policy for target distributions with different $p_{\text{target}}$, aggregated over 6 random trials, where the randomness is over dataset generation.}
\label{fig:gain_treat_1d}
\end{figure}

\begin{figure}
\centering
\includegraphics[width=0.5\textwidth]{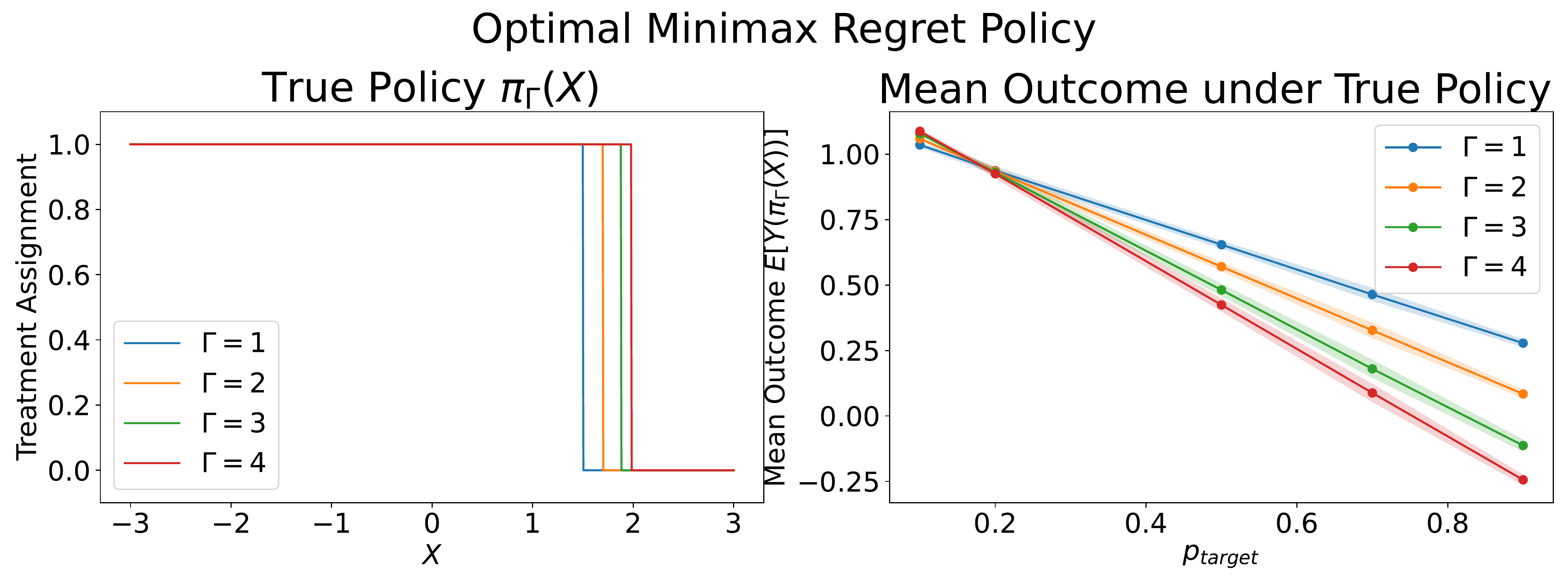}
\caption{Minimax regret policy results for one-dimensional toy example. \textbf{Left}: We visualize the true minimax regret policies, computed using the data model \eqref{eq:dgp_1d}. \textbf{Right}: We visualize the mean outcome of the learned policy for target distributions with different $p_{\text{target}}$, aggregated over 6 random trials, where the randomness is over dataset generation.}
\label{fig:regret_1d}
\end{figure}

We use the data model in \eqref{eq:dgp_1d} to compute the nuisance parameters $\tau(\cdot), H_{\Gamma}(\cdot), H_{\Gamma}^{+}(\cdot), H_{\Gamma}^{-}(\cdot)$ and plug them into \eqref{eq:opt_maxmin} and \eqref{eq:opt_gain} to compute the true optimal policies. A comparison of the left and middle left plots of Figures \ref{fig:maxmin_1d}, \ref{fig:gain_control_1d}, \ref{fig:gain_treat_1d} reveals that our learned policies match the true policies. A comparison of the middle right and right plots of Figures \ref{fig:maxmin_1d}, \ref{fig:gain_control_1d}, \ref{fig:gain_treat_1d} demonstrates that over 6 random trials, the learned policies and the true policies achieve similar mean outcome.

Note that $\pi_{\text{non-robust}}$ from \eqref{eq:nonrobust} is equal to the optimal max-min, max-min gain, or minimax regret policies when $\Gamma=1$. The non-robust policy recommends to treat units with nonnegative CATE. This strategy performs well when $p_{\text{target}} < 0.3$ but it is outperformed by the max-min policy and the max-min gain over always control policy when $p_{\text{target}} > 0.3.$ We note that the max-min gain over always treat policy does not deviate from the baseline for $\Gamma> 1.$ Somewhat surprisingly, in this particular example, the minimax regret policy is less conservative that $\pi_{\text{non-robust}}$ (Figure \ref{fig:regret_1d}) because it recommends treatment to a higher proportion of the population than $\pi_{\text{non-robust}}$ as $\Gamma$ increases. 
\end{subsubsection}

\end{subsection}

\begin{subsection}{High-Dimensional Experiment}
Second, we evaluate our methods in a high-dimensional $(d=10)$ simulation.

\begin{subsubsection}{Data}
We consider a high-dimensional $(d=10)$ example where the study and target potential outcome distribution are generated as follows. Let $U_{i}$ be an unobserved variable that impacts $Y_{i}(1).$
\begin{equation}
\label{eq:dgp_high_dim}
\begin{split}
X_{i} &\sim \text{Uniform}[-3, 3]^{d} \\
U_{i} &\sim \text{Bernoulli}(p) \\
Y_{i}(0) \mid X_{i} &\sim N(\sin(a^{T} X_{i}) \,, \sigma^{2}) \\ 
Y_{i}(1) \mid X_{i} &\sim N(\sin(a^{T}X_{i}) + 1.5 - U_{i} \cdot (2 (X_{i, 1} + X_{i, 2} + X_{i, 3}))_{+} \,, \sigma^{2}),
\end{split}
\end{equation}
where $\sigma=0.2$ and $a$ is a constant vector defined in Section \ref{sec:exp_details}.
\end{subsubsection}

\begin{subsubsection}{Policies}
We evaluate the following policies.
\begin{enumerate}
	\item Max-Min (learned via RU Regression)
	\item Max-Min Gain over Baseline Policy (learned via RU Regression)
	\[ \pi_{0}(X) = \mathbb{I}(X_{1} \leq 0).\]
\end{enumerate}
\end{subsubsection}

\begin{subsubsection}{Results}

\begin{figure}
\centering
\includegraphics[width=0.5\textwidth]{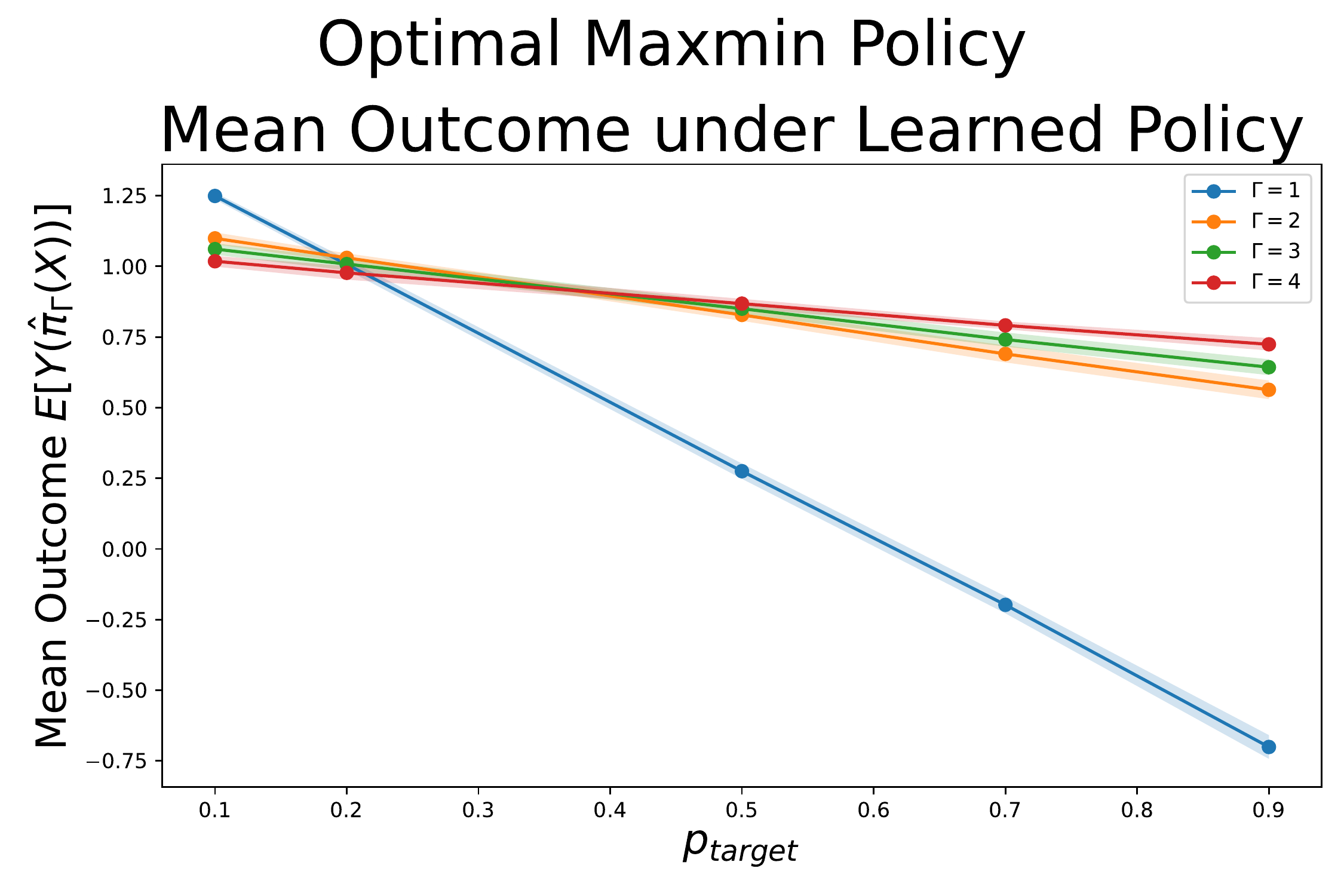}
\caption{Max-min policy results from high-dimensional simulation.}
\label{fig:maxmin_high_dim}
\end{figure}

\begin{figure}
\centering
\includegraphics[width=0.5\textwidth]{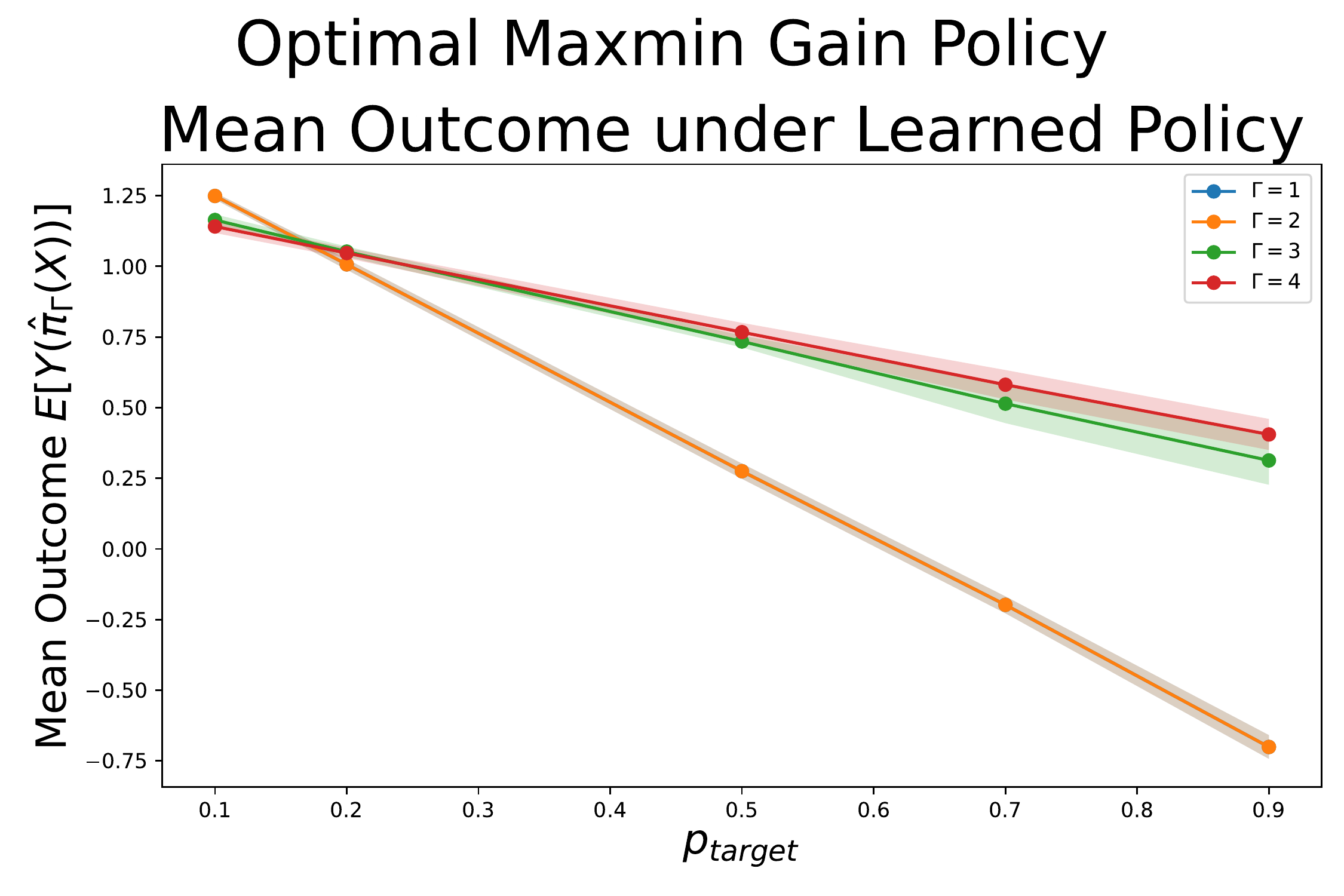}
\caption{Max-min gain policy results from high-dimensional simulation. We note that when $\Gamma=1, 2$, the max-min gain policy does not deviate from $\pi_{\text{non-robust}}$. For higher values of $\Gamma$, the max-min gain policy outperforms $\pi_{\text{non-robust}}$ for high values of $p_{\text{target}}.$}
\label{fig:gain_high_dim}
\end{figure}

As in the toy experiments, we observe that when $\Gamma=1$ the optimal robust policies corresponds to $\pi_{\text{non-robust}}.$ As before, we observe that $\pi_{\text{non-robust}}$ performs well when $p_{\text{target}}$ is small but deteriorates in performance for larger values of $p_{\text{target}}.$ In contrast, when $\Gamma> 1$, the max-min (Figure \ref{fig:maxmin_high_dim}) and max-min gain (Figure \ref{fig:gain_high_dim}) policies are more robust to changes in $p_{\text{target}}$.
\end{subsubsection}

\end{subsection}

\begin{subsection}{Semi-Synthetic Experiment with Voting Dataset}
We compare the behavior of the max-min and max-min gain policies in a semi-synthetic experiment, where we use the voting dataset of \citet{gerber2008social}.

\begin{subsubsection}{Data}
The voting dataset of \citet{gerber2008social} was collected from a randomized
controlled trial-style study that aimed to effect of various actions on voter turnout. The researchers designed one control and four treatment actions that involved mailing the selected units a letter ahead of the 2006 Michigan primary election. In the original field experiment, the probability of the control action was $\frac{5}{9}$ and the probability of each treatment action was $\frac{1}{9}.$ 

In our experiment we focus on only on the ``Control'' action $(W_{i}=0)$ and the ``Neighbors'' action ($W_{i}=1$). Note that this implies that the probability of treatment is $e=\frac{1}{9} / (\frac{1}{9} + \frac{5}{9}) = \frac{1}{6}$. The ``Neighbors'' action entailed mailing a letter to the individual that contained voting participation records of the individual, the other members of the individual’s household, as well as the neighbors. The letter also mentioned a follow-up letter will be sent after the election with everyone’s updated participation, so the individual’s participation will be made known among the neighbors. The ``Control'' action sends no letter to the individual. 

In our experiment, the covariates $X_{i}$ include the household size of the individual, age, sex, and whether the individual $i$ voted in the previous primary elections from 2000-2002 and the previous general elections from 2000-2004.  The outcome $Y_{i}$ is whether individual $i$ voted in the 2006 primary election. Although the dataset also contains information on whether individual $i$ voted in the 2004 primary election, we treat this as an unobserved variable $U_{i}$ and use it generate synthetic target and study populations.

Through a dataset generation procedure described in Appendix \ref{sec:voting}, we create a combined training and validation dataset where 67$\%$ of the units have $U_{i}=1$. In contrast, units with $U_{i}=1$ make up the only 18$\%$ of the test dataset. The training, validation, and test sets consist of 62044, 41364, and  126036 samples. By design, units who voted in the 2004 election are over-represented in the study population compared to the target population. We fit policies using data from the synthetic study population and evaluate the mean outcome under the synthetic target population.
\end{subsubsection}

\begin{subsubsection}{Policies}
We evaluate the following policies.

\begin{enumerate}
	\item Max-Min (learned via RU Regression)
	\item Max-Min Gain with Always Control Baseline (learned via RU Regression)
	\[ \pi_{0}(X) = 0.\]
	\item Max-Min Gain with Always Treat Baseline (learned via RU Regression)
	\[ \pi_{0}(X) = 1.\]
\end{enumerate}

\end{subsubsection}

\begin{subsubsection}{Results}
Recall that as before, $\Gamma=1$ corresponds to $\pi_{\text{non-robust}}$, so we report results for $\pi_{\text{non-robust}}$ (learned via RU Regression with $v_{\text{maxmin}}$ and $\Gamma=1$). We evaluate the robust policies learned for $\Gamma=1.1, 1.2, 1.3, 1.5, 2, 3, 4$. We simply report results for $\Gamma=1.1, 1.2, 1.3, 1.5$ because when $\Gamma \geq 1.5$ we observe the same results as when $\Gamma=1.5$. This can be explained by the fact that the outcomes $Y_{i}(0), Y_{i}(1)$ are binary-valued, so $q_{\zeta(\Gamma)}(Y(w) \mid X=x)$ may take on the same value for many values of $\Gamma$. Note that our theoretical results only apply to the case where the conditional study potential outcome distribution is absolutely continuous with respect to Lebesgue measure, but in this experiment, the study potential outcome distribution is a discrete distribution. Nevertheless, we observe results that are in line with our results for continuous-valued outcomes. We report the target policy value and the proportion of target population treated under the different learned policies in Table \ref{tab:voting}. We find that the non-robust policy recommends recommends to treat about 66\% of the population. We find that the max-min and max-min gain over always treat baseline policy recommends to treat the entire population. In contrast, the max-min gain over the always control baseline recommends to treat a decreasing fraction of the population as $\Gamma$ increases. In this particular example, we observe that different robust objective functions can yield very different policies.

\begin{table}
\scriptsize
\centering
\begin{tabular}{|c|c|c|}
\hline
Method & \multicolumn{1}{c}{Target Policy Value}  & Proportion of Target Population Treated\\
\hline 
Non-Robust & 0.3106 $\pm$ 0.003 & 0.66\\
\hline
Max-Min $(\Gamma=1.1)$ & 0.3272 $\pm$ 0.003 & 0.87 \\
Max-Min $(\Gamma=1.2)$ & 0.3375 $\pm$ 0.004  & 1.0 \\
Max-Min $(\Gamma=1.3)$ & 0.3375 $\pm$ 0.004  & 1.0 \\
Max-Min $(\Gamma=1.5)$ & 0.3375 $\pm$ 0.004  & 1.0 \\
\hline
Max-Min Gain over Always Treat $(\Gamma=1.1)$ & 0.3375 $\pm$ 0.004 & 1.0 \\
Max-Min Gain over Always Treat $(\Gamma=1.2)$ & 0.3375 $\pm$ 0.004 & 1.0\\
Max-Min Gain over Always Treat $(\Gamma=1.3)$ & 0.3375 $\pm$ 0.004 & 1.0\\
Max-Min Gain over Always Treat $(\Gamma=1.5)$ & 0.3375 $\pm$ 0.004 & 1.0\\
\hline
Max-Min Gain over Always Control $(\Gamma=1.1)$ & 0.2934 $\pm$ 0.003 &  0.44 \\
Max-Min Gain over Always Control $(\Gamma=1.2)$ & 0.2780 $\pm$ 0.002 & 0.25\\  
Max-Min Gain over Always Control $(\Gamma=1.3)$ & 0.2705 $\pm$ 0.002 & 0.15\\ 
Max-Min Gain over Always Control $(\Gamma=1.5)$ & 0.2653 $\pm$ 0.001 & 0.0\\ 
\hline
\end{tabular}
\caption{The non-robust policy recommends to treat 66\% of the population. The max-min and max-min gain over always treat policies recommend to treat the entire population. In contrast, the max-min gain over always control policies recommends treatment to a much smaller proportion of the population.}
\label{tab:voting}
\end{table}

\end{subsubsection}

\end{subsection}

%% file: 04b-additional-results.tex
\begin{table}
\scriptsize
\centering
\begin{tabular}{|c|c|c|c|c|c|}
\hline
Method & \multicolumn{5}{c|}{Target Policy Value} \\
\hline 
& $p=0.1$ & $p=0.2$ & $p=0.5$ & $p=0.7$ & $p=0.9$ \\
\hline
RU Regression & \multirow{2}{*}{1.029 $\pm$  0.016} & \multirow{2}{*}{0.930 $\pm$  0.009} & \multirow{2}{*}{0.659 $\pm$  0.015} & \multirow{2}{*}{ 0.491 $\pm$  0.025} & \multirow{2}{*}{0.308 $\pm$  0.062} \\
$(\Gamma = 1)$ & & & & & \\
True & \multirow{2}{*}{1.035 $\pm$  0.010} &  \multirow{2}{*}{0.937 $\pm$  0.016} & \multirow{2}{*}{0.654 $\pm$  0.015}  & \multirow{2}{*}{0.464 $\pm$  0.025} &\multirow{2}{*}{0.278 $\pm$  0.015} \\
$(\Gamma=1)$ & & & & & \\
\hline
RU Regression & \multirow{2}{*}{0.908 $\pm$  0.015} &  \multirow{2}{*}{ 0.882 $\pm$  0.007} &  \multirow{2}{*}{0.819 $\pm$  0.003} & \multirow{2}{*}{0.780 $\pm$  0.017} &  \multirow{2}{*}{0.740 $\pm$  0.019} \\
$(\Gamma = 2)$ & & & & & \\
True & \multirow{2}{*}{0.899 $\pm$  0.005} & \multirow{2}{*}{0.881 $\pm$  0.007} &  \multirow{2}{*}{0.824 $\pm$  0.005} & \multirow{2}{*}{0.787 $\pm$  0.006} & \multirow{2}{*}{0.749 $\pm$  0.004} \\
$(\Gamma = 2)$ & & & & & \\
\hline
RU Regression & \multirow{2}{*}{0.874 $\pm$  0.008} & \multirow{2}{*} {0.861 $\pm$  0.009} &  \multirow{2}{*}{0.823 $\pm$  0.003} & \multirow{2}{*}{0.801 $\pm$  0.002} & \multirow{2}{*}{0.776 $\pm$ 0.009} \\
$(\Gamma = 3)$ & & & & & \\
True & \multirow{2}{*}{0.873 $\pm$  0.003} & \multirow{2}{*}{0.863 $\pm$ 0.005} &    \multirow{2}{*}{0.824 $\pm$  0.005} & \multirow{2}{*}{0.802 $\pm$  0.004} &  \multirow{2}{*}{0.778 $\pm$  0.003} \\
$(\Gamma = 3)$ & & & & & \\
\hline
RU Regression & \multirow{2}{*}{0.856 $\pm$  0.010} & \multirow{2}{*}{0.847 $\pm$  0.009} & \multirow{2}{*}{0.819 $\pm$  0.001} &  \multirow{2}{*}{0.804 $\pm$  0.004} & \multirow{2}{*}{0.785 $\pm$  0.008} \\
$(\Gamma = 4)$ & & & & & \\
True & \multirow{2}{*}{0.850 $\pm$  0.004} &  \multirow{2}{*}{0.844 $\pm$  0.006} & \multirow{2}{*}{0.820 $\pm$  0.005} &   \multirow{2}{*}{0.806 $\pm$  0.004} &  \multirow{2}{*}{0.790 $\pm$  0.003} \\
$(\Gamma = 4)$ & & & & & \\
\hline
\end{tabular}
\caption{Max-min policy results from one-dimensional simulation experiment. We report the mean and standard deviation of the target policy value from 6 random trials, where the randomness is over the dataset generation. Note that RU Regression with $\Gamma=1$ yields a policy that is equivalent to $\pi_{\text{non-robust}}$ and incurs low policy value for low values of $p$. The maxmin policies $\hat{\pi}_{\Gamma}$ for $\Gamma >1$ yield higher policy values for high values of $p$. }
\end{table}

\begin{table}
\scriptsize
\centering
\begin{tabular}{|c|c|c|c|c|c|}
\hline
Method & \multicolumn{5}{c|}{Target Policy Value} \\
\hline 
& $p=0.1$ & $p=0.2$ & $p=0.5$ & $p=0.7$ & $p=0.9$ \\
\hline
RU Regression &  \multirow{2}{*}{1.031 $\pm$  0.016} & \multirow{2}{*}{0.930 $\pm$  0.010} & \multirow{2}{*}{0.656 $\pm$  0.019} & \multirow{2}{*}{0.484 $\pm$  0.024} &    \multirow{2}{*}{0.301 $\pm$  0.057}\\
$(\Gamma = 1)$ & & & & & \\
True & \multirow{2}{*}{1.035 $\pm$  0.010} & \multirow{2}{*}{0.937 $\pm$  0.016} & \multirow{2}{*}{0.654 $\pm$  0.015} &  \multirow{2}{*}{0.464 $\pm$ 0.025} &  \multirow{2}{*}{0.278 $\pm$  0.015} \\
$(\Gamma=1)$ & & & & & \\
\hline
RU Regression &  \multirow{2}{*}{0.897 $\pm$  0.011} &  \multirow{2}{*}{0.876 $\pm$  0.005} & \multirow{2}{*}{0.821 $\pm$  0.003} & \multirow{2}{*}{0.788 $\pm$  0.010} & \multirow{2}{*}{0.753 $\pm$  0.011} \\
$(\Gamma = 2)$ & & & & & \\
True &  0.894 $\pm$  0.004 & \multirow{2}{*}{0.878 $\pm$  0.006} &  \multirow{2}{*}{0.825 $\pm$  0.005} & \multirow{2}{*}{0.791 $\pm$  0.005} &  \multirow{2}{*}{0.755 $\pm$  0.003} \\
$(\Gamma = 2)$ & & & & & \\
\hline
RU Regression &  \multirow{2}{*}{0.854 $\pm$  0.009} &  \multirow{2}{*}{0.846 $\pm$  0.008} & \multirow{2}{*}{0.818 $\pm$  0.002} & \multirow{2}{*}{0.804 $\pm$  0.004} & \multirow{2}{*}{0.788 $\pm$  0.008}  \\
$(\Gamma = 3)$ & & & & & \\
True &  \multirow{2}{*}{0.850 $\pm$  0.004} &  \multirow{2}{*}{0.844 $\pm$  0.006} &  \multirow{2}{*}{0.820 $\pm$  0.005} &  \multirow{2}{*}{0.806 $\pm$  0.004} & \multirow{2}{*}{0.790 $\pm$  0.003} \\
$(\Gamma = 3)$ & & & & & \\
\hline
RU Regression &  \multirow{2}{*}{0.834 $\pm$  0.006} & \multirow{2}{*}{0.827 $\pm$  0.005} & \multirow{2}{*}{0.812 $\pm$  0.001} & \multirow{2}{*}{0.801 $\pm$  0.005} & \multirow{2}{*}{0.792 $\pm$  0.004}\\
$(\Gamma = 4)$ & & & & & \\
True &  \multirow{2}{*}{0.829 $\pm$  0.005} & \multirow{2}{*}{0.824 $\pm$  0.005} &  \multirow{2}{*}{0.811 $\pm$  0.004} &  \multirow{2}{*}{0.802 $\pm$  0.004} &  \multirow{2}{*}{0.793 $\pm$  0.003} \\
$(\Gamma = 4)$ & & & & & \\
\hline
\end{tabular}
\caption{Max-min gain over always control policy results from one-dimensional simulation experiment. We report the mean and standard deviation of the target policy value from 6 random trials, where the randomness is over the dataset generation. Note that RU Regression with $\Gamma=1$ yields a policy that is equivalent to $\pi_{\text{non-robust}}$ and incurs low policy value for low values of $p$. The maxmin gain over always control policies $\hat{\pi}_{\Gamma}$ for $\Gamma >1$ yield higher policy values for high values of $p$.}
\end{table}

\begin{table}
\scriptsize
\centering
\begin{tabular}{|c|c|c|c|c|c|}
\hline
Method & \multicolumn{5}{c|}{Target Policy Value} \\
\hline 
& $p=0.1$ & $p=0.2$ & $p=0.5$ & $p=0.7$ & $p=0.9$ \\
\hline
RU Regression &  \multirow{2}{*}{1.028 $\pm$  0.016}& \multirow{2}{*}{0.929 $\pm$  0.009} & \multirow{2}{*}{0.660 $\pm$  0.020} & \multirow{2}{*}{0.492 $\pm$  0.029} & \multirow{2}{*}{0.311 $\pm$  0.066} \\
$(\Gamma = 1)$ & & & & & \\
True &  \multirow{2}{*}{1.035 $\pm$  0.010} & \multirow{2}{*}{0.937 $\pm$  0.016} & \multirow{2}{*}{0.654 $\pm$  0.015} & \multirow{2}{*}{0.464 $\pm$  0.025} & \multirow{2}{*}{0.278 $\pm$  0.015} \\
$(\Gamma=1)$ & & & & & \\
\hline
RU Regression & \multirow{2}{*}{1.124 $\pm$ 0.021} & \multirow{2}{*}{0.755 $\pm$  0.019} & \multirow{2}{*}{-0.355 $\pm$  0.035} & \multirow{2}{*}{-1.131 $\pm$  0.043} & \multirow{2}{*}{ -1.856 $\pm$  0.032} \\
$(\Gamma = 2)$ & & & & & \\
True &   \multirow{2}{*}{1.126 $\pm$  0.022}  &  \multirow{2}{*}{0.760 $\pm$  0.019} & \multirow{2}{*}{-0.358 $\pm$  0.031} & \multirow{2}{*}{-1.139 $\pm$  0.037} & \multirow{2}{*}{-1.875 $\pm$  0.039} \\
$(\Gamma = 2)$ & & & & & \\
\hline
RU Regression & \multirow{2}{*}{1.124 $\pm$  0.021} & \multirow{2}{*}{0.755 $\pm$  0.019} & \multirow{2}{*}{-0.355 $\pm$  0.035} & \multirow{2}{*}{-1.131 $\pm$  0.043} & \multirow{2}{*}{-1.856 $\pm$  0.032} \\
$(\Gamma = 3)$ & & & & & \\
True &  \multirow{2}{*}{1.126 $\pm$  0.022} & \multirow{2}{*}{0.760 $\pm$  0.019} & \multirow{2}{*}{-0.358 $\pm$  0.031} & \multirow{2}{*}{-1.139 $\pm$  0.037} & \multirow{2}{*}{-1.875 $\pm$ 0.039}\\
$(\Gamma = 3)$ & & & & & \\
\hline
RU Regression & \multirow{2}{*}{1.121 $\pm$  0.025} & \multirow{2}{*}{0.747 $\pm$  0.013} & \multirow{2}{*}{-0.356 $\pm$  0.043} & \multirow{2}{*}{-1.127 $\pm$  0.052} & \multirow{2}{*}{-1.841 $\pm$  0.012} \\
$(\Gamma = 4)$ & & & & & \\
True &  \multirow{2}{*}{1.126 $\pm$ 0.022} & \multirow{2}{*}{0.760 $\pm$  0.019} & \multirow{2}{*}{-0.358 $\pm$  0.031} & \multirow{2}{*}{-1.139 $\pm$  0.037} & \multirow{2}{*}{-1.875 $\pm$  0.039} \\
$(\Gamma = 4)$ & & & & & \\
\hline
\end{tabular}
\caption{Max-min gain over always treat policy results from one-dimensional simulation experiment. We report the mean and standard deviation of the target policy value from 6 random trials, where the randomness is over the dataset generation. Note that RU Regression with $\Gamma=1$ yields a policy that is equivalent to $\pi_{\text{non-robust}}$. This max-min gain policy defaults to the baseline policy $\pi_{0}(x) = 1$ for $\Gamma > 1$.}
\end{table}

\begin{table}
\scriptsize
\centering
\begin{tabular}{|c|c|c|c|c|c|}
\hline
Method & \multicolumn{5}{c|}{Target Policy Value} \\
\hline 
& $p=0.1$ & $p=0.2$ & $p=0.5$ & $p=0.7$ & $p=0.9$ \\
\hline
RU Regression &  \multirow{2}{*}{1.249 $\pm$  0.012} &  \multirow{2}{*}{1.007 $\pm$  0.019} & \multirow{2}{*}{0.275 $\pm$ 0.027} & \multirow{2}{*}{-0.198 $\pm$  0.030} & \multirow{2}{*}{-0.701 $\pm$  0.042} \\
$(\Gamma = 1)$ & & & & & \\
\hline
RU Regression & \multirow{2}{*}{1.249 $\pm$  0.012} &  \multirow{2}{*}{1.007 $\pm$  0.019} & \multirow{2}{*}{0.275 $\pm$ 0.027} & \multirow{2}{*}{-0.198 $\pm$  0.030} & \multirow{2}{*}{-0.701 $\pm$  0.042} \\
$(\Gamma = 2)$ & & & & & \\
\hline
RU Regression & \multirow{2}{*}{1.164 $\pm$ 0.020} & \multirow{2}{*}{1.052 $\pm$  0.018} &  \multirow{2}{*}{0.734 $\pm$  0.021} & \multirow{2}{*}{0.514 $\pm$  0.069} & \multirow{2}{*}{0.313 $\pm$  0.086} \\
$(\Gamma = 3)$ & & & & & \\
\hline
RU Regression &  \multirow{2}{*}{1.141 $\pm$  0.023} & \multirow{2}{*}{1.047 $\pm$  0.018} & \multirow{2}{*}{ 0.767 $\pm$  0.033} & \multirow{2}{*}{0.581 $\pm$  0.052} & \multirow{2}{*}{0.405 $\pm$  0.055} \\
$(\Gamma = 4)$ & & & & & \\
\hline
\end{tabular}
\caption{Max-min gain policy results for high-dimensional simulation. }
\end{table}

\begin{table}
\scriptsize
\centering
\begin{tabular}{|c|c|c|c|c|c|}
\hline
Method & \multicolumn{5}{c|}{Target Policy Value} \\
\hline 
& $p=0.1$ & $p=0.2$ & $p=0.5$ & $p=0.7$ & $p=0.9$ \\
\hline
RU Regression &  \multirow{2}{*}{1.249 $\pm$  0.012} &  \multirow{2}{*}{1.007 $\pm$  0.019} & \multirow{2}{*}{0.275 $\pm$ 0.027} & \multirow{2}{*}{-0.198 $\pm$  0.030} & \multirow{2}{*}{-0.701 $\pm$  0.042} \\
$(\Gamma = 1)$ & & & & & \\
\hline
RU Regression & \multirow{2}{*}{1.099 $\pm$  0.021} & \multirow{2}{*}{1.030 $\pm$  0.016} & \multirow{2}{*}{  0.828 $\pm$  0.021} & \multirow{2}{*}{0.690 $\pm$  0.030} & \multirow{2}{*}{ 0.563 $\pm$  0.033} \\
$(\Gamma = 2)$ & & & & & \\
\hline
RU Regression & \multirow{2}{*}{1.061 $\pm$  0.020} & \multirow{2}{*}{1.008 $\pm$  0.022} & \multirow{2}{*}{ 0.850 $\pm$  0.020} & \multirow{2}{*}{0.741 $\pm$  0.025} & \multirow{2}{*}{0.643 $\pm$  0.028} \\
$(\Gamma = 3)$ & & & & & \\
\hline
RU Regression &  \multirow{2}{*}{1.018 $\pm$  0.019} & \multirow{2}{*}{0.977 $\pm$  0.024} & \multirow{2}{*}{ 0.868 $\pm$  0.017} & \multirow{2}{*}{0.791 $\pm$  0.014} & \multirow{2}{*}{0.724 $\pm$  0.022} \\
$(\Gamma = 4)$ & & & & & \\
\hline
\end{tabular}
\caption{Max-min gain policy results for high-dimensional simulation. }
\end{table}

%% file: 05-experiment-details.tex
\begin{subsection}{Toy Example}
\begin{subsubsection}{Model}
\label{sec:toy_details_model}

For the RU Regression model, we jointly train two neural networks to learn the regression function $h$ and the quantile function $\alpha$, respectively. A visualization of the model architecture for RU Regression is provided in Figure \ref{fig:architecture}. Each of the neural networks has 3 hidden layers and 64 units per layer and ReLU activation. The RU Regression model has 17.1K
trainable parameters.

\end{subsubsection}

\begin{subsubsection}{Dataset Splits}
\label{sec:toy_details_data}
For all methods, the train, validation, and test sets consists of 20000, 10000, and 10000 samples, respectively. The train and validation sets are generated via the data model specified in \eqref{eq:dgp_1d} with $p = 0.2.$ All methods are evaluated on the same test sets, which are generated via the data model in \eqref{eq:dgp_1d} parameter $p$ taking value in $[0.1, 0.2, 0.5, 0.7, 0.9]$. For each of 6 random seeds $[0, 1, 2, 3, 4, 5]$, a new dataset (train, validation, test set) is generated.
\end{subsubsection}

\begin{subsubsection}{Training Procedure}
\label{sec:toy_details_training}
The models are trained for a maximum of $50$ epochs with batch size equal to $4000$ and we use the Adam optimizer with learning rate $1e-2$. Each epoch we check the loss obtained on the validation set and select the model that minimizes the loss on the validation set.
\end{subsubsection}

\end{subsection}

\begin{subsection}{High-Dimensional Experiment}

\begin{subsubsection}{Data}
In the data model from \eqref{eq:dgp_high_dim}, we set 
\begin{align*} a =&[ 0.50067509, -0.67108696, -2.22006362, -0.37834265,  1.05841302,\\ &-0.4509034, 1.15857361, 0.62236239, -0.77458079, -0.74790281].
\end{align*}
\end{subsubsection}
\begin{subsubsection}{Model}
We use the same models as in toy experiment (See Section \ref{sec:toy_details_model}).
\end{subsubsection}

\begin{subsubsection}{Dataset Splits}
We use the same dataset splits as in the toy experiment (See Section \ref{sec:toy_details_data})
\end{subsubsection}

\begin{subsubsection}{Training Procedure}
We use the same training procedure as in the toy experiment(See Section \ref{sec:toy_details_training}).
\end{subsubsection}
\end{subsection}

\begin{subsection}{Voting Semi-Synthetic Experiment}
\label{sec:voting}
\begin{subsubsection}{Model}
We use the same models as in toy experiment (See Section \ref{sec:toy_details_model}).
\end{subsubsection}

\begin{subsubsection}{Dataset Splits}
We randomly select 75$\%$ of units with $U_{i}=1$ and 25$\%$ of units with $U_{i}=0$ to create the combined training and validation dataset, which represents the study population. The remaining units make up the test set, which represents the target population.
\end{subsubsection}

\begin{subsubsection}{Training Procedure}
We use the same training procedure as in the toy experiment(See Section \ref{sec:toy_details_training}).
\end{subsubsection}

\end{subsection}

%% file: 06-standard-results.tex
\begin{lemm}
\label{lemm:cvar_properties}
Let $\zeta \in (0, 1).$ CVaR has the following properties:
\begin{enumerate}
\item If $Y$ has a density, then
\[ \EE{Y} = (1- \zeta) \mathrm{CVaR}_{\zeta}(Y) - \zeta \mathrm{CVaR}_{1-\zeta}(-Y),\]
\item $\mathrm{CVaR}$ is positively homogenenous, i.e. 
\[ \mathrm{CVaR}_{\zeta}(cY) = c \mathrm{CVaR}_{\zeta}(Y)\]
if $c > 0,$
\end{enumerate}
\citep{pflug2000some}.
\end{lemm}

\begin{lemm}
\label{lemm:cvar_decomposition}
For probability measures with finite second moment, CVaR is a strongly coherent risk measure, meaning it is convex, continuous, and for random variables $X, Y$ with absolutely continuous distributions $F_{X}, F_{Y}$ 
\[ \mathrm{CVaR}(X) + \mathrm{CVaR}(Y) = \sup \{\mathrm{CVaR}(\tilde{X} + \tilde{Y}): (\tilde{X}, \tilde{Y}) \sim \mu, \mu \in \mathcal{T}(F_{X}, F_{Y})\}\]
where $\mathcal{T}$ is the set of all bivariate distributions with marginals $F_{X}, F_{Y}$ \citep{ekeland2012comonotonic}.
\end{lemm}

\begin{prop}
\label{prop:sup_extremal_point}
Let $f$ be a convex function, and let $C$ be a closed bounded convex set contained in $\mathrm{dom } f$. Then the supremum of $f$ relative to $C$ is finite and it is attained at some extreme point of $C$ \citep{rockafellar1997convex}.
\end{prop}

\begin{lemm}
\label{lemm:proper_scoring_rule}
Let $J(z) = A \log(1 + \exp\{2z - 1\}) + B \log (1 + \exp\{-2z + 1\}).$ Let $z^{*} =\argmax_{z \in [0, 1]} J(z)$. $A \geq B$ iff $z^{*} \geq \frac{1}{2}.$
 \hyperref[subsec:proper_scoring_rule]{Proof in Appendix \ref{subsec:proper_scoring_rule}.}
\end{lemm}

%% file: 07-main-proofs.tex
\begin{subsection}{Notation}
We use the following notation in the proofs. Here, we assume that $Y(0), Y(1)$ are distributed according to a potential outcome distribution whose marginals match those of $P_{\text{obs}}$
\begin{align*}
c_{1}^{+}(x) &= \text{CVaR}_{\zeta(\Gamma)}(Y(1) \mid x) \\
c_{1}^{-}(x) &= \text{CVaR}_{\zeta(\Gamma)}(-Y(1) \mid x) \\
c_{0}^{+}(x) &= \text{CVaR}_{\zeta(\Gamma)}(Y(0) \mid x) \\
c_{0}^{-}(x) &= \text{CVaR}_{\zeta(\Gamma)}(-Y(0) \mid x)\\
g_{1}(x) &= \Gamma \tau(x) + (1- \Gamma) (c_{1}^{+}(x) + c_{0}^{-}(x)) \\
g_{0}(x) &= -\Gamma \tau(x) + (1- \Gamma)(c_{0}^{+}(x) +c_{1}^{-}(x)). 
       \end{align*}
 Note that
 \begin{align*}
   H_{\Gamma}^{+}(x) &= (1 - \Gamma^{-1})\cdot (c_{1}^{+}(x) + c_{0}^{-}(x))\\
   H_{\Gamma}^{-}(x) &= (1 - \Gamma^{-1})\cdot (-c_{1}^{-}(x) - c_{0}^{+}(x))\\
   \tau(x) \ge H_{\Gamma}^{+}(x) &\Longleftrightarrow g_{1}(x) \ge 0 \\
  \tau(x) \ge H_{\Gamma}^{-}(x) &\Longleftrightarrow g_{0}(x) \le 0.
 \end{align*}
 Throughout this section, we assume that $P_{(Y(1), Y(0))\mid X = x}$ is absolutely continuous with respect to Lebesgue measure almost surely for any $x$. 

\end{subsection}

\begin{subsection}{Technical Lemmas}

\begin{lemm}
\label{lemm:joint_sampling_bias}
Let $P, Q$ be potential outcome distributions. $Q$ generates $P$ via $\Gamma$-biased sampling if and only if
\begin{equation} \label{eq:joint_likelihood_ratio} \Gamma^{-1} \leq \frac{dQ_{Y(0), Y(1) \mid X=x}(y_{0}, y_{1})}{dP_{Y(0), Y(1) \mid X=x}(y_{0}, y_{1})} \leq \Gamma \quad \forall x \in \mathcal{X}, y_{0}, y_{1} \in \mathbb{R}\end{equation}
and $ \sup_{x \in \mathcal{X}} \frac{dP_{X}(x)}{dQ_{X}(x)} < \infty.$
 \hyperref[subsec:joint_sampling_bias]{Proof in Appendix \ref{subsec:conditional_sampling_bias}.}
\end{lemm}

\begin{lemm}
\label{lemm:np}
A policy $\pi$ solves \eqref{eq:potential_outcomes_problem} for every $Q_{X}$ such that $Q_{X} \ll P_{\mathrm{obs}, X}$ and $\sup_{x \in \mathcal{X}} \frac{dP_{\mathrm{obs}, X}(x)}{dQ_{X}(x)} < \infty$ iff $\pi$ solves
\begin{equation} \label{eq:np_cond} \max_{\pi(x) \in \{0, 1\}}  \inf_{P \in \mathcal{T}(P_{\mathrm{obs}})} \Gamma \EE[P]{v^{*}(\pi(x); X, Y(0), Y(1)) \mid X=x} + (1- \Gamma)\mathrm{CVaR}_{\zeta(\Gamma)}(v^{*}(\pi(x); X, Y(0), Y(1)) \mid x) \end{equation}
for every $x \in \mathrm{supp}(P_{\mathrm{obs}, X}).$ Similarly, a function $h$ solves \eqref{eq:continuous_potential_outcomes}
for every $Q_{X}$ such that $Q_{X} \ll P_{\mathrm{obs}, X}$ and $\sup_{x \in \mathcal{X}} \frac{dP_{\mathrm{obs}, X}(x)}{dQ_{X}(x)} < \infty$ if and only if it solves
\begin{equation} \label{eq:np_continuous_cond} \max_{h(x) \in [0, 1]}  \inf_{P \in \mathcal{T}(P_{\mathrm{obs}})} \Gamma \EE[P]{v^{*}(h(x); X, Y(0), Y(1)) \mid X=x} + (1- \Gamma)\mathrm{CVaR}_{\zeta(\Gamma)}(v^{*}(h(x); X, Y(0), Y(1)) \mid x) \end{equation}
for every $x \in \mathrm{supp}(P_{\mathrm{obs}, X}).$
\hyperref[subsec:np]{Proof in Appendix \ref{subsec:np}.}
\end{lemm}

\begin{lemm}
\label{lemm:wc_gain}
Let $v^{*}(z; x, y_{0}, y_{1}) = (z - \pi_{0}(x)) \cdot (y_{1} - y_{0}).$ Suppose that $Q_{X}$ satisfies $Q_{X} \ll P_{\mathrm{obs}, X}$ and $\sup_{x \in \mathcal{X}} \frac{dP_{\mathrm{obs}, X}(x)}{dQ_{X}(x)} < \infty.$ Then,
\begin{equation}\label{eq:gain_conditional}\begin{aligned} &\inf_{Q \in \mathcal{S}^{\Gamma}(P_{\mathrm{obs}}, Q_{X})} \EE[Q]{v^{*}(\pi(X); X, Y(0), Y(1)) \mid X=x} \\
&=(\pi(x) - \pi_{0}(x))_{+} g_{1}(x) +  (\pi_{0}(x) - \pi(x))_{+}g_{0}(x),\end{aligned}\end{equation}
for any $x \in \mathrm{supp}(P_{\mathrm{obs}, X}).$
\hyperref[subsec:wc_gain]{Proof in Appendix \ref{subsec:wc_gain}.}
\end{lemm}


\begin{lemm}
\label{lemm:g}
For $x \in \mathrm{supp}(P_{\mathrm{obs}, X})$, $\min\{g_{1}(x), g_{0}(x)\} \leq 0.$ In addition, if $g_{1}(x) = 0$, then $g_{0}(x) \leq 0.$
\hyperref[subsec:g]{Proof in Appendix \ref{subsec:g}.}
\end{lemm}


\begin{lemm}
\label{lemm:potential_robustness_implies_observed_robustness}
Let $P_{\mathrm{obs}}$ be an observed data distribution where $P_{\mathrm{obs}, W \mid X} = \mathrm{Bernoulli}(e)$ for a treatment probability $e$. Let $Q \in \mathcal{S}^{\Gamma}(P_{\mathrm{obs}}, Q_{X})$. Suppose that under an RCT with treatment probability $e$, the potential outcome distribution $Q$ yields the observed data distribution $Q_{\mathrm{obs}}.$ Then, $Q_{\mathrm{obs}}  \in \mathcal{S}^{\Gamma}_{\mathrm{obs}}(P_{\mathrm{obs}}, Q_{X}).$
\hyperref[subsec:potential_robustness_implies_observed_robustness]{Proof in Appendix \ref{subsec:potential_robustness_implies_observed_robustness}.}
\end{lemm}

\begin{lemm}
\label{lemm:np_obs_data}
Let $\zeta(\Gamma) = \frac{1}{\Gamma + 1}.$ Let $q_{\zeta}(v(h(x); X, Y, W) \mid x, w)$ be the $\zeta$-th quantile of $v(h(x); x, Y, w)$ where $Y \sim P_{\mathrm{obs}, Y \mid X=x, W=w}.$ The distribution $dQ^{*}_{\mathrm{obs}}$ that solves
\begin{equation} \label{eq:wc_dist} \argmin_{Q_{\mathrm{obs}} \in \mathcal{S}^{\Gamma}_{\mathrm{obs}}(P_{\mathrm{obs}}, Q_{X})} \EE[Q_{\mathrm{obs}}]{v(h(x); X, Y, W)} \end{equation}
is given by $dQ_{X} \cdot dP_{\mathrm{obs}, W \mid X} \cdot dQ^{*}_{\mathrm{obs}, Y\mid X, W}$, where
\begin{equation}
\label{eq:worst_case_distribution}
dQ_{\mathrm{obs}, Y\mid X=x, W=w}^{*}(y) = \begin{cases} \Gamma^{-1} \cdot dP_{\mathrm{obs}, Y \mid X=x, W=w}(y) & v(h(x); x, y, w) \geq q_{\zeta(\Gamma)}(v(h(x); X, Y, W) \mid w, x) \\
\Gamma \cdot dP_{\mathrm{obs}, Y \mid X=x, W=w}(y) &  \mathrm{o.w.}.\end{cases}
\end{equation}
\hyperref[subsec:np_obs_data]{Proof in Appendix \ref{subsec:np_obs_data}.}
\end{lemm}

\begin{lemm}
\label{lemm:conditional_sampling_bias}
Let $P, Q$ be potential outcome distributions. If $Q$ generates $P$ via $\Gamma$-biased sampling, then
\begin{equation}
\label{eq:likelihood_ratio_potential_outcomes_single_conditional}
\Gamma^{-1} \leq \frac{dQ_{Y(w) \mid X=x}(y)}{dP_{Y(w) \mid X=x}(y)} \leq \Gamma \quad \forall x \in \mathcal{X}, y \in \mathbb{R}, w \in \{0, 1\}
 \end{equation}
 and $\sup_{x \in \mathcal{X}} \frac{dP_{\mathrm{obs}, X}(x)}{dQ_{X}(x)} < \infty$.
 \hyperref[subsec:conditional_sampling_bias]{Proof in Appendix \ref{subsec:conditional_sampling_bias}.}
\end{lemm}

\end{subsection}

\begin{subsection}{Proof of Theorem \ref{theo:opt_maxmin}}
\label{subsec:opt_maxmin}
Define $v^{*}(\pi(X); X, Y(0), Y(1)) = Y(\pi(X)) = \pi(X) \cdot Y(1) + (1- \pi(X)) Y(0).$ We can apply Lemma \ref{lemm:np} to see that solving \eqref{eq:maxmin} for any $Q_{X}$ such that $Q_{X} \ll P_{\text{obs}, X}$ and $\sup_{x \in \mathcal{X}} \frac{dP_{\text{obs}, X}(x)}{dQ_{X}(x)} < \infty$ is equivalent to solving the following conditional worst-case risk minimization problem 
\begin{equation} \label{eq:np} \max_{\pi(x) \in \{0, 1\}} \inf_{P \in \mathcal{T}(P_{\text{obs}})} \Gamma \EE[P]{Y(\pi(x)) \mid X=x} + (1- \Gamma)\text{CVaR}_{\zeta(\Gamma)}(Y(\pi(x)) \mid x)\end{equation}
for every $x \in \text{supp}(P_{\text{obs}, X})$.

We note that 
\begin{align*}
\EE[P]{Y(\pi(x)) \mid X=x} = \pi(x) \EE[P]{Y(1) \mid X=x} + (1- \pi(x)) \EE[P]{Y(0) \mid X=x}, 
\end{align*}
so this term is identified under $P_{\text{obs}}$ and does not depend on the choice of $P \in \mathcal{T}(P_{\text{obs}}).$ Let $P$ be any potential outcome distribution $\mathcal{T}(P_{\text{obs}}).$ So, \eqref{eq:np} can be written as 
\begin{equation}\label{eq:a1} \max_{\pi(x) \in \{0, 1\}} \Gamma (\pi(x) \EE[P]{Y(1) \mid X=x} + (1- \pi(x)) \EE[P]{Y(1) \mid X=x}) + \inf_{P' \in \mathcal{T}(P_{\text{obs}})} (1- \Gamma)\text{CVaR}_{\zeta(\Gamma)}(Y(\pi(x)) \mid x). \end{equation}
Simplifying the term on the right, we have that
\begin{align}
&\inf_{P' \in \mathcal{T}(P_{\text{obs}})} (1- \Gamma)\text{CVaR}_{\zeta(\Gamma)}(Y(\pi(x)) \mid x) \\
&= (1- \Gamma) \sup_{P' \in \mathcal{T}(P_{\text{obs}})} \text{CVaR}_{\zeta(\Gamma)}(Y(\pi(x)) \mid x) \\
&= (1- \Gamma) \sup_{P' \in \mathcal{T}(P_{\text{obs}})} \text{CVaR}_{\zeta(\Gamma)}(\pi(x) Y(1) + (1- \pi(x)) Y(0) \mid x) \\
&= (1- \Gamma) \Big( \text{CVaR}_{\zeta(\Gamma)}(\pi(x) Y(1)) + \text{CVaR}_{\zeta(\Gamma)}((1- \pi(x))Y(0)) \Big) \label{eq:decomp} \\
&= (1- \Gamma) \Big(\pi(x) \text{CVaR}_{\zeta(\Gamma)}(Y(1) \mid x) + (1- \pi(x)) \text{CVaR}_{\zeta(\Gamma)}(Y(0) \mid x) \Big) \label{eq:pos} \\
&= (1- \Gamma) \Big(\pi(x) c_{1}^{+}(x) + (1- \pi(x)) c_{0}^{+}(x) \Big).
\end{align}
\eqref{eq:decomp} follows from Lemma \ref{lemm:cvar_decomposition}. \eqref{eq:pos} follows from Lemma \ref{lemm:cvar_properties}.

Plugging in the above expression into \eqref{eq:a1} yields
\begin{equation}
\max_{\pi(x) \in \{0, 1\}} \pi(x) (\Gamma \EE[P]{Y(1) \mid X=x} + (1- \Gamma) c_{1}^{+}(x)) + (1- \pi(x)) (\Gamma \EE[P]{Y(0) \mid X=x} + (1- \Gamma) c_{0}^{+}(x)).
\end{equation}
Given the form of the worst-case risk above, the optimal policy is equal to 
\[\mathbb{I}(\Gamma \EE[P]{Y(1) \mid X=x} + (1- \Gamma) c_{1}^{+}(x) \geq \Gamma \EE[P]{Y(0) \mid X=x} + (1- \Gamma) c_{0}^{+}(x)).\]
Rearranging terms, we have that this policy is equal to
\begin{align*}
\mathbb{I}( \tau(x) \geq (1 - \Gamma^{-1}) (c_{1}^{+}(x) - c_{0}^{+}(x))).
\end{align*}
Note that this is equal to \eqref{eq:opt_maxmin}.

\end{subsection}

\begin{subsection}{Proof of Lemma \ref{lemm:maxmin_lower_bounds}}
\label{subsec:maxmin_lower_bounds}
We have that 
\begin{align}
\inf_{Q \in \mathcal{S}^{\Gamma}(P_{\text{obs}}, Q_{X})}  \EE[Q]{Y(w) \mid X=x} 
&= \inf_{P \in \mathcal{T}(P_{\text{obs}})} \inf_{Q \in \mathcal{R}^{\Gamma}(P, Q_{X})}  \EE[Q]{Y(w) \mid X=x} \\
&= \inf_{P \in \mathcal{T}(P_{\text{obs}})} \Gamma \EE[P]{Y(w) \mid X=x} + (1-\Gamma) \text{CVaR}_{\zeta(\Gamma)}(Y(1) \mid X=x) \label{eq:np_pot_outcome} \\
&= \Gamma \EE[P]{Y(w) \mid X=x} + (1-\Gamma) c_{w}^{+}(x) \label{eq:cvar_plus},
\end{align}
where $P$ is any study potential outcome distribution in $\mathcal{T}(P_{\text{obs}}).$ In the above derivation, \eqref{eq:np_pot_outcome} follows from Lemma \ref{lemm:np} and \eqref{eq:cvar_plus} follows from Lemma \ref{lemm:cvar_decomposition}.
We examine the policy from \eqref{eq:lower_bounds} and find that it can be written as follows
\begin{align*}
&\mathbb{I}\left(\inf_{Q \in \mathcal{S}^{\Gamma}(P_{\text{obs}}, Q_{X})} \EE[Q]{Y(1) \mid X=x} \geq \inf_{Q \in \mathcal{S}^{\Gamma}(P_{\text{obs}}, Q_{X})} \EE[Q]{Y(0) \mid X=x})\right) \\
&= \mathbb{I}\left(\Gamma \EE[P]{Y(1) \mid X=x} + (1 - \Gamma) c_{1}^{+}(x) \geq \Gamma \EE[P]{Y(0) \mid X=x} + (1 - \Gamma) c_{0}^{+}(x)\right) \\
&= \mathbb{I}(\Gamma \tau(x) \geq (\Gamma- 1)(c_{1}^{+}(x) - c_{0}^{+}(x))) \\
&= \mathbb{I}(\tau(x) \geq (1 - \Gamma^{-1} )(c_{1}^{+}(x) - c_{0}^{+}(x))) \\
&= \mathbb{I}(\tau(x) \geq H_{\Gamma}(x)) \\
&= \pi_{\Gamma, \text{maxmin}}^{*}(x).
\end{align*}
Thus, we conclude that the policy that compares the lower bounds on $\EE[Q]{Y(1) \mid X=x}$ and $\EE[Q]{Y(0) \mid X=x}$ over the robustness set $\mathcal{S}^{\Gamma}(P_{\text{obs}}, Q_{X})$ is the max-min policy $\pi_{\Gamma, \text{maxmin}}^{*}.$

\end{subsection}

\begin{subsection}{Proof of Theorem \ref{theo:opt_gain}}
\label{subsec:opt_gain}
We note that $\EE[Q]{Y(\pi(X))} - \EE[Q]{Y(\pi_{0}(X))} = \EE[Q]{(\pi(X) -\pi_{0}(X)) \cdot (Y(1) - Y(0))}$, so \eqref{eq:gain} is equivalent to 
\[ \sup_{\pi \in \Pi} \inf_{Q \in \mathcal{S}^{\Gamma}(P_{\text{obs}}, Q_{X})}  \EE[Q]{v^{*}(\pi(X); X, Y(0), Y(1))}\]
for $v^{*}(z; x, y_{0}, y_{1}) = (z - \pi_{0}(x)) \cdot (y_{1} - y_{0}).$ Solving the above problem for every $Q_{X}$ such that $Q_{X} \ll P_{\text{obs}, X}$ and $\sup_{x \in \mathcal{X}} \frac{dP_{\text{obs}, X}(x)}{dQ_{X}(x)} < \infty$ is equivalent to solving the conditional worst-case gain problem (left side of \eqref{eq:gain_conditional}) for every $x \in \text{supp}(P_{\text{obs}, X}).$ Thus, we can apply Lemma \ref{lemm:wc_gain} to see that a policy $\pi$ that solves \eqref{eq:gain} solves the right side of \eqref{eq:gain_conditional} for every $x$ in $\text{supp}(P_{\text{obs}, X}).$

First, we consider the case where $\pi_{0}(x) = 0.$ In this case, \eqref{eq:gain_conditional} reduces to
\[ \max_{\pi(x) \in \{0, 1\}}  (\pi(x))_{+} (\Gamma \tau(x) + (1- \Gamma) (c_{1}^{+}(x) + c_{0}^{-}(x))).\]
So, the optimal policy when $\pi_{0}(x) = 0$ is equivalent to $\mathbb{I}(\Gamma \tau(x) + (1- \Gamma) (c_{1}^{+}(x) + c_{0}^{-}(x)) \geq 0).$ Rearranging terms, this is equivalent to $\mathbb{I}(\tau(x) \geq H_{\Gamma}^{+}(x)).$ 

Second, we consider the case where $\pi_{0}(x) = 1.$ In this case, \eqref{eq:gain_conditional} reduces to
\[ \max_{\pi(x) \in \{0, 1\}}  (1 - \pi(x))_{+} (-\Gamma \tau(x) + (1- \Gamma)(c_{0}^{+}(x) +c_{1}^{-}(x))). \] 
So, the optimal policy when $\pi_{0}(x) = 1$ is equivalent to $\mathbb{I}( -\Gamma \tau(x) + (1- \Gamma)(c_{0}^{+}(x) +c_{1}^{-}(x)) \leq 0)$. Rearranging terms, this is equivalent to $\mathbb{I}(\tau(x) \geq  H_{\Gamma}^{-}(x)).$

Combining these two results yields the policy in \eqref{eq:opt_gain}.

\end{subsection}

\begin{subsection}{Proof of Lemma \ref{lemm:threshold_comparison}}
\label{subsec:threshold_comparison}
Let $q_{\zeta}(Z)$ denote the $\zeta$-th quantile of a random variable $Z$. We note that for a random variable $Z$, $q_{\zeta}(Z) = -q_{1-\zeta}(-Z).$  

We note that 
\begin{align*}
-c_{1}^{-}(x) &= -\EE[P]{-Y(1) \mid -Y(1) \geq q_{\zeta(\Gamma)}(-Y(1) \mid x), X=x} \\
&= \EE[P]{Y(1) \mid Y(1) \leq - q_{\zeta(\Gamma)}(-Y(1) \mid x), X=x} \\
&= \EE[P]{Y(1) \mid Y(1) \leq q_{1 - \zeta(\Gamma)} (Y(1) \mid x), X=x} \\
&\leq \EE[P]{Y(1) \mid Y(1) \geq q_{\zeta(\Gamma)}(Y(1) \mid x), X=x} \\
&= c_{1}^{+}(x).
\end{align*}
This implies that
\begin{align*}
H^{-}_{\Gamma}(x) &= (1 - \Gamma^{-1}) \cdot (-c_{1}^{-}(x) - c_{0}^{+}(x)) \\
&\leq (1 - \Gamma^{-1}) \cdot (c_{1}^{+}(x) - c_{0}^{+}(x)) \\
&= H_{\Gamma}(x).
\end{align*}

Also, we have that
\begin{align*}
-c_{0}^{+}(x) &= -\EE[P]{Y(0) \mid Y(0) \geq q_{\zeta(\Gamma)}(Y(0) \mid x), X=x} \\
&= \EE[P]{-Y(0) \mid -Y(0) \leq -q_{\zeta(\Gamma)}(Y(0)\mid x), X=x} \\
&= \EE[P]{-Y(0) \mid -Y(0) \leq q_{1 -\zeta(\Gamma)}(-Y(0)\mid x), X=x } \\
&\leq \EE[P]{-Y(0) \mid -Y(0) \geq q_{\zeta(\Gamma)}(-Y(0)\mid x), X=x } \\
&= c_{0}^{-}(x).
\end{align*}

This implies that 
\begin{align*}
H_{\Gamma}(x) &= (1 - \Gamma^{-1}) (c_{1}^{+}(x) - c_{0}^{+}(x))  \\
&\leq  (1 - \Gamma^{-1}) (c_{1}^{+}(x) + c_{0}^{-}(x)) \\
&= H^{+}_{\Gamma}(x).
\end{align*}

Combining these two results yields that $H^{-}_{\Gamma}(x) \leq H_{\Gamma}(x) \leq H^{+}_{\Gamma}(x).$

\end{subsection}

\begin{subsection}{Proof of Theorem \ref{theo:opt_regret}}
\label{subsec:opt_regret}
We note that \eqref{eq:minmax_regret} can be expressed as follows.
\begin{align*}
\inf_{\pi \in \Pi} \sup_{Q \in \mathcal{S}^{\Gamma}(P_{\text{obs}}, Q_{X})} R_{Q}(\pi(X)) &= \inf_{\pi \in \Pi} \sup_{Q \in \mathcal{S}^{\Gamma}(P_{\text{obs}}, Q_{X})} \sup_{\pi'} \EE[Q]{Y(\pi'(X))} - \EE[Q]{Y(\pi(X))} \\
&= \sup_{\pi \in \Pi} \inf_{Q \in \mathcal{S}^{\Gamma}(P_{\text{obs}}, Q_{X})} \EE[Q]{Y(\pi(X))} - \inf_{\pi' \in \Pi} \EE[Q]{Y(\pi'(X))} \\
&=  \sup_{\pi \in \Pi} \inf_{Q \in \mathcal{S}^{\Gamma}(P_{\text{obs}}, Q_{X})} \inf_{\pi' \in \Pi} \EE[Q]{Y(\pi(X))} -  \EE[Q]{Y(\pi'(X))} \\
&= \sup_{\pi \in \Pi} \inf_{\pi' \in \Pi} \inf_{Q \in \mathcal{S}^{\Gamma}(P_{\text{obs}}, Q_{X})}  \EE[Q]{Y(\pi(X))} -  \EE[Q]{Y(\pi'(X))}.
\end{align*}
We note that solving the above problem for any $Q_{X}$ such that $Q_{X} \ll P_{\text{obs}, X}$ and $\sup_{x \in \mathcal{X}} \frac{dP_{\text{obs}, X}(x)}{ dQ_{X}(x)} < \infty$ is equivalent to solving the conditional problem for every $x \in \text{supp}(P_{\text{obs}, X}).$
\begin{equation} \label{eq:regret_conditional} \sup_{\pi(x) \in \{0, 1\}} \inf_{\pi'(x) \in \{0, 1\}} \inf_{Q \in \mathcal{S}^{\Gamma}(P_{\text{obs}}, Q_{X})}  \EE[Q]{Y(\pi(X)) - Y(\pi'(X)) \mid X=x}. \end{equation}
Since $\EE[Q]{Y(\pi(X))} - \EE[Q]{Y(\pi'(X))} = \EE[Q]{(\pi(X) -\pi'(X)) \cdot (Y(1) - Y(0))}$, we can apply Lemma \ref{lemm:wc_gain} to \eqref{eq:regret_conditional}, which yields
\begin{equation}
\label{eq:game}
\begin{aligned}
 \sup_{\pi(x) \in \{0, 1\}} \inf_{\pi'(x) \in \{0, 1\}} &(\pi(x) - \pi'(x))_{+} g_1(x) + (\pi'(x) - \pi(x))_{+} g_0(x).
\end{aligned}
\end{equation}

When $\pi(x) = 1$, the objective function reduces to $(1 - \pi'(x))g_1(x)$ and hence the infimum over $\pi'$ is given by $\min\{g_1(x), 0\} = -(g_{1}(x))_{-}$. Similarly, when $\pi(x) = 0$, the infimum over $\pi'$ is given by $-(g_0(x))_{-}$.
Thus, the optimal policy $\pi$ is given by $\mathbb{I}(-(g_{1}(x))_{-} \geq -(g_{0}(x))_{-}).$

To simplify the form of this policy, we consider the four cases where (a) $g_{1}(x) \leq 0$ and $g_{0}(x) \leq 0,$ (b) $g_{1}(x) \leq 0$ and $g_{0}(x) > 0$, (c) $g_{1}(x) > 0$ and $g_{0}(x) \leq 0,$ and (d) $g_{1}(x) > 0$ and $g_{0}(x) > 0.$ From Lemma \ref{lemm:g}, we realize that we do not have to consider case (d) because $\min(g_{1}(x), g_{0}(x)) \leq 0$. In addition, we do not have to consider the part of case (b) where $g_{1}(x) = 0$ and $g_{0}(x) > 0$ because this cannot occur.

We consider case (a), where $g_{1}(x) \leq 0$ and $g_{0}(x) \leq 0.$ Then
\begin{align*}
\pi(x) &= \mathbb{I}(-(g_{1}(x))_{-} \geq -(g_{0}(x))_{-}) \\
&= \mathbb{I}(g_{1}(x) \geq g_{0}(x)).
\end{align*}

In case (b), we only need to consider $g_{1}(x) < 0$  and $g_{0}(x) > 0$ (because by Lemma \ref{lemm:g} $g_{1}(x) = 0, g_{0}(x) > 0$ does not occur). So, we have that
\begin{align*}
\pi(x) &= \mathbb{I}(-(g_{1}(x))_{-} \geq -(g_{0}(x))_{-}) \\
&= \mathbb{I}(g_{1}(x) \geq 0) \\
&= \mathbb{I}(g_{1}(x) \geq g_{0}(x)).
\end{align*}

In case (c), where $g_{1}(x) > 0$ and $g_{0}(x) \leq 0$. Then 
\begin{align*}
\pi(x) &= \mathbb{I}(-(g_{1}(x))_{-} \geq -(g_{0}(x))_{-}) \\
&= \mathbb{I}(0 \geq g_{0}(x)) \\
&= \mathbb{I}(g_{1}(x) \geq g_{0}(x)).
\end{align*}

We note case (d) does not occur by Lemma \ref{lemm:g}.

~\\

Combining the results, we have that $\pi(x) = \mathbb{I}(g_{1}(x) \geq g_{0}(x))$. We can express $\mathbb{I}(g_{1}(x) \geq g_{0}(x))$ as threshold rule on the CATE, as follows.

\begin{align*}
\mathbb{I}(g_{1}(x) \geq g_{0}(x)) &= \mathbb{I}(\Gamma \tau(x) + (1 - \Gamma) (c_{1}^{+}(x) + c_{0}^{-}(x)) \geq - \Gamma \tau(x) +(1- \Gamma)(c_{0}^{+}(x) + c_{1}^{-}(x)) ) \\
&= \mathbb{I}\Bigg(2\Gamma \tau(x) \geq (1- \Gamma) \Big( c_{0}^{+}(x) + c_{1}^{-}(x) - c_{1}^{+}(x) - c_{0}^{-}(x)\Big)\Bigg) \\
&= \mathbb{I} \Big( \tau(x) \geq \frac{1}{2}(1 - \Gamma^{-1}) \cdot (c_{1}^{+}(x) +c_{0}^{-}(x)-c_{0}^{+}(x) - c_{1}^{-}(x)) \Big) \\
&= \mathbb{I} \Big(\tau(x) \geq \frac{H_{\Gamma}^{+}(x) + H_{\Gamma}^{-}(x)}{2}\Big).
\end{align*}

\end{subsection}

\begin{subsection}{Proof of Theorem \ref{theo:po_regret}}
\label{subsec:po_regret}
By Lemma \ref{lemm:np}, if $\pi$ solves \eqref{eq:wc_value_regret} for any $Q_{X}$ such that $Q_{X} \ll P_{\text{obs}, X}$ and $\sup_{x \in \mathcal{X}} \frac{dP_{\text{obs}, X}(x)}{dQ_{X}(x)} < \infty$ then it solves
\begin{equation} \label{eq:cond_po_regret} \max_{\pi(x) \in \{0, 1\}} \inf_{P \in \mathcal{T}(P_{\text{obs}})} \Gamma \EE[P]{\tilde{v}_{\text{regret}}(\pi(x); X, Y(0), Y(1)) \mid X=x} + (1- \Gamma) \text{CVaR}_{\zeta(\Gamma)}(\tilde{v}_{\text{regret}}(\pi(x); X, Y(0), Y(1)) \mid x).
\end{equation}
for every $x \in \text{supp}(P_{\text{obs}, X}).$

We note that
\begin{align*}
\EE[P]{\tilde{v}_{\text{regret}}(\pi(x); X, Y(0), Y(1)) \mid X=x} &= (2\pi(x) - 1) \tau(x) \\
&= (2\pi(x) - 1)_{+}\tau(x) +(-2\pi(x) +1)_{+}(-\tau(x)),
\end{align*}
and is identified under $P_{\text{obs}}$ and does not depend on the choice of $P \in \mathcal{T}(P_{\text{obs}}).$

In addition, we have that

\begin{align}
&\inf_{P \in \mathcal{T}(P_{\text{obs}})}  (1 - \Gamma) \text{CVaR}_{\zeta(\Gamma)}(\tilde{v}_{\text{regret}}(\pi(x); X, Y(0), Y(1)) \mid x) \\
&= (1 - \Gamma) \sup_{P \in \mathcal{T}(P_{\text{obs}})} \text{CVaR}_{\zeta(\Gamma)}(\tilde{v}_{\text{regret}}(\pi(x); X, Y(0), Y(1)) \mid x) \\
&= (1 - \Gamma) \sup_{P \in \mathcal{T}(P_{\text{obs}})} \text{CVaR}_{\zeta(\Gamma)}( (2\pi(x) - 1)(Y(1) - Y(0)) \mid x)  \\
&= (1 - \Gamma) \sup_{P \in \mathcal{T}(P_{\text{obs}})} (2\pi(x) -1)_{+}\text{CVaR}_{\zeta(\Gamma)}(Y(1) - Y(0) \mid x) + (-2\pi(x) + 1)_{+}\text{CVaR}_{\zeta(\Gamma)}(Y(0) - Y(1) \mid x) \label{eq:pos2} \\
&= (1 - \Gamma) \Big( (2\pi(x) -1)_{+}(c_{1}^{+}(x) + c_{0}^{-}(x)) + (-2\pi(x) + 1)_{+}(c_{0}^{+}(x) + c_{1}^{-}(x)) \Big). \label{eq:decomp2}
\end{align}

\eqref{eq:pos2} follows from Lemma \ref{lemm:cvar_properties} and \eqref{eq:decomp2} follows from Lemma \ref{lemm:cvar_decomposition}.

Thus, we can write \eqref{eq:cond_po_regret} as 
\begin{equation} \max_{\pi(x) \in \{0, 1\}} (2\pi(x) - 1)_{+}(\Gamma \tau(x) + (1- \Gamma)(c_{1}^{+}(x) + c_{0}^{-}(x))) + (-2\pi(x) + 1)_{+}(-\Gamma\tau(x) + (1 - \Gamma)(c_{0}^{+}(x) + c_{1}^{-}(x))).
\end{equation}

Since $\pi(x)$ that maximizes the above expression must be either $0$ or $1$, we realize that $\pi(x) = 1$ when $\Gamma \tau(x) + (1- \Gamma)(c_{1}^{+}(x) + c_{0}^{-}(x))) \geq (-\Gamma\tau(x) + (1 - \Gamma)(c_{0}^{+}(x) + c_{1}^{-}(x)))$ and is $0$ otherwise. Thus, $\pi$ that maximizes the above expression is equal to \eqref{eq:opt_regret}.
\end{subsection}

\begin{subsection}{Proof of Theorem \ref{theo:equivalence}}
\label{subsec:equivalence}

We aim to show that any policy that solves \eqref{eq:observed_data_problem} also solves \eqref{eq:potential_outcomes_problem}. Lemma \ref{lemm:potential_robustness_implies_observed_robustness} implies that for any $Q \in \mathcal{S}^{\Gamma}(P_{\text{obs}}, Q_{X}),$ there is a $Q_{\text{obs}} \in \mathcal{S}^{\Gamma}(P_{\text{obs}}, Q_{X})$ where $Q$ is consistent with $Q_{\text{obs}}$ under an RCT. So, by Assumption \ref{assumption:equal_value} we have that
\begin{equation}
  \label{eq:inf_comparison}
   \inf_{Q \in \mathcal{S}^{\Gamma}(P_{\text{obs}}, Q_{X})} \EE[Q]{v^{*}(\pi(X); X, Y(0), Y(1))} \geq \inf_{Q \in \mathcal{S}^{\Gamma}_{\text{obs}}(P_{\text{obs}}, Q_{X})} \EE[Q_{\text{obs}}]{v(\pi(X); X, Y, W)}.
\end{equation}
This yields that
\begin{equation} \label{eq:greater_than} \sup_{\pi \in \Pi} \inf_{Q \in \mathcal{S}^{\Gamma}(P_{\text{obs}}, Q_{X})} \EE[Q]{v^{*}(\pi(X); X, Y(0), Y(1))} \geq \sup_{\pi \in \Pi} \inf_{Q \in \mathcal{S}^{\Gamma}_{\text{obs}}(P_{\text{obs}}, Q_{X})} \EE[Q_{\text{obs}}]{v(\pi(X); X, Y, W)}. \end{equation}

It remains to show
\begin{equation}
\label{eq:observed_data_proof_main}
\sup_{\pi \in \Pi} \inf_{Q \in \mathcal{S}^{\Gamma}(P_{\text{obs}}, Q_{X})} \EE[Q]{v^{*}(\pi(X); X, Y(0), Y(1))} \leq \sup_{\pi \in \Pi} \inf_{Q \in \mathcal{S}^{\Gamma}_{\text{obs}}(P_{\text{obs}}, Q_{X})} \EE[Q_{\text{obs}}]{v(\pi(X); X, Y, W)}.
\end{equation}
In fact, if we denote by $f_{L}(\pi)$ and $f_{R}(\pi)$ the LHS and RHS of \eqref{eq:inf_comparison}, $f_{L}(\pi)\ge f_{R}(\pi)$ for every $\pi\in \Pi$. Let $\pi_{R}^{*}$ be any maximizer of $f_{R}(\pi)$.   Then \eqref{eq:greater_than} and \eqref{eq:observed_data_proof_main} imply $\sup_{\pi\in \Pi}f_{L}(\pi) = f_{R}(\pi_{R}^{*})$.  If there exists $\pi \in \Pi$ such that $f_{L}(\pi) > f_{L}(\pi_{R}^{*})$, then
  \[f_{L}(\pi) > f_{L}(\pi_{R}^{*})\ge f_{R}(\pi_{R}^{*}) = \sup_{\pi\in \Pi}f_{L}(\pi),\]
  which yields contradiction. Thus, $\pi_{R}^{*}$ must be a maximier of $f_{L}(\pi)$.

  Let \[\mathcal{A}_{x, w} = \{ y \in \mathbb{R} \mid v(\pi(x); x, y, w) \geq q_{\zeta(\Gamma)}(v(\pi(x); X, Y, W) \mid x, w) \}.\] Recall that from Lemma \ref{lemm:np_obs_data}, $q_{\zeta(\Gamma)}(v(\pi(X); X, Y, W) \mid x, w)$ is the $\zeta(\Gamma)$-th quantile of $v(\pi(x); x, Y, w)$ where $Y \sim P_{\text{obs}, Y \mid X=x, W=w}$. Since $P_{\text{obs}, Y\mid X=x, W=w}$ is absolutely continuous with respect to Lebesgue measure, we must have that
\begin{equation}
\label{eq:A_mass} 
\PP[P_{\text{obs}, Y \mid X, W = w}]{Y \in \mathcal{A}_{X, w} \mid X=x} = \frac{\Gamma}{\Gamma + 1} \quad \forall x \in \mathcal{X}, w \in \{0, 1\}. 
\end{equation}
By Lemma \ref{lemm:np_obs_data}, the distribution $dQ^{*}_{\text{obs}}$ that achieves the infimum on the right side of \eqref{eq:observed_data_proof_main} is given by $dQ_{X} \cdot dP_{\text{obs}, W \mid X} \cdot dQ^{*}_{\text{obs}, Y \mid X, W},$ where $dQ^{*}_{\text{obs}, Y \mid X, W} = \theta^{(W)}(Y \mid X)\cdot  dP_{\text{obs}, Y \mid X, W}$ and 
\begin{equation}
\label{eq:weight}
\theta^{(w)}(y \mid x) = \Gamma^{-1}\cdot \mathbb{I}(y\in \mathcal{A}_{x, w}) + \Gamma \cdot \mathbb{I}(y\not\in \mathcal{A}_{x, w}).
\end{equation}

To show \eqref{eq:observed_data_proof_main}, we show that there exists a potential outcome distribution $Q \in \mathcal{S}_{\Gamma}(P_{\text{obs}}, Q_{X})$ that yields the same worst-case value as $Q^{*}_{\text{obs}}$ for all $\pi$. Under Assumption \ref{assumption:equal_value}, it is sufficient to show that there is a $Q \in \mathcal{S}_{\Gamma}(P_{\text{obs}}, Q_{X})$ that is consistent with $Q_{\text{obs}}^{*}$ under an RCT with treatment propensity $e$. To do so, we first construct a potential outcome distribution $P$ that is consistent with $P_{\text{obs}}$ under an RCT and has the special property that for $(X, Y(0), Y(1)) \sim P$, $\theta^{(0)}(Y(0) \mid X)= \theta^{(1)}(Y(1) \mid X)$. After that, we define $Q$ in terms of $P$ and check that $Q \in \mathcal{S}^{\Gamma}(P_{\text{obs}}, Q_{X})$. 

We construct $P$ as follows. We first sample $X\sim P_{X}$ and $Y(0)\sim P_{\text{obs}, Y\mid X, W = 0}$. Then we sample $Y(1)$ as follows.
\begin{itemize}
\item If $Y(0)\in \mathcal{A}_{X, 0}$, sample
  \[Y(1) \sim P_{\text{obs}, Y\mid X, W = 1, Y \in \mathcal{A}_{X, 1}}.\]
\item If $Y(0)\not\in \mathcal{A}_{X, 0}$, sample
  \[Y(1) \sim P_{\text{obs}, Y\mid X, W = 1, Y \not\in \mathcal{A}_{X, 1}}.\]
\end{itemize}
Now we prove $P_{(X, Y(0), Y(1))}$ is consistent with $P_{\text{obs}}$. By construction,
  \[P_{X} = P_{\text{obs}, X}, \quad P_{Y(0)\mid X} = P_{\text{obs}, Y\mid X, W = 0},\]
  and
  \begin{align*}
    P_{Y(1)\mid X} &= \PP{Y(0)\in \mathcal{A}_{X, 0}} P_{\text{obs}, Y \mid X, W = 1, Y\in \mathcal{A}_{X, 1}} + \PP{Y(0)\not\in \mathcal{A}_{X, 0}}P_{\text{obs}, Y \mid X, W = 1, Y\not\in \mathcal{A}_{X, 1}}\\
                   & = \PP[P_{\text{obs}, Y\mid X, W = 0}]{Y\in \mathcal{A}_{X, 0}} P_{\text{obs}, Y \mid X, W = 1, Y\in \mathcal{A}_{X, 1}} + \PP[P_{\text{obs}, Y\mid X, W = 0}]{Y\not\in \mathcal{A}_{X, 0}}P_{\text{obs}, Y \mid X, W = 1, Y\not\in \mathcal{A}_{X, 1}}\\
                   & = \PP[P_{\text{obs}, Y\mid X, W = 1}]{Y\in \mathcal{A}_{X, 1}} P_{\text{obs}, Y \mid X, W = 1, Y\in \mathcal{A}_{X, 1}} + \PP[P_{\text{obs}, Y\mid X, W = 1}]{Y\not\in \mathcal{A}_{X, 1}}P_{\text{obs}, Y \mid X, W = 1, Y\not\in \mathcal{A}_{X, 1}}\\
    & = P_{\text{obs}, Y\mid X, W = 1},
  \end{align*}
  where the applies \eqref{eq:A_mass}. Thus, $P$ is consistent with $P_{\text{obs}}$ under an RCT.

Given such a potential outcome distribution $P$, we can define a potential outcome distribution $Q$ with covariate distribution equal to $Q_{\text{obs}, X}^{*}$ and
\begin{equation}
\label{eq:Q} 
\frac{dQ_{Y(0), Y(1) \mid X=x}(y_{0}, y_{1})}{dP_{Y(0), Y(1) \mid X=x}(y_{0}, y_{1})} = \theta^{(0)}(y_{0}\mid x).
\end{equation}
We can show that $Q$ defined by \eqref{eq:Q} is a probability measure for any $x$.
\begin{align*}
\int dQ_{Y(0), Y(1) \mid X=x} &= \int  \theta^{(0)}(y_{0} \mid x) dP_{Y(0), Y(1) \mid X = x}(y_{0}, y_{1})\\
&= \int \theta^{(0)}(y_{0} \mid x) dP_{Y(0) \mid X = x}(y_{0}) \\
&=  \int \theta^{(0)}(y_{0} \mid x) dP_{\text{obs}, Y \mid X=x, W=0}(y_{0})\\
&= \int dQ^{*}_{\text{obs}, Y \mid X=x, W=0}(y_{0})\\
&= 1.
\end{align*}
In addition, because $\theta^{(0)}(Y(0) \mid X) = \theta^{(1)}(Y(1) \mid X)$ on the support of $P,$ we also have that 
\begin{equation}
\label{eq:Q_joint} 
\frac{dQ_{Y(0), Y(1) \mid X=x}(y_{0}, y_{1})}{dP_{Y(0), Y(1) \mid X=x}(y_{0}, y_{1})} = \theta^{(0)}(y_{0}\mid x) = \theta^{(1)}(y_{1} \mid x).
\end{equation}

Marginalizing over $Y(0)$ and $Y(1)$, respectively, we realize that 
\[ \frac{dQ_{Y(0) \mid X=x}(y_{0})}{dP_{Y(0) \mid X=x}(y_{0})} = \theta^{(0)}(y_{0} \mid x), \quad  \frac{dQ_{Y(1) \mid X=x}(y_{1})}{dP_{Y(1) \mid X=x}(y_{1})} = \theta^{(1)}(y_{1} \mid x).\]
Since $P_{Y(w) \mid X=x} = P_{\text{obs}, Y\mid X =x,W=w}$, \eqref{eq:weight} implies 
\[Q_{Y(w) \mid X=x} = Q_{\text{obs}, Y\mid X=x, W=w}^{*}.\]
This demonstrates that $Q$ is consistent with $Q_{\text{obs}}$ under an RCT. 

Furthermore, we can check that $Q \in \mathcal{S}^{\Gamma}(P_{\text{obs}}, Q_{X})$. We have already shown that $P$ is consistent with $P_{\text{obs}}$ under an RCT, and we can verify that $Q$ generates $P$ via $\Gamma$-biased sampling. We have that for all $x \in \mathcal{X}$ and $y_{0}, y_{1} \in \mathbb{R},$
\begin{align} \frac{dQ_{Y(0), Y(1) \mid X=x}(y_{0}, y_{1})}{dP_{Y(0), Y(1) \mid X=x}(y_{0}, y_{1})} &= \theta^{(0)}(y_{0}\mid x) \label{eq:q_1} \\
&= \frac{dQ_{\text{obs}, Y \mid X=x, W=0}^{*}(y_{0}) }{dP_{\text{obs}, Y \mid X=x, W=0}(y_{0})} \label{eq:q_2} \\
&\in [\Gamma^{-1}, \Gamma]. \label{eq:q_3}
\end{align}
 \eqref{eq:q_1} follows from \eqref{eq:Q_joint}. \eqref{eq:q_2} holds by the definition of $\theta^{(w)}(Y \mid X)$ in \eqref{eq:weight}. Lastly, \eqref{eq:q_3} holds because $Q^{*}_{\text{obs}} \in \mathcal{S}^{\Gamma}_{\text{obs}}(P_{\text{obs}}, Q_{X}).$ 
By Assumption \ref{assumption:equal_value}, this result implies that there is a distribution $Q \in\mathcal{S}^{\Gamma}(P_{\text{obs}}, Q_{X})$ that yields the same policy value as $Q^{*}_{\text{obs}}$ for all $h$. So, we must have that
 \begin{align*} \inf_{Q \in \mathcal{S}^{\Gamma}(P_{\text{obs}}, Q_{X})} \EE[Q]{v^{*}(\pi(X); X, Y(0), Y(1))} &\leq  \EE[Q^{*}_{\text{obs}}]{v(\pi(X); X, Y, W)} \\
 &= \inf_{Q_{\text{obs}} \in \mathcal{S}^{\Gamma}_{\text{obs}}(P_{\text{obs}}, Q_{X})} \EE[Q_{\text{obs}}]{v(\pi(X); X, Y, W)}.
 \end{align*}
So, we must have \eqref{eq:observed_data_proof_main}. Finally, \eqref{eq:greater_than} and \eqref{eq:observed_data_proof_main} yield the desired claim.

\end{subsection}

\begin{subsection}{Proof of Theorem \ref{theo:ru_reg}}
\label{subsec:ru_reg}
We note that any $h$ that solves \eqref{eq:observed_data_problem} solves the conditional worst-case value maximization problem
\begin{equation}
\label{eq:observed_cond}
\sup_{h(x) \in [0, 1]} \inf_{Q_{\text{obs}} \in \mathcal{S}^{\Gamma}_{\text{obs}}(P_{\text{obs}}, Q_{X}) } \EE[Q_{\text{obs}}]{v(h(x); X, Y, W) \mid X=x}
\end{equation}
for every $x \in \text{supp}(P_{\text{obs}, X}).$

By Lemma \ref{lemm:np_obs_data}, we can write that \eqref{eq:observed_cond} is equivalent to
\begin{equation}
\label{eq:np3}
\begin{aligned}
\sup_{h(x) \in [0, 1]} &\sum_{w \in \{0, 1\}} \Gamma \PP{W=w \mid X=x} \cdot \EE[P_{\text{obs}}]{v(h(x); X, Y, W) \mid X=x, W=w} \\
&\indent+ (\Gamma^{-1} - \Gamma) \EE[P_{\text{obs}}]{v(h(x); X, Y, W) \mathbb{I}(v(h(x); X, Y, W) \geq q_{\zeta(\Gamma)}(v(h(x); X, Y, W) \mid x, w)) \mid X=x, W=w}.
\end{aligned}
\end{equation}

We note that 
\begin{align*}
&\EE[P_{\text{obs}}]{v(h(x); X, Y, W) \mathbb{I}(v(h(x); X, Y, W) \geq q_{\zeta(\Gamma)}(v(h(x); X, Y, W) \mid x, w)) \mid X=x, W=w}\\
&= (1 - \zeta(\Gamma)) \cdot \text{CVaR}_{\zeta(\Gamma)}(v(h(x); X, Y, W) \mid X=x, W=w).
\end{align*}

Thus, we have that \eqref{eq:np3} can be written as

\begin{equation}
\begin{aligned}
\sup_{h(x) \in [0, 1]} \sum_{w \in \{0, 1\}} \PP{W= w \mid X=x} &\Big(\Gamma \EE[P_{\text{obs}}]{v(h(x); X, Y, W) \mid X=x, W=w} \\
&\indent + (1 - \Gamma) \cdot \text{CVaR}_{\zeta(\Gamma)}(v(h(x); X, Y, W) \mid x, w) \Big). \\
\end{aligned}
\end{equation}

We can apply Lemma \ref{lemm:cvar_properties} to rewrite this expression as
\begin{equation}
\begin{aligned}
\sup_{h(x) \in [0, 1]} \sum_{w \in \{0, 1\}} \PP{W= w \mid X=x} &\Big(\Gamma^{-1} \EE[P_{\text{obs}}]{v(h(x); X, Y, W) \mid X=x, W=w} \\
&+ (\Gamma^{-1} - 1) \cdot \text{CVaR}_{1-\zeta(\Gamma)}(-v(h(x); X, Y, W) \mid x, w)\Big).\\
\end{aligned}
\end{equation}

We can transform this problem into a minimization problem as follows.

\begin{equation}
\label{eq:minimization}
\begin{aligned}
\inf_{h(x) \in [0, 1]} \sum_{w \in \{0, 1\}} \PP{W= w \mid X=x} &\Big(\Gamma^{-1}\EE[P_{\text{obs}}]{-v(h(x); X, Y, W) \mid X=x, W=w} \\
&+ (1- \Gamma^{-1} ) \cdot \text{CVaR}_{1-\zeta(\Gamma)}(-v(h(x); X, Y, W) \mid X=x, W=w)\Big) \\
\end{aligned}
\end{equation}

We can apply Theorem 2 of \cite{rockafellar2000optimization} to see that \eqref{eq:minimization} can be written as

\begin{equation}
\label{eq:ru}
\begin{aligned}
\inf_{h(x) \in [0, 1]} &\Gamma^{-1} \EE[P_{\text{obs}}]{-v(h(x); X, Y, W) \mid X=x} \\
&+ e \cdot (1 - \Gamma^{-1}) \cdot \inf_{\alpha(x, 1) \in \mathbb{R}} \Big(\alpha(x, 1) + \zeta(\Gamma)^{-1} \EE[P_{\text{obs}}]{(-v(h(x); X, Y, W) - \alpha(x, 1))_{+} \mid X=x, W=1} \Big) \\
&+(1 - e) \cdot (1 - \Gamma^{-1}) \cdot \inf_{\alpha(x, 0) \in \mathbb{R}} \Big(\alpha(x, 0) + \zeta(\Gamma)^{-1} \EE[P_{\text{obs}}]{(-v(h(x); X, Y, W) - \alpha(x, 0))_{+} \mid X=x, W=0} \Big).
\end{aligned}
\end{equation}

Thus, we have the joint minimization problem

\begin{equation}
\label{eq:ru_loss_terms}
\begin{aligned}
\inf_{h(x) \in [0, 1], \alpha(x, w) \in \mathbb{R}} &\Gamma^{-1} \EE[P_{\text{obs}}]{-v(h(x); X, Y, W) \mid X=x} \\
&+ e \cdot (1 - \Gamma^{-1}) \cdot \Big(\alpha(x, 1) + \zeta(\Gamma)^{-1} \EE[P_{\text{obs}}]{(-v(h(x); X, Y, W) - \alpha(x, 1))_{+} \mid X=x, W=1} \Big) \\
&+(1 - e) \cdot (1 - \Gamma^{-1}) \cdot \Big(\alpha(x, 0) + \zeta(\Gamma)^{-1} \EE[P_{\text{obs}}]{(-v(h(x); X, Y, W) - \alpha(x, 0))_{+} \mid X=x, W=0} \Big).
\end{aligned}
\end{equation}

This resembles the conditional RU regression problem.
\begin{equation}
\label{eq:ru_cond}
\begin{aligned}
\inf_{h(x) \in [0, 1], \alpha(x, w) \in \mathbb{R}} &e \EE[P_{\text{obs}}]{L_{\text{RU}}(h(X), \alpha(X, W); X, Y, W) \mid X=x, W=1} \\
&+ (1 - e) \EE[P_{\text{obs}}]{L_{\text{RU}}(h(X), \alpha(X, W); X, Y, W) \mid X=x, W=0}.
\end{aligned}
\end{equation}

Finally, any functions $h, \alpha(\cdot, 1), \alpha(\cdot, 0)$ that solve the conditional RU regression problem \eqref{eq:ru_cond} for every $x \in \text{supp}(P_{\text{obs}, X})$ must solve \eqref{eq:ru_regression}. Thus, any solution \eqref{eq:ru_regression} is a solution to \eqref{eq:observed_data_problem}.

\end{subsection}

\begin{subsection}{Proof of Theorem \ref{theo:po_maxmin}}
\label{subsec:po_maxmin}
By Lemma \ref{lemm:np}, if $h$ solves \eqref{eq:wc_value_maxmin} for every $Q_{X}$ such that $Q_{X} \ll P_{\text{obs}, X}$ and $\sup_{x \in \mathcal{X}} \frac{dP_{\text{obs}, X}(x)}{dQ_{X}(x)} < \infty$ then it solves
\begin{equation} \label{eq:cond_po_maxmin} \max_{h(x) \in [0, 1]} \inf_{P \in \mathcal{T}(P_{\text{obs}})} \Gamma \EE[P]{v^{*}_{\text{maxmin}}(h(x); X, Y(0), Y(1)) \mid X=x} + (1- \Gamma) \text{CVaR}_{\zeta(\Gamma)}(v^{*}_{\text{maxmin}}(h(x); X, Y(0), Y(1)) \mid x)\end{equation}
for every $x \in \text{supp}(P_{\text{obs}, X}).$
We note that 
\begin{align*}
&\EE[P]{v^{*}_{\text{maxmin}}(h(x); X, Y(0), Y(1)) \mid X=x} \\
&= \EE[P]{\log(1 + \exp(2h(x) -1)) Y(1) + \log(1 + \exp(-2h(x) + 1))Y(0) \mid X=x} \\
&= \log(1 + \exp(2h(x) -1)) \EE[P]{Y(1) \mid X=x} +  \log(1 + \exp(-2h(x) + 1)) \EE[P]{Y(0) \mid X=x}.
\end{align*}
So, the first term of \eqref{eq:cond_po_gain} is identifiable under $P_{\text{obs}}$ and does not depend on the choice of $P \in \mathcal{T}(P_{\text{obs}}).$ Next, applying Lemma \ref{lemm:cvar_decomposition} and Lemma \ref{lemm:cvar_properties}, we realize that
\begin{align*}
&\inf_{P \in \mathcal{T}(P_{\text{obs}})}  (1 - \Gamma) \text{CVaR}_{\zeta(\Gamma)}(v^{*}_{\text{maxmin}}(h(x); X, Y(0), Y(1)) \mid x) \\
&= (1 - \Gamma) \sup_{P \in \mathcal{T}(P_{\text{obs}})} \text{CVaR}_{\zeta(\Gamma)}(v^{*}_{\text{maxmin}}(h(x); X, Y(0), Y(1)) \mid x) \\
&= (1 - \Gamma) \sup_{P \in \mathcal{T}(P_{\text{obs}})} \text{CVaR}_{\zeta(\Gamma)}\Big(\log(1 + \exp(2h(x) -1)) Y(1) + \log(1 + \exp(-2h(x) + 1))Y(0) \mid x\Big) \\
&= (1 - \Gamma) \Big( \log(1 + \exp(2h(x) - 1)) c_{1}^{+}(x) + \log(1 + \exp(-2h(x) + 1)) c_{0}^{+}(x) \Big).
\end{align*}

Let $P$ be a potential outcome distribution in $\mathcal{T}(P_{\text{obs}})$. \eqref{eq:cond_po_maxmin} can be written as
\begin{equation}\label{eq:simple_cond_po_maxmin} \begin{aligned}\max_{h(x) \in [0, 1]} &\log(1 + \exp(2h(x) - 1)) \cdot (\Gamma \EE[P]{Y(1) \mid X=x} + (1 - \Gamma) c_{1}^{+}(x)) \\
&+ \log (1 + \exp(-2h(x) + 1)) \cdot (\Gamma \EE[P]{Y(0) \mid X=x} + (1 - \Gamma) c_{0}^{+}(x)).
\end{aligned}
\end{equation}
By Lemma \ref{lemm:proper_scoring_rule}, $h^{*}(x)$ that solves \eqref{eq:simple_cond_po_maxmin} satisfies $h^{*}(x) \geq \frac{1}{2}$ when $\Gamma \EE[P]{Y(1) \mid X=x} + (1 - \Gamma) c_{1}^{+}(x) \geq \Gamma \EE[P]{Y(0) \mid X=x} + (1 - \Gamma) c_{0}^{+}(x)$. In other words, $\pi(x) = \mathbb{I}(h^{*}(x) \geq \frac{1}{2})$ is equivalent to $\mathbb{I}(\tau(x) \geq (1 - \Gamma^{-1}) \cdot (c_{1}^{+}(x) - c_{0}^{+}(x))$, which matches \eqref{eq:opt_maxmin}.

\end{subsection}

\begin{subsection}{Proof of Theorem \ref{theo:po_gain}}
\label{subsec:po_gain}
By Lemma \ref{lemm:np}, if $h$ solves \eqref{eq:wc_value_gain} for any $Q_{X}$ such that $Q_{X} \ll P_{\text{obs}, X}$ and $\sup_{x \in \mathcal{X}} \frac{dP_{\text{obs}, X}}{dQ_{X}(x)} < \infty$ then it solves
\begin{equation} \label{eq:cond_po_gain} \max_{h(x) \in [0, 1]} \inf_{P \in \mathcal{T}(P_{\text{obs}})} \Gamma \EE[P]{v^{*}_{\text{gain}}(h(x); X, Y(0), Y(1)) \mid X=x} + (1- \Gamma) \text{CVaR}_{\zeta(\Gamma)}(v^{*}_{\text{gain}}(h(x); X, Y(0), Y(1)) \mid x). \end{equation}
for every $x \in \text{supp}(Q_{X}).$

We note that 
\begin{align*}
&\EE[P]{v^{*}_{\text{gain}}(h(x); X, Y(0), Y(1)) \mid X=x} \\
&= \EE[P]{(1-\pi_{0}(X)) \log(1 + \exp(2h(x) -1)) (Y(1) - Y(0)) + \pi_{0}(X) \log(1 + \exp(-2h(x)))(Y(0) - Y(1)) \mid X=x} \\
&= (1-\pi_{0}(x)) \log(1 + \exp(2h(x) -1)) \tau(x) + \pi_{0}(x) \cdot \log(1 + \exp(-2h(x) + 1)) \cdot (-\tau(x))
\end{align*}
so the first term is identifiable under $P_{\text{obs}}$ and does not depend on the choice of $P \in \mathcal{T}(P_{\text{obs}}).$ Next, applying Lemma \ref{lemm:cvar_decomposition} and Lemma \ref{lemm:cvar_properties} we realize that
\begin{align*}
&\inf_{P \in \mathcal{T}(P_{\text{obs}})}  (1 - \Gamma) \text{CVaR}_{\zeta(\Gamma)}(v^{*}_{\text{gain}}(h(x); X, Y(0), Y(1)) \mid x) \\
&= (1 - \Gamma) \sup_{P \in \mathcal{T}(P_{\text{obs}})} \text{CVaR}_{\zeta(\Gamma)}(v^{*}_{\text{gain}}(h(x); X, Y(0), Y(1)) \mid x) \\
&= (1 - \Gamma) \sup_{P \in \mathcal{T}(P_{\text{obs}})} \text{CVaR}_{\zeta(\Gamma)}( (1 - \pi_{0}(x)) \log(1 + \exp(2h(x) -1)) (Y(1) - Y(0))  \\
&\indent\indent\indent\indent\indent\indent\indent\indent+ \pi_{0}(x) \log(1 + \exp(-2h(x)))(Y(0) - Y(1)) \mid x) \\
&= (1 - \Gamma) \Big( (1 - \pi_{0}(x)) \cdot \log(1 + \exp(2h(x) - 1)) (c_{1}^{+}(x) + c_{0}^{-}(x)) +  \pi_{0}(x)\cdot \log(1 + \exp(-2h(x) + 1)) (c_{0}^{+}(x) + c_{1}^{-}(x)) \Big).
\end{align*}
Thus, we can rewrite \eqref{eq:cond_po_gain} as
\begin{equation}
\begin{aligned}
\max_{h(x) \in [0, 1]} &(1 - \pi_{0}(x)) \log(1 + \exp(2h(x) -1)) \cdot \Big(\Gamma \tau(x) +  (1 - \Gamma)(c_{1}^{+}(x) + c_{0}^{-}(x))\Big) \\
&+ \pi_{0}(x) \log(1 + \exp(-2h(x) +1)) \Big(-\Gamma \tau(x) +  (1 - \Gamma)(c_{1}^{-}(x) + c_{0}^{+}(x))\Big).
\end{aligned}
\end{equation}

When $\pi_{0}(x) = 0,$ the above problem reduces to
\[ \max_{h(x) \in [0, 1]}  \log(1 + \exp(2h(x) -1)) \cdot \Big(\Gamma \tau(x) +  (1 - \Gamma)(c_{1}^{+}(x) + c_{0}^{-}(x))\Big).\]
Applying Lemma \ref{lemm:proper_scoring_rule}, $h^{*}(x)$ that maximizes the above equation satisfies $h^{*}(x) \geq \frac{1}{2}$ when $\Big(\Gamma \tau(x) +  (1 - \Gamma)(c_{1}^{+}(x) + c_{0}^{-}(x))\Big) \geq 0.$ As a consequence, $\mathbb{I}(h^{*}(x) \geq \frac{1}{2})$ is equal to $\mathbb{I}(\tau(x) \geq H^{+}(x))$ when $\pi_{0}(x) = 0.$

When $\pi_{0}(x) = 1,$ the above problem reduces to 
\[ \max_{h(x) \in [0, 1]}  \log(1 + \exp(-2h(x) +1)) \cdot \Big(-\Gamma \tau(x) +  (1 - \Gamma)(c_{1}^{-}(x) + c_{0}^{+}(x))\Big).\]
Applying Lemma \ref{lemm:proper_scoring_rule}, $h^{*}(x)$ that maximizes the above equation satisfies $h^{*}(x) \geq \frac{1}{2}$ when $0 \geq -\Gamma \tau(x) +  (1 - \Gamma)(c_{1}^{-}(x) + c_{0}^{+}(x)).$ As a consequence $\mathbb{I}(h^{*}(x) \geq \frac{1}{2})$ is equal to $\mathbb{I}(\tau(x) \geq H^{-}_{\Gamma}(x))$ when $\pi_{0}(x) = 1.$
Thus, $\pi(x) =  \mathbb{I}(h^{*}(x) \geq \frac{1}{2})$ is equivalent to \eqref{eq:opt_gain}.

\end{subsection}

\begin{subsection}{Proof of Theorem \ref{theo:po_regret_loss}}
\label{subsec:po_regret_loss}
By Lemma \ref{lemm:np}, if $h$ solves \eqref{eq:wc_value_regret_loss} for any $Q_{X}$ such that $Q_{X} \ll P_{\text{obs}, X}$ and $\sup_{x \in \mathcal{X}} \frac{dP_{\text{obs}, X}(x)}{dQ_{X}(x)} < \infty$ then it solves
\begin{equation} \label{eq:cond_po_regret_loss} \max_{h(x) \in [0, 1]} \inf_{P \in \mathcal{T}(P_{\text{obs}})} \Gamma \EE[P]{v^{*}_{\text{regret}}(h(x); X, Y(0), Y(1)) \mid X=x} + (1- \Gamma) \text{CVaR}_{\zeta(\Gamma)}(v^{*}_{\text{regret}}(h(x); X, Y(0), Y(1)) \mid x). \end{equation}
for every $x \in \text{supp}(Q_{X}).$

We note that 
\begin{align*}
&\EE[P]{v^{*}_{\text{regret}}(h(x); X, Y(0), Y(1)) \mid X=x} \\
&= \EE[P]{(2h(x)-1) \cdot (Y(1) - Y(0)) + \log(|2h(x) - 1|) \mid X=x} \\
&= (2h(x) - 1)\tau(x) + \log(|2h(x) - 1|)
\end{align*}
so the first term is identifiable under $P_{\text{obs}}$ and does not depend on the choice of $P \in \mathcal{T}(P_{\text{obs}}).$ Next, applying Lemma \ref{lemm:cvar_decomposition} and Lemma \ref{lemm:cvar_properties} we realize that
\begin{align*}
&\inf_{P \in \mathcal{T}(P_{\text{obs}})}  (1 - \Gamma) \text{CVaR}_{\zeta(\Gamma)}(v^{*}_{\text{regret}}(h(x); X, Y(0), Y(1)) \mid x) \\
&= (1 - \Gamma) \sup_{P \in \mathcal{T}(P_{\text{obs}})} \text{CVaR}_{\zeta(\Gamma)}(v^{*}_{\text{regret}}(h(x); X, Y(0), Y(1)) \mid x) \\
&= (1 - \Gamma) \Big(\sup_{P \in \mathcal{T}(P_{\text{obs}})} \text{CVaR}_{\zeta(\Gamma)}( (2h(x) -1) \cdot (Y(1) - Y(0)) \mid x) + \log(|2h(x) - 1|) \Big) \\
&= (1 - \Gamma) \Big( (2h(x) -1)_{+} (c_{1}^{+}(x) + c_{0}^{-}(x)) + (-2h(x) + 1)_{+}(c_{0}^{+} + c_{1}^{-}(x)) + \log(|2h(x) - 1|) \Big).
\end{align*}
We define
\begin{equation}
J(z; x) = (2z -1)_{+} \cdot g_{1}(x) + (-2z+1)_{+} \cdot g_{0}(x) + \log(|2z - 1|), 
\end{equation}
where $g_{1}(x), g_{0}(x)$ are as defined in Lemma \ref{lemm:g}. When $h(x) \geq \frac{1}{2},$ $J(h(x), x) = (2h(x) - 1) g_{1}(x) + \log(|2h(x) - 1|).$ When $h(x) \leq \frac{1}{2},$ $J(h(x), x) = (-2h(x) + 1) g_{0}(x) + \log(|2h(x) - 1|).$ We can rewrite \eqref{eq:cond_po_regret_loss} as
\begin{equation}
\begin{aligned}
\label{eq:regret_obj}
\max_{h(x) \in [0, 1]} J(h(x); x).
\end{aligned}
\end{equation}
Let $\tilde{h}(x)$ maximize \eqref{eq:regret_obj}. We must have that $\tilde{h}(x) \neq \frac{1}{2}$ because
\begin{align*} 
\lim_{h(x) \rightarrow \frac{1}{2}^{+}} J(h(x); x) &= -\infty \\
\lim_{h(x) \rightarrow \frac{1}{2}^{-}} J(h(x); x) &= -\infty.
\end{align*}
We can show that $\tilde{h}(x) > \frac{1}{2}$ if and only if  $g_{1}(x) > g_{0}(x).$ Note that a sufficient condition for $\tilde{h}(x) > \frac{1}{2}$ is that for any $z \in [0, \frac{1}{2})$, we can find $\tilde{z} \in (\frac{1}{2}, 1]$ such that $J(\tilde{z}, x) > J(z, x).$ We can show that $g_{1}(x) > g_{0}(x)$ implies this sufficient condition. Let $z \in [0, \frac{1}{2})$, then
\begin{align*}
J(z, x) &= (-2z + 1) \cdot g_{0}(x) + \log(|2z - 1|) \\
&= (2 (1 - z) - 1) \cdot g_{0}(x) + \log(|2(1-z) - 1|) \\
&< (2 (1 - z) - 1) \cdot g_{1}(x) + \log(|2(1-z) - 1|) \\
&= J( 1- z, x).
\end{align*}
We can define $\tilde{z} = 1-z.$ Thus, if $g_{1}(x) > g_{0}(x)$, then we have that $\tilde{h}(x) > \frac{1}{2}.$ We can also show that $\tilde{h}(x) > \frac{1}{2}$ implies that $g_{1}(x) > g_{0}(x).$ If we have that $\tilde{h}(x) > \frac{1}{2}$, then we have that $J(\tilde{h}(x); x) > J(1- \tilde{h}(x); x)$. This means that 
\begin{align*}
(2\tilde{h}(x) -1)_{+}g_{1}(x) + \log(|2\tilde{h}(x) - 1|) > (-2(1-\tilde{h}(x)) -1)_{+} g_{0}(x) + \log(|2(1-\tilde{h}(x)) - 1|).
\end{align*}
Cancelling terms and simplifying yields that $g_{1}(x) > g_{0}(x).$
\end{subsection}

\begin{subsection}{Proof of Lemma \ref{lemm:equal_value}}
\label{subsec:equal_value}

Let $P$ be any potential outcome distribution that is consistent with $P_{\text{obs}}$ under an RCT. We check each value function, realizing under $P_{\text{obs}}$ SUTVA and random treatment assignment hold.

First, we show that $v_{\text{maxmin}}, v_{\text{maxmin}}^{*}$ satisfy Assumption \ref{assumption:equal_value}.
\begin{align*}
&\EE[P_{\text{obs}}]{v_{\text{maxmin}}(z; X, Y, W)} \\
&= \EE[P_{\text{obs}}]{\log\left(1 + \exp\{(2z - 1) \cdot (2W-1)\}\right) \cdot \Big( \frac{Y}{W e + (1-W)(1-e)} \Big)} \\
&= \EE[P_{\text{obs}, X}]{\EE[P_{\text{obs}, Y, W\mid X}]{\log(1 + \exp\{(2z - 1) \cdot (2W-1)\}) \cdot \Big( \frac{Y}{W e + (1-W)(1-e)} \Big) \mid X}} \\
&= \EE[P_{\text{obs}, X}]{e \cdot \EE[P_{\text{obs}, Y \mid X, W}]{\log\left(1 + \exp\{2z - 1\}\right) \cdot \frac{Y(1)}{e} \mid X=x, W=1}} \\
&\indent+  \cdot \EE[P_{\text{obs}}, X]{(1-e) \cdot \EE[P_{\text{obs}, Y \mid X, W}]{\log\left(1 + \exp\{-2z + 1\}\right) \cdot \frac{Y(0)}{1-e}   \mid X=x, W=0} } \\
&= \EE[P_{\text{obs}, X}]{\log\left(1 + \exp\{2z - 1\} \right)\EE[P_{Y(1) \mid X}]{ Y(1) \mid X=x} + \log\left(1 + \exp\{-2z + 1\}\right) \EE[P_{Y(0) \mid X}]{Y(0) \mid X=x} } \\
&= \EE[P]{v^{*}_{\text{maxmin}}(z; X, Y(0), Y(1))}.
\end{align*}
Second, we show that $v_{\text{gain}}, v^{*}_{\text{gain}}$ satisfy Assumption \ref{assumption:equal_value}.
\begin{align*}
&\EE[P_{\text{obs}}]{v_{\text{gain}}(z; X, Y, W)} \\
&= \EE[P_{\text{obs}}]{ (1 - \pi_{0}(X)) \log(1 + \exp\{2z-1\}) \cdot \Big(\frac{Y \cdot W}{e} - \frac{Y \cdot (1 - W)}{1 -e}  \Big)} \\
&\indent+  \EE[P_{\text{obs}}]{\pi_{0}(X) \log(1 + \exp\{-2z+1\}) \cdot \Big(\frac{Y \cdot (1-W)}{1 -e} - \frac{Y \cdot W}{e}\Big)} \\
&= \EE[P_{\text{obs}, X}]{ e \cdot  \EE[P_{\text{obs}, Y \mid X, W}]{(1 - \pi_{0}(X)) \log(1 + \exp\{2z-1\}) \cdot \frac{Y(1)}{e} \mid X=x, W=1}} \\
&\indent- \EE[P_{\text{obs}, X}]{(1- e) \cdot \EE[P_{\text{obs}, Y \mid X, W}]{(1 - \pi_{0}(X)) \log(1 + \exp\{2z-1\}) \cdot \frac{Y(0)}{1 -e}\mid X=x, W=0}} \\
&\indent - \EE[P_{\text{obs}, X}]{ e \cdot  \EE[P_{\text{obs}, Y \mid X, W}]{\pi_{0}(X) \log(1 + \exp\{-2z+1\}) \cdot \frac{Y(1)}{e} \mid X=x, W=1}} \\
&\indent+ \EE[P_{\text{obs}, X}]{(1- e) \cdot \EE[P_{\text{obs}, Y \mid X, W}]{\pi_{0}(X) \log(1 + \exp\{-2z+1\}) \cdot \frac{Y(0)}{1-e} \mid X=x, W=0}} \\
&= \EE[P_{\text{obs}, X}]{(1-\pi_{0}(X))\log(1 + \exp\{2z-1\})\cdot \tau(X) - \pi_{0}(X) \cdot \log(1 + \exp\{-2z+1\})\cdot \tau(X) } \\
&= \EE[P]{v^{*}_{\text{gain}}(z; X, Y(0), Y(1))}
\end{align*}
Third, we show that $v_{\text{regret}}, v^{*}_{\text{regret}}$ satisfy Assumption \ref{assumption:equal_value}.
\begin{align*}
&\EE[P_{\text{obs}}]{v_{\text{regret}}(z; X, Y, W)} \\
&= \EE[P_{\text{obs}}]{(2z-1) \cdot \Big(\frac{Y\cdot W}{e} - \frac{Y \cdot (1-W)}{1 -e} \Big) + \log(|2z- 1|)} \\
&= \EE[P_{\text{obs}, X}]{ e \cdot \EE[P_{\text{obs}, Y \mid X, W}]{ (2z-1) \cdot \frac{Y(1)}{e} \mid X=x, W=1} } \\
&\indent- \EE[P_{\text{obs}, X}]{ (1- e) \cdot \EE[P_{\text{obs}, Y \mid X, W}] {(2z-1) \cdot \frac{Y(0)}{1 -e}\mid X=x, W=0} } + \log(|2z- 1|) \\
&= \EE[P_{\text{obs}, X}]{(2z-1)\tau(X)} + \log(|2z- 1|) \\
&= \EE[P]{v^{*}_{\text{regret}}(z; X, Y(0), Y(1))}
\end{align*}

\end{subsection}

%% file: 08-technical-proofs.tex
\begin{subsection}{Proof of Lemma \ref{lemm:proper_scoring_rule}}
  \label{subsec:proper_scoring_rule}
Let $f(z) = \log(1 + \exp\{2z - 1\})$. Then
    \[J(z) = A f(z) + Bf(1 - z).\]
    Clearly, $f(z)$ is strictly increasing in $z$. If $A > B$, for any $z > 1/2$, $J(z) > J(1 - z)$ and hence
    \[\max_{z > 1/2}J(z) > \max_{z < 1/2}J(z)\Longrightarrow z^{*} > 1/2.\]
    Similarly, if $A < B$,
    \[\max_{z > 1/2}J(z) < \max_{z < 1/2}J(z)\Longrightarrow z^{*} < 1/2.\]
    When $A = B$,
    \[J(z) = A\left( f(z) + f(1 - z)\right).\]
    It is easy to verify that $J$ is concave in $z$ and symmetric around $1/2$. Thus, $z^{*} = 1/2$.

\end{subsection}

\begin{subsection}{Proof of Lemma \ref{lemm:joint_sampling_bias}}
\label{subsec:joint_sampling_bias}
The proof of this result is identical to the proof of Lemma 1 from \cite{sahoo2022learning}.

\end{subsection}

\begin{subsection}{Proof of Lemma \ref{lemm:np}}
\label{subsec:np}
We note that any policy $\pi$ that solves \eqref{eq:potential_outcomes_problem} for every $Q_{X}$ such that $Q_{X} \ll P_{X}$ must solve the conditional problem for every $x$, i.e.,
\begin{equation} \label{eq:cond_pot} \sup_{\pi(x) \in \{0, 1\}} \inf_{Q \in \mathcal{S}^{\Gamma}(P_{\text{obs}}, Q_{X})} \EE[Q]{v^{*}(\pi(x); x, Y(0), Y(1)) \mid X=x}. \end{equation}

By the definition of $\mathcal{S}^{\Gamma}(P_{\text{obs}}, Q_{X})$, we have that
\begin{align*}
  &\inf_{Q \in \mathcal{S}^{\Gamma}(P_{\text{obs}}, Q_{X})} \EE[Q]{v^{*}(\pi(x); x, Y(0), Y(1)) \mid X=x}\\
  & =  \inf_{P \in \mathcal{T}(P_{\text{obs}})} \inf_{Q \in \mathcal{R}^{\Gamma}(P, Q_{X})} \EE[Q]{v^{*}(\pi(x); x, Y(0), Y(1)) \mid X=x}.
\end{align*}

Recall that $Q \in \mathcal{R}^{\Gamma}(P, Q_{X})$ if $Q$ has covariate distribution $Q_{X}$ and generates $P$ via $\Gamma$-biased sampling. For every $Q_{X}$ such that $Q_{X} \ll P_{X}$ and $\sup_{x \in \mathcal{X}} \frac{dP_{\text{obs}, X}(x)}{dQ_{X}(x)} < \infty$, we have that \eqref{eq:joint_likelihood_ratio} holds. Using this result, we observe that finding the worst-case distribution $Q^{*}_{Y(0), Y(1) \mid X}$ that solves \[\inf_{Q \in \mathcal{R}^{\Gamma}(P, Q_{X})} \EE[Q]{v^{*}(\pi(x); x, Y(0), Y(1)) \mid X=x}\] amounts to minimizing a concave (linear) function in $Q$ subject to convex (linear) constraints in $Q$. By Proposition \ref{prop:sup_extremal_point}, the solution must occur at an extremal point of the feasible set.
Since $P_{Y(0), Y(1)\mid X = x}$ is absolutely continuous with respect to the Lebesgue measure, we can show that $\frac{dQ^{*}_{Y(0), Y(1) \mid X=x}(y_{0}, y_{1})}{dP_{Y(0), Y(1) \mid X=x}(y_{0}, y_{1})} \in \{ \Gamma^{-1}, \Gamma \}.$  

To determine the structure of the optimal solution $dQ^{*}_{Y(0), Y(1) \mid X},$ we consider the distribution over $v^{*}(\pi(x); x, Y(0), Y(1))$ where $Y(0), Y(1) \sim P_{Y(0), Y(1) \mid X=x}.$ We must assign a weight $\Gamma^{-1}$ to values of $Y(0), Y(1)$ that correspond to $v^{*}(\pi(x); x, Y(0), Y(1))$ falling above the $\zeta$-th quantile for some $\zeta \in (0, 1)$ and a weight of $\Gamma$ to values of $Y(0), Y(1)$ that correspond to $v^{*}(\pi(x);x, Y(0), Y(1))$ falling below a the $\zeta$-th quantile. Otherwise, there would exist a $Q$ that obtains lower policy value than $Q^{*}.$ The choice of quantile $\zeta$ is set to ensure that $dQ^{*}_{Y(0), Y(1) \mid X}$ is a valid probability distribution. We pick $\zeta$ that satisifes
\[ \Gamma^{-1} \zeta + \Gamma (1 - \zeta) = 1.\]
So, we express $\zeta$ as a function of $\Gamma$ as follows: $\zeta(\Gamma) = \frac{1}{\Gamma + 1}.$ Let the $\zeta(\Gamma)$-th quantile of $v^{*}(\pi(x); x, Y(0), Y(1))$ where $Y(0), Y(1) \sim P_{Y(0), Y(1) \mid X=x}$ be given by $q_{\zeta(\Gamma)}(v^{*}(\pi(x); X, Y(0), Y(1)) \mid X=x).$ Thus, we can conclude that the worst-case distribution is given by 

\begin{equation}
dQ^{*}_{Y(0), Y(1) \mid X=x}(y_{0}, y_{1}) = \begin{cases} \Gamma^{-1} \cdot dP_{Y(0), Y(1) \mid X=x}(y_{0}, y_{1}) & v^{*}(\pi(x); x, y_{0}, y_{1}) \geq q_{\zeta(\Gamma)}(v^{*}(\pi(x); X, Y(0), Y(1))\mid x) \\ \Gamma \cdot dP_{Y(0), Y(1) \mid X=x}(y_{0}, y_{1}) & \text{o.w.}\end{cases}
\end{equation}

We can write that
\begin{align*}
&\inf_{Q \in \mathcal{R}^{\Gamma}(P, Q_{X})} \EE[P]{v^{*}(\pi(x); x, Y(0), Y(1)) \mid X=x} \\
&= \Gamma^{-1} \EE[Q]{v^{*}(\pi(x); x, Y(0), Y(1)) \mathbb{I}(v^{*}(\pi(x); x, Y(0), Y(1)) \geq q_{\zeta(\Gamma)}(v^{*}(\pi(x); X, Y(0), Y(1))) \mid x) \mid X=x}  + \\
&\indent+ \Gamma \EE[P]{v^{*}(\pi(x); x, Y(0), Y(1)) \mathbb{I}(v^{*}(\pi(x); x, Y(0), Y(1)) \leq q_{\zeta(\Gamma)}(v^{*}(\pi(x); X, Y(0), Y(1))) \mid x) \mid X=x} \\
&= \Gamma \EE[P]{v^{*}(\pi(x); x, Y(0), Y(1)) \mid X=x} + (1- \Gamma) \cdot \text{CVaR}_{\zeta(\Gamma)}(v^{*}(\pi(x); x, Y(0), Y(1)) \mid x).
\end{align*}

Thus, $\pi$ that solves must solve \eqref{eq:cond_pot} must solve
\[ \sup_{\pi(x) \in \{0, 1\}} \inf_{P \in \mathcal{T}(P_{\text{obs}})} \Gamma \EE[P]{v^{*}(\pi(x); x, Y(0), Y(1)) \mid X=x} + (1- \Gamma) \cdot \text{CVaR}_{\zeta(\Gamma)}(v^{*}(\pi(x); x, Y(0), Y(1)) \mid x).\]
as desired. 

We demonstrate that the converse of this result also holds. Suppose that $\tilde{\pi}$ solves \eqref{eq:np_cond} for every $x \in \text{supp}(P_{\text{obs}, X})$. By the previous argument, this implies that $\tilde{\pi}$ solves \eqref{eq:cond_pot} for every $x \in \text{supp}(P_{\text{obs}, X})$. Let $\mathcal{T}$ be any set with nonzero measure with respect to $Q_{X}$. Then $\mathcal{T}$ must have nonzero measure with respect to $P_{\text{obs}}$ because $Q_{X} \ll P_{X}.$ So, 
\[ \inf_{Q \in \mathcal{S}^{\Gamma}(P_{\text{obs}}, Q_{X})} \EE[Q]{v^{*}(\tilde{\pi}(X); X, Y(0), Y(1)) \mid X \in \mathcal{T}} \geq \inf_{Q \in \mathcal{S}^{\Gamma}(P_{\text{obs}}, Q_{X})} \EE[Q]{v^{*}(\pi(X); X, Y(0), Y(1)) \mid X \in \mathcal{T}} \]
for any $\pi \in \Pi$ and any set $\mathcal{T}$ with nonzero measure with respect to $Q_{X}$. This is sufficient to show that $\tilde{\pi}$ must solve \eqref{eq:potential_outcomes_problem}.

The same argument also applies to show that $h$ solves \eqref{eq:continuous_potential_outcomes} for every $Q_{X}$ such that $Q_{X} \ll P_{\text{obs}, X}$ and $\sup_{x \in \mathcal{X}} \frac{dP_{X}(x)}{dQ_{X}(x)} < \infty$ iff it solves \eqref{eq:np_continuous_cond} for every $x \in \text{supp}(P_{\text{obs}, X}).$

\end{subsection}

\begin{subsection}{Proof of Lemma \ref{lemm:wc_gain}}
\label{subsec:wc_gain}
By Lemma \ref{lemm:np}, we have that  
\begin{equation} \label{eq:np2} \inf_{P \in \mathcal{T}(P_{\text{obs}})} \Gamma \EE[P]{v^{*}(\pi(x), \pi_{0}(X); X, Y(0), Y(1)) \mid X=x} + (1- \Gamma) \text{CVaR}(v^{*}(\pi(x), \pi_{0}(X); X, Y(0), Y(1)) \mid x).\end{equation}
We note that 
\begin{align*}
\EE[P]{v^{*}(\pi(x), \pi_{0}(X); X, Y(0), Y(1)) \mid X=x} &= (\pi(x) - \pi_{0}(x)) \cdot \tau(x) \\
&= (\pi(x) - \pi_{0}(x))_{+} \tau(x) - (\pi_{0}(x) - \pi(x))_{+} \tau(x),
\end{align*}
which is identified under $P_{\text{obs}}$ and does not depend on the choice of $P \in \mathcal{T}(P_{\text{obs}}).$ So, we can rewrite \eqref{eq:np2} as
\begin{equation}
\label{eq:a2}
(\pi(x) - \pi_{0})_{+} \tau(x) - (\pi_{0}(x) - \pi(x))_{+} \tau(x) + \inf_{P \in \mathcal{T}(P_{\text{obs}})} (1- \Gamma) \text{CVaR}_{\zeta(\Gamma)}(v^{*}(\pi(x), \pi_{0}(X); X, Y(0), Y(1)) \mid x).
\end{equation}
We have that
\begin{align}
&\inf_{P \in \mathcal{T}(P_{\text{obs}})} (1- \Gamma) \text{CVaR}_{\zeta(\Gamma)}(v^{*}(\pi(x), \pi_{0}(X); X, Y(0), Y(1)) \mid x) \\
&= (1- \Gamma) \sup_{P \in \mathcal{T}(P_{\text{obs}})}  \text{CVaR}_{\zeta(\Gamma)}(v^{*}(\pi(x), \pi_{0}(X); X, Y(0), Y(1)) \mid x) \\
&= (1- \Gamma) \sup_{P \in \mathcal{T}(P_{\text{obs}})}  \text{CVaR}_{\zeta(\Gamma)}((\pi(x) - \pi_{0}(x)) \cdot(Y(1) - Y(0)) \mid x) \\
&= (1- \Gamma) \sup_{P \in \mathcal{T}(P_{\text{obs}})}  (\pi(x) - \pi_{0}(x))_{+}\text{CVaR}_{\zeta(\Gamma)}(Y(1) - Y(0)) \mid x) + (\pi_{0}(x) - \pi(x))_{+} \text{CVaR}_{\zeta(\Gamma)}(Y(0) - Y(1) \mid x) \label{eq:pos3} \\
&= (1- \Gamma) \Big((\pi(x) - \pi_{0}(x))_{+} (c_{1}^{+}(x) + c_{0}^{-}(x)) + (\pi_{0}(x) - \pi(x))_{+} (c_{0}^{+}(x) + c_{1}^{-}(x)) \Big). \label{eq:decomp3}
\end{align}
\eqref{eq:pos3} follows from Lemma \ref{lemm:cvar_properties} and \eqref{eq:decomp3} follows from Lemma \ref{lemm:cvar_decomposition}. Plugging this expression into \eqref{eq:a2} yields \eqref{eq:gain_conditional}.
\end{subsection}

\begin{subsection}{Proof of Lemma \ref{lemm:g}}
\label{subsec:g}
We have that 
\begin{align*}
g_{1}(x) &= \Gamma \tau(x) + (1- \Gamma) (c_{1}^{+}(x) + c_{0}^{-}(x))  \\
&= \Gamma \tau(x) + ( 1- \Gamma) \Big( \EE[P]{Y(1) \mid Y(1) \geq q_{\zeta(\Gamma)}(Y(1) \mid x), X=x} + \EE[P]{-Y(0) \mid -Y(0) \geq q_{\zeta(\Gamma)}(-Y(0) \mid x), X=x} \Big) \\
&\leq \Gamma  \tau(x) + (1-\Gamma) \EE[P]{Y(1) - Y(0) \mid X=x} \\
&= \tau(x).
\end{align*}
Furthermore, we also have that
\begin{align*}
g_{0}(x) &= -\Gamma \tau(x) + (1- \Gamma) (c_{1}^{-}(x) + c_{0}^{+}(x))  \\
&= -\Gamma \tau(x) + (1- \Gamma) \Big( \EE[P]{Y(0) \mid Y(0) \geq q_{\zeta(\Gamma)}(Y(1) \mid x), X=x} + \EE[P]{-Y(1) \mid -Y(1) \geq q_{\zeta(\Gamma)}(-Y(1) \mid x), X=x} \Big) \\
&\leq -\Gamma \tau(x) +  (1 - \Gamma) \EE[P]{Y(0) - Y(1) \mid X=x} \\
&= -\Gamma \tau(x) + (1- \Gamma) (-\tau(x)) \\
&= -\tau(x).
\end{align*}
Thus, we have that $\min\{g_{1}(x), g_{0}(x)\} \leq \min\{\tau(x), -\tau(x)\} \leq 0.$ Moreover, if $g_{1}(x) = 0,$ then $\tau(x) \ge 0$ and hence $g_{0}(x)\le 0$.

\end{subsection}

\begin{subsection}{Proof of Lemma \ref{lemm:potential_robustness_implies_observed_robustness}}
\label{subsec:potential_robustness_implies_observed_robustness}
To show that $Q_{\text{obs}} \in \mathcal{S}_{\text{obs}}^{\Gamma}(P_{\text{obs}}, Q_{X})$, we first verify \eqref{eq:likelihood_ratio_observed_data}. Since $Q \in \mathcal{S}^{\Gamma}(P_{\text{obs}}, Q_{X})$, $Q$ generates a potential outcome distribution $P$ via $\Gamma$-biased sampling which is consistent with $P_{\text{obs}}$ under an RCT with treatment propensity function $e$. Since $Q$ generates $P$ under $\Gamma$-biased sampling, then \eqref{eq:likelihood_ratio_potential_outcomes_single_conditional} holds by Lemma \ref{lemm:conditional_sampling_bias}. Since $P$ is consistent with $P_{\text{obs}}$ under an RCT, we have that  
\[ P_{\text{obs}, Y \mid X=x, W=w} = P_{Y(w) \mid X=x} \quad \forall x \in \mathcal{X}, w\in \{0, 1\}.\]
In addition, because $Q$ is consistent with $Q_{\text{obs}}$ under an RCT, we have that 
\[ Q_{\text{obs}, Y \mid X=x, W=w} = Q_{Y(w) \mid X=x} \quad \forall x \in \mathcal{X}, w \in \{0, 1\}.\]
Substituting the above two equations into \eqref{eq:likelihood_ratio_potential_outcomes_single_conditional} yields \eqref{eq:likelihood_ratio_observed_data}, as desired. 

Next, we verify that $Q_{\text{obs}, W \mid X} = P_{\text{obs}, W \mid X}.$ By assumption, $P_{\text{obs}, W \mid X} = \text{Bernoulli}(e).$ In addition, since $Q_{\text{obs}}$ is generated with treatment propensity function $e$, we have that $Q_{\text{obs}, W \mid X} = \text{Bernoulli}(e).$ Thus, $Q_{\text{obs}, W\mid X} = P_{\text{obs}, W \mid X}.$

Lastly, we note that $Q_{\text{obs}, X} = Q_{X}$ because running an RCT does not affect the covariate distribution of the data.

Thus, we conclude that $Q_{\text{obs}} \in \mathcal{S}^{\Gamma}_{\text{obs}}(P_{\text{obs}}, Q_{X}).$ 

\end{subsection}

\begin{subsection}{Proof of Lemma \ref{lemm:np_obs_data}}
\label{subsec:np_obs_data}
We can solve \eqref{eq:wc_dist} conditionally for every $x$:
\begin{equation}
\label{eq:cond_prob}
\inf_{Q_{\text{obs}} \in \mathcal{S}_{\text{obs}}^{\Gamma}(P_{\text{obs}}, Q_{X})} \EE[Q_{\text{obs}, Y, W \mid X=x}]{v(h(X); X, Y, W) \mid X=x}. \\
\end{equation}

Taking the optimization variable to be $dQ_{\text{obs}, Y, W \mid X},$ we can view the above problem as the minimization of a linear function subject to convex constraints. These constraints include that 
\begin{enumerate}
	\item $Q_{\text{obs}, Y, W \mid X}$ is a valid probability distribution, so $dQ_{\text{obs}, Y, W}(y, w) \geq 0$ and $\int dQ_{\text{obs}, Y, W}(y, w) = 1$.
	\item Treatment assignment is random. $dQ_{\text{obs}, W \mid X} = dP_{\text{obs}, W \mid X} = \text{Bernoulli}(e).$
	\item $\Gamma^{-1} \leq \frac{dQ_{\text{obs}, Y \mid X=x, W=w}(y)}{dP_{\text{obs}, Y \mid X=x, W=w}(y)} \leq \Gamma$ for all $w \in \{0, 1\}, x \in \mathcal{X}$.
\end{enumerate}
The supremum of a convex function over a closed, bounded, convex set exists and is achieved at some extreme point of the feasible set (Proposition \ref{prop:sup_extremal_point}). Since $P_{Y(0), Y(1)\mid X = x}$ is absolutely continuous with respect to the Lebesgue measure, the optimal solution must have $\frac{dQ^{*}_{\text{obs}, Y \mid X=x, W=w}(y)}{dP_{\text{obs}, Y \mid X=x, W=w}(y)} \in \{\Gamma^{-1}, \Gamma\}$ and satisfy the remaining constraints, in particular that $dQ^{*}_{\text{obs}, W \mid X}= \text{Bernoulli}(e)$.

To determine of the structure of the optimal solution $dQ^{*}_{\text{obs}, Y \mid X, W}$, we can consider the distribution over $v(h(x); x, Y, w)$ where $Y \sim P_{\text{obs}, Y \mid X=x, W=w}$. We must assign a weight of $\Gamma^{-1}$ to values of $Y$ that correspond to values $v(h(x); x, Y, w)$ that fall above a particular $\zeta$-th quantile for some $\zeta \in (0, 1)$ and a weight of $\Gamma$ to values of $Y$ that correspond to values $v(h(x); x, Y, w)$ that fall below this threshold. Otherwise, there would exist a $dQ_{\text{obs}}$ under that attains lower policy value than $dQ^{*}_{\text{obs}}.$ To choice of quantile $\zeta$ is set to ensure that $dQ^{*}_{\text{obs}}$ is a valid probability distribution. We pick $\zeta$ that satisfies
\[ \Gamma^{-1} \zeta + \Gamma (1 - \zeta) = 1.\]
So, we express $\zeta$ as a function of $\Gamma$ as follows $\zeta(\Gamma) = \frac{1}{\Gamma + 1}.$
Furthermore, let the $\zeta(\Gamma)$-th quantile of $v(h(x); x, Y, w)$ be given by $q_{\zeta(\Gamma)}(v(h(X); X, Y, W) \mid x, w).$ Thus, we can conclude that the worst-case conditional distribution is given by \eqref{eq:worst_case_distribution}
. 
Thus, the distribution $dQ_{\text{obs}}^{*}$ that yields the infimum in \eqref{eq:wc_dist} is given by $dQ_{X} \cdot dP_{\text{obs} W \mid X} \cdot dQ^{*}_{\text{obs}, Y \mid X, W},$ where $dQ^{*}_{\text{obs}, Y \mid X, W}$ is as defined in  \eqref{eq:worst_case_distribution}.
\end{subsection}

\begin{subsection}{Proof of Lemma \ref{lemm:conditional_sampling_bias}}
\label{subsec:conditional_sampling_bias}
First, we assume that $Q$ generates $P$ via $\Gamma$-biased sampling. By Lemma \ref{lemm:joint_sampling_bias}, we have that the density ratio between the covariate distributions is bounded. In addition, we also have that \begin{align}
\frac{dP_{Y(0) \mid X=x}(y)}{dQ_{Y(0) \mid X=x}(y)} &= \frac{\PP[\tilde{Q}]{S=1 \mid X=x, Y(0) = y}}{\PP[\tilde{Q}]{S=1 \mid X=x}} \label{eq:prev_result} \\
&= \EE[\tilde{Q}_{Y(1) \mid X=x, Y(0)=y}]{\frac{\PP[\tilde{Q}]{S=1 \mid X=x, Y(0) = y, Y(1)}}{\PP[\tilde{Q}]{S=1 \mid X=x}}} \label{eq:tower}\\
&\in [\Gamma^{-1}, \Gamma] \label{eq:apply_sampling_bias_2}.
\end{align}
Using an identical argument as in the proof of Lemma 1 of \cite{sahoo2022learning}, we arrive at \eqref{eq:prev_result}. We apply \eqref{eq:sampling_bias_intro} to the quantity within in the expectation in \eqref{eq:tower} to get \eqref{eq:apply_sampling_bias_2}. The same argument applies to $\frac{dQ_{Y(1) \mid X=x}}{dP_{Y(1) \mid X=x}}$. 
\end{subsection}